\documentstyle[11pt]{article}
\newcommand{\room}{\rule[-0.1cm]{0cm}{0.6cm}}

\newcommand{\vsp}{\vspace*{3mm}}
\newcommand{\be}{\begin{equation}}
\newcommand{\ee}{\end{equation}}
\newcommand{\bd}{\begin{displaymath}}
\newcommand{\ed}{\end{displaymath}}
\newcommand{\bdm}{\begin{displaymath}}
\newcommand{\edm}{\end{displaymath}}
\newcommand{\bea}{\begin{eqnarray}}
\newcommand{\eea}{\end{eqnarray}}
\newcommand{\sgn}{~{\rm sgn}}
\newcommand{\inn}{{\vspace*{-0.1mm}\cdot\vspace*{-0.1mm}}}

\newcommand{\pprime}{{\prime\prime}}

\newcommand{\bra}{\langle}
\newcommand{\ket}{\rangle}
\newcommand{\bigbra}{\left\langle\room}
\newcommand{\bigket}{\right\rangle\room}

\newcommand{\order}{{\cal O}}
\newcommand{\minus}{\!-\!}
\newcommand{\plus}{\!+\!}
\newcommand{\erf}{{\rm erf}}
\newcommand{\bnul}{\mbox{\boldmath $0$}}

\newcommand{\bk}{\mbox{\boldmath $k$}}

\newcommand{\bq}{\mbox{\boldmath $q$}}

\newcommand{\bx}{\mbox{\boldmath $x$}}
\newcommand{\by}{\mbox{\boldmath $y$}}

\newcommand{\bB}{\mbox{\boldmath $B$}}
\newcommand{\bF}{\mbox{\boldmath $F$}}

\newcommand{\bJ}{\mbox{\boldmath $J$}}

\newcommand{\bQ}{\mbox{\boldmath $Q$}}

\newcommand{\hQ}{\hat{Q}}
\newcommand{\hR}{\hat{R}}

\newcommand{\hbq}{\hat{\mbox{\boldmath $q$}}}

\newcommand{\hbJ}{\hat{\mbox{\boldmath $J$}}}
\newcommand{\hbQ}{\hat{\mbox{\boldmath $Q$}}}
\newcommand{\hbR}{\hat{\mbox{\boldmath $R$}}}

\newcommand{\bsigma}{\mbox{\boldmath $\sigma$}}

\newcommand{\bOmega}{\mbox{\boldmath $\Omega$}}

\newcommand{\bpsi}{\mbox{\boldmath $\psi$}}

\newcommand{\bxi}{\mbox{\boldmath $\xi$}}

\newcommand{\A}{{\cal A}}
\newcommand{\cA}{{\cal A}}

\newcommand{\cD}{{\cal D}}

\newcommand{\cF}{{\cal F}}

\newcommand{\cL}{{\cal L}}
\newcommand{\opt}{{\rm opt}}
\newcommand{\set}{{\tilde{D}}}

\begin{document}

\title{\bf Statistical Mechanical Analysis of the\\ Dynamics of Learning in Perceptrons}
\author{C.W.H. Mace~~and~~A.C.C. Coolen\\[3mm]
Department of Mathematics\\
King's College London\\
Strand, London WC2R 2LS, U.K.}
\maketitle

\begin{abstract}\noindent
We describe the application of tools from statistical mechanics to analyse 
the dynamics of various classes of supervised  
learning rules in perceptrons. The character of this paper is mostly that 
of a cross between a  biased non-encyclopedic review and lecture notes: 
we try to present a coherent and self-contained picture of the basics
of this field, to explain the 
ideas and tricks, to show how the predictions of the 
theory compare with (simulation) experiments, and to bring together 
scattered results. 
Technical details are given explicitly in an appendix. 
In order to avoid distraction we concentrate the 
references in a final section. 
In addition this paper contains some new results:  
$(i)$ explicit solutions of the macroscopic equations that describe the 
error  evolution for on-line and batch learning rules, $(ii)$ 
an analysis of 
the dynamics of arbitrary macroscopic observables (for complete and incomplete training
sets), leading to a general Fokker-Planck equation, and $(iii)$ 
the macroscopic laws describing batch learning with complete training
sets. We close the paper with 
a preliminary expos\'{e} of ongoing research on the 
dynamics of learning for the case where the training set is 
incomplete
(i.e. where the number of examples scales linearly with the network size).
\end{abstract}
\vspace*{20mm}
\begin{center}
{\em (to be published in `Statistics and Computing')}
\end{center}

\pagebreak
\tableofcontents


\pagebreak
\section{Introduction}

\subsection{Supervised Learning in Neural Networks}

In this paper we study the dynamics of supervised
learning in artificial neural networks. The basic scenario is as follows. A 
`student' neural network
executes a certain known operation $S:D\to R$,
which is parametrised by a vector $\bJ$, 
usually representing synaptic weights and/or neuronal thresholds.  
Here $D$ denotes the set of all possible `questions' and
$R$ denotes the set of all possible `answers'. 
The student is being trained to emulate a given `teacher', which executes 
some as yet
unknown 
operation $T:D\to R$. 
In order to achieve the objective the student network $S$ tries to gradually 
improve its performance by adapting its parameters 
$\bJ$ according to an iterative procedure, using only 
examples of input vectors (or `questions') 
$\bxi\in\Re^N$ which are drawn at random from a fixed training set
$\tilde{D}\subseteq D$ of size $|\set|$, and the corresponding values of the 
teacher outputs $T(\bxi)$ (the `correct answers'). The 
iterative procedure (the `learning rule') is not allowed to involve any 
further knowledge of the operation $T$.
\begin{figure}[h]
\setlength{\unitlength}{0.09mm}
\newcommand{\sq}{\thinlines\framebox(10,10)}
\newcommand{\here}{\makebox(0,0)}
\begin{picture}(1300,750)(-100,350)
\newcommand{\Smallsize}{130}
\newcommand{\Size}{270}
\put(50,1000){\here{$\xi_1$}}
\put(50,950){\here{$.$}}
\put(50,900){\here{$.$}}
\put(50,850){\here{$.$}}
\put(50,800){\here{$.$}}
\put(50,750){\here{$.$}}
\put(50,700){\here{$.$}}
\put(50,650){\here{$.$}}
\put(50,600){\here{$.$}}
\multiput(100,390)(0,50){13}{\sq}
\put(50,550){\here{$.$}}
\put(50,500){\here{$.$}}
\put(50,450){\here{$.$}}
\put(50,400){\here{$\xi_N$}}
\put(400,800){\framebox(600,200)}
\put(700,950){\here{\large Student (Neural Network)}}
\put(700,850){\here{$S(\bxi)=f[\bxi;\bJ]$}}
\put(1010,900){\line(1,0){220}}
\put(1010,900){\vector(1,0){110}}
\put(1300,900){\here{$S(\bxi)$}}
\put(120,400){\line(3,1){\Size}}
\put(120,500){\line(3,4){\Size}}
\put(120,600){\line(1,1){\Size}}
\put(120,700){\line(3,2){\Size}}
\put(120,800){\line(3,1){\Size}}
\put(120,900){\line(1,0){\Size}}
\put(120,1000){\line(3,-1){\Size}}
\put(120,400){\line(3,5){\Size}}
\put(120,500){\line(1,0){\Size}}
\put(120,600){\line(3,-1){\Size}}
\put(120,700){\line(3,-2){\Size}}
\put(120,800){\line(1,-1){\Size}}
\put(120,900){\line(3,-4){\Size}}
\put(120,1000){\line(3,-5){\Size}}
\put(400,400){\framebox(600,200)}
\put(700,550){\here{\large Teacher}}
\put(700,450){\here{\large ?}}
\put(1010,500){\line(1,0){220}}
\put(1010,500){\vector(1,0){110}}
\put(1300,500){\here{$T(\bxi)$}}
\put(120,400){\vector(3,1){\Smallsize}}
\put(120,500){\vector(3,4){\Smallsize}}
\put(120,600){\vector(1,1){\Smallsize}}
\put(120,700){\vector(3,2){\Smallsize}}
\put(120,800){\vector(3,1){\Smallsize}}
\put(120,900){\vector(1,0){\Smallsize}}
\put(120,1000){\vector(3,-1){\Smallsize}}
\put(120,500){\vector(1,0){\Smallsize}}
\put(120,600){\vector(3,-1){\Smallsize}}
\put(120,700){\vector(3,-2){\Smallsize}}
\put(120,800){\vector(1,-1){\Smallsize}}
\put(120,900){\vector(3,-4){\Smallsize}}
\end{picture}
\caption{The general scenario of supervised learning: a 
`student network' $S$ is being `trained' to perform an operation $T:D\to\ R$ 
by updating its control parameters $\bJ$ according to an iterative 
procedure, the `learning rule'. This rule is allowed to make use 
only of examples 
of `question/answer pairs' $(\bxi,T(\bxi))$, where $\bxi\in\set\subseteq D$. 
The actual `teacher operation' $T$ that generated the answers $T(\bxi)$, on the other hand, cannot 
be observed directly. The goal is to arrive at a situation where $S(\bxi)=T(\bxi)$ for all $\bxi\in D$.}
\label{fig:scenario}
\end{figure}
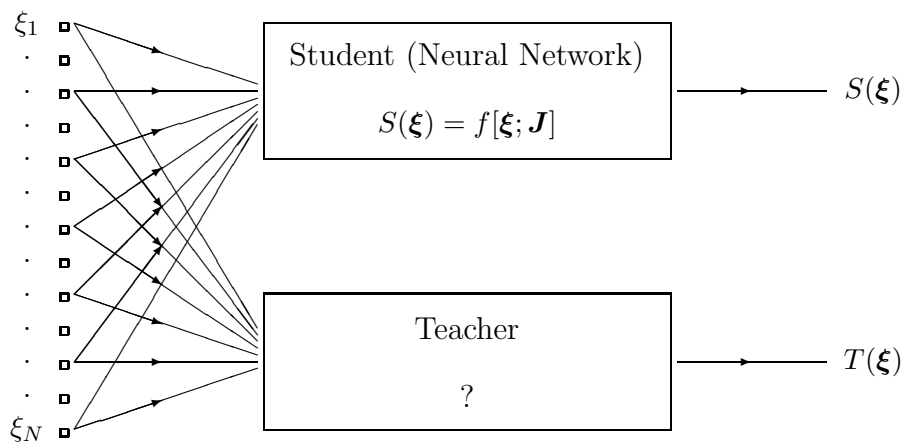
As far as the student is concerned 
the teacher is an `oracle', or `black box'; the only information available 
about the inner workings of the black box is contained in the various 
answers $T(\bxi)$ it provides. 
See figure 
\ref{fig:scenario}. 
For simplicity we will assume each `question' $\bxi$ to be equally likely 
to occur (generalization of what follows to the case where the questions $\bxi$ carry non-uniform probabilities or probability densities $p(\bxi)$ 
is straightforward). 

We will consider the following two 
classes of learning rules, i.e. of recipes for the iterative modification of the student's control parameters $\bJ$, which we will refer to as  
on-line learning rules and batch learning rules, respectively:
\be
\begin{array}{ll}
{\rm On\!\!-\!Line:} & 
\bJ(t\plus 1)=
\bJ(t)+
\bF\left[\room\bxi(t),\bJ(t),T(\bxi(t))\right]
\\[3mm]
{\rm Batch:} & 
\bJ(t\plus 1)=
\bJ(t)+
\bra \bF\left[\room\bxi,\bJ(t),T(\bxi)\right]\ket_\set
\end{array}
\label{eq:weightdynamics}
\ee
The integer variable $t=0,1,2,3,\ldots$ labels the iteration steps. 
In the case of on-line learning an input  vector $\bxi(t)$ is drawn independently 
at each iteration step from the training set $\set$, followed by a modification of the control parameters $\bJ$. Therefore this process is stochastic (Markovian). 
In the case of batch learning the modification that would have been made in the 
on-line version  
is averaged over the input vectors in the training set $\set$, 
at each iteration step. 
This process is therefore a deterministic iterative
map\footnote{Clearly one could define an infinite number of
intermediate classes of learning rules (e.g. learning with
`momentum'); the present two are just the extreme cases.  Note also
that the term `batch' unfortunately means different things to
different scientists. The definition used here is  
sometimes described as `off-line'.}.
Both rules in  (\ref{eq:weightdynamics}) 
can formally be written in the general form 
of a Markovian stochastic process. 
We introduce the 
probability density $\hat{p}_{t}(\bJ)$ 
to find parameter vector $\bJ$ at discrete iteration step  $t$. In terms
of this microscopic probability density the processes
(\ref{eq:weightdynamics}) can be written as:
\be
\hat{p}_{t+1}(\bJ)=
\int\!d\bJ^\prime~W[\bJ;\bJ^\prime]\hat{p}_t(\bJ^\prime)
\label{eq:markovprocess}
\ee
with the transition probability densities
\be
\begin{array}{ll}
{\rm On\!\!-\!Line:} & 
W[\bJ;\bJ^\prime]=\bra \delta\left\{
\bJ\minus \bJ^\prime\minus 
\bF\left[\room\bxi,\bJ^\prime,T(\bxi)\right]\right\}\ket_\set
\\[3mm]
{\rm Batch:} & 
W[\bJ;\bJ^\prime]=\delta\left\{
\bJ\minus \bJ^\prime\minus 
\bra \bF\left[\room\bxi,\bJ^\prime,T(\bxi)\right]\ket_\set\right\}
\end{array}
\label{eq:transitionmatrix}
\ee
(in which $\delta[z]$ denotes the delta-distribution). 
The advantage of using the on-line version of the learning rule is a reduction in the amount of 
calculations that have to be done at each iteration step; the price paid for this 
reduction is the presence of fluctuations, with as yet unknown impact on the performance of the system. 
\vsp

 We will denote averages over the probability density $\hat{p}_t(\bJ)$, averages over the full set $D$ of possible input vectors 
 and averages over the training set $\set$  
in the following way:
\bd
\bra g(\bJ)\ket=\int\!d\bJ~\hat{p}_t(\bJ) g(\bJ)
~~~~~~~~~~
\bra K(\bxi)\ket_{D}=\frac{1}{|D|}\sum_{\bxi\in D} K(\bxi)
~~~~~~~~~~
\bra K(\bxi)\ket_{\set}=\frac{1}{|\set|}\sum_{\bxi\in \set} K(\bxi)
\ed
The average $\bra K(\bxi)\ket_\set$ will in general 
depend on the microscopic
realisation of the training set $\set$. 
To quantify the goal and the progress of the student one finally defines an error 
$E[T(\bxi),S(\bxi)]=E[T(\bxi),f[\bxi;\bJ]]$, which measures the mismatch between student answers and correct (teacher) answers for individual questions. The two key quantities of interest in supervised learning are the (time-dependent) averages of this error measure, calculated over the training set $\set$ and the full question set $D$, respectively:
\be
\begin{array}{lll}
{\rm Training~ Error:} &&
E_{\rm t}(\bJ)=\bra E[T(\bxi),f[\bxi;\bJ]]\ket_\set
\\[3mm]
{\rm Generalization ~Error:} &&
E_{\rm g}(\bJ)=\bra E[T(\bxi),f[\bxi;\bJ]]\ket_D
\end{array}
\label{eq:errors}
\ee
These quantities are stochastic observables, since they are functions of the stochastically evolving vector $\bJ$. Their expectation values over the stochastic process 
(\ref{eq:markovprocess}) are given by
\be
\begin{array}{lll}
{\rm Mean~Training~ Error:} &&
\bra E_{\rm t}\ket =\bra \bra E[T(\bxi),f[\bxi;\bJ]]\ket\ket_\set
\\[3mm]
{\rm Mean~Generalization ~Error:} &&
\bra E_{\rm g}\ket =\bra \bra E[T(\bxi),f[\bxi;\bJ]]\ket\ket_D
\end{array}
\label{eq:expectederrors}
\ee
Note that the prefix `mean' refers to the 
stochasticity in the vector $\bJ$; both $\bra E_{\rm t}\ket$ and $\bra E_{\rm g}\ket$ will in general still depend on the realisation of the training set $\set$. 

The training error measures the performance 
of the student on the questions it could have been confronted with during the learning stage (in the case of on-line learning the student need not have seen all of them). The generalization error measures the student's performance on the full question set and its minimisation is therefore the main target of the process. The quality of a theory describing the dynamics of supervised learning can be measured by the degree to which it succeeds in predicting 
the values of $\bra E_{\rm t}\ket$ and $\bra E_{\rm g}\ket$ as a function of the iteration time $t$ and for arbitrary choices made for the function $F[\ldots]$ that determines the details of the learning rules
 (\ref{eq:weightdynamics}).

\subsection{Statistical Mechanics and Its Applicability}

Statistical mechanics deals with large systems of stochastically  
interacting microscopic 
elements (particles, magnets, polymers, etc.). 
The general strategy of 
statistical mechanics is to abandon any ambition to solve models of such systems at the 
microscopic level of individual elements, 
but to use the microscopic laws to 
calculate laws describing the behaviour of a suitably choosen set of {\em macroscopic} observables. 
The toolbox of statistical mechanics consists of various methods and tricks to perform this reduction from the microscopic to a macroscopic level, which are based on clever ways to do the bookkeeping of probabilities. The experience and intuition that has been built up over the last century tells us what to expect (e.g. phase transitions), and serves as a guide in choosing  the macroscopic observables and in seeing the difference between relevant mathematical subtleties and irrelevant ones. 
As in any statistical theory, clean and transparent mathematical laws can be expected to emerge only for large (preferably infinitely large) systems. 

Supervised learning processes as described in the previous subsection appear to meet the
criteria for statistical mechanics to apply, provided we are happy to restrict ourselves to
large systems ($N\to\infty$). Here the microscopic stochastic dynamical variables are the components of the vector $\bJ$, and one is as little interested in knowing all individual components of $\bJ$  as one would be in 
knowing all position coordinates of the molecules in a bucket of water. We are rather after the
generalization and training errors, which are indeed {\em macroscopic} observables.  
\vsp

\begin{figure}[t]
\vspace*{90mm}\hbox to \hsize{\hspace*{-15mm}\includegraphics{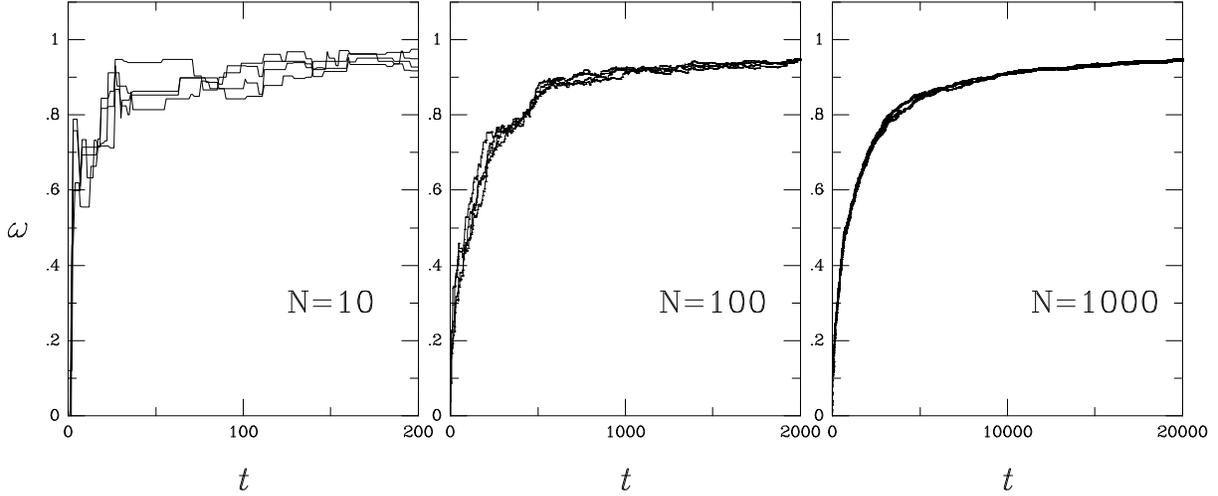}\hspace*{15mm}}
\vspace*{-20mm}
\caption{Evolution in time of the observable
$\omega=\bJ\!\cdot\!\bB/|\bJ|$ during numerical simulations of 
the standard perceptron learning rule (with a randomly
drawn normalised teacher weight vector $\bB$), following random 
initialisations  
of the student weight vector $\bJ$.} 
\label{fig:standardperceptron}
\end{figure}

Further support for the applicability of statistical mechanics is provided by numerical simulations. 
Consider, for instance, the example of the ordinary perceptron learning rule. For simplicity we choose $\set=D=\{-1,1\}^N$, with a task $T$ generated by a `teacher perceptron', corresponding to the following choices in the language of the previous subsection:
\bd
S(\bxi)=\sgn(\bJ\inn\bxi)
~~~~~~~~~~
T(\bxi)=\sgn(\bB\inn\bxi)
\ed
with $\bJ,\bB\in\Re^N$.
The teacher weight vector $\bB$  
is choosen at random, and  
normalised according to $|\bB|=1$. 
The (on-line) perceptron learning rule is 
\bd
\bJ(t\plus 1)=
\bJ(t)+\bxi(t)~\theta\left[\room \minus
T(\bxi(t))(\bJ(t)\inn\bxi(t))\right]
\ed
with the step function $\theta[z>0]=1$, $\theta[z<0]=0$. 
An educated guess for a possibly relevant 
macroscopic observable is the object that also played a central role in the original perceptron convergence proof:  
$\omega(t)=\bJ(t)\inn\bB/|\bJ(t)|$. 
The result of measuring the value of $\omega(t)$ during the execution of the above 
learning rule is shown in figure 
\ref{fig:standardperceptron}. 
These experiments clearly suggest 
(keeping in mind that for specifically
constructed pathological teacher vectors the picture might be
different),  that  
if viewed on the relevant $N$-dependent time-scale (as in the figure), the
fluctuations in $\omega$ 
become negligible as $N\rightarrow\infty$, and a clean deterministic law emerges. 
This is the type of situation we need in order to use statistical mechanics, and finding an analytical expression for this 
deterministic law will be our goal. 

\begin{figure}[t]
\vspace*{90mm}\hbox to \hsize{\hspace*{-15mm}\includegraphics{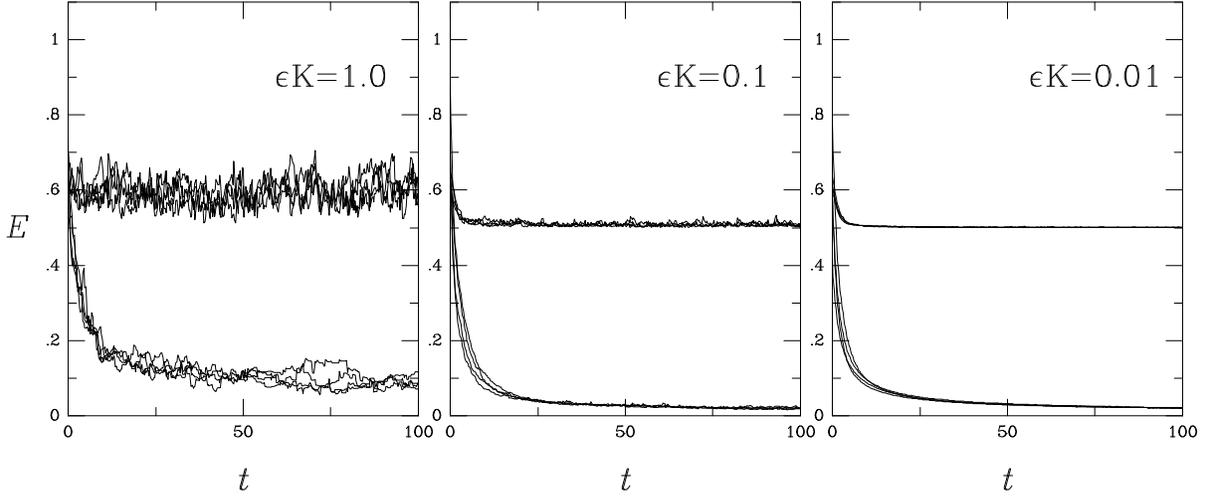}\hspace*{15mm}}
\vspace*{-20mm}
\caption{Evolution of the overall error $E$ in a two-layer
feed-forward network, trained by error backpropagation
(with $K=15$ input neurons, $L=10$ hidden neurons, and a single output
neuron). The results refer to independent experiments involving either
a linearly separable task (with random teacher vector, lower curves) or the parity
operation (upper curves), following random initialisation.}
\label{fig:bpvariouse}
\end{figure}

As a second example we will choose a two-layer network, trained
according to the error back-propagation rule. Here the microscopic
stochastic variables are both the weights $\{W_{ij}\}$ from the input
to the hidden layer (of $L$ neurons) and the weights $\{J_i\}$ from the hidden layer to the output layer. 
We define $\set=D=\{-1,1\}^K$ and  
\bd
S(\bxi)=\tanh\left[\sum_{i=1}^L J_i y_i(\bxi)\right]
~~~~~~~~~~
y_i(\bxi)=\tanh\left[\sum_{j=1}^K
W_{ij}\xi_j\right]
\ed
We consider two types of tasks, a linearly separable one (which is always learnable by the present student), and the parity operation (which is learnable only for $L\geq K$): 
\bd
\bxi\in\{-1,1\}^K:~~~~~~~~
\begin{array}{ll}
{\rm task~ I}: & T(\bxi)=\sgn(\bB\inn\bxi)\in\{-1,1\} \\[3mm]
{\rm task~II}: & T(\bxi)=\prod_{i=1}^K \xi_i \in\{-1,1\} \room
\end{array}
\ed
Our macroscopic observable will be the 
mean error (since $\set=D$ the generalization error and the training error are here identical):
\bd
E= \bra E[T(\bxi),S(\bxi)]\ket_D,
~~~~~~~~~~
E[T(\bxi),S(\bxi)]=\frac{1}{2}\left[T(\bxi)-S(\bxi)\right]^2
\ed
Perfect performance would correspond to $E=0$. On the other hand, 
a
trivial perceptron with zero weights 
throughout  
would give $S(\bxi)=0$ so $E=\frac{1}{2}\bra T^2(\bxi)\ket_D=\frac{1}{2}$. 
The learning rule used is the discretised on-line version (with learning rate $\epsilon$) of the error backpropagation rule:
\bd
J_i(t\plus \epsilon)=J_i(t)-\epsilon 
\frac{\partial}{\partial J_i}E[T(\bxi(t)),S(\bxi(t))]  
~~~~~~~~~~
W_{ij}(t\plus \epsilon)=W_{ij}(t)-\epsilon 
\frac{\partial}{\partial W_{ij}}E[T(\bxi(t)),S(\bxi(t))]  
\ed
The results of doing several such simulations, for $K=15$ and $L=10$ (so that the parity operation is an unlearnable task for the student network) and following random initialisation of the various weights,  
are shown in figure \ref{fig:bpvariouse}. 
These experiments again clearly suggest 
that also for multi-layer networks statistical mechanics will be a natural tool to analyse the dynamics of learning. Provided we scale our parameters 
appropriately and take a suitable limit 
(there will be different equivalent ways of
doing this) 
the fluctuations in suitably chosen 
macroscopic observables can be made to vanish, such that 
transparent deterministic laws emerge. 

\subsection{A Preview}

There are two main classes of situations in the supervised learning arena, 
which differ fundamentally in their dynamics and in the degree to which 
we can analyse them mathematically. The first class is the one where
the training set $\set$ is what we call `complete':  
sufficiently large and sufficiently diverse to 
lead to a learning dynamics which in the limit $N\to\infty$ 
is identical to that of the situation where $\set=D$. For example: 
in single perceptrons and in multi-layer perceptrons with a finite number of hidden nodes one finds, 
for the case where $D=\{-1,1\}^N$ and where the members of the 
training set $\set$ are drawn at random from $D$, that completeness of the training set amounts to $\lim_{N\to\infty}N/|\set|=0$. This makes sense:
it means that for $N\to\infty$ there will be an infinite number of training examples per degree of freedom.  
For this class of models it is fair to say that the dynamics 
of learning can be fully analysed in a reasonably simple  
way\footnote{In these models one can still study interesting new
phenomena, such as the effects of having noisy teachers etc., but at least the route to be followed to solve them is well-defined and guaranteed to work.}.   
Because this situation is now so nicely under control and admits for 
analytical solutions, it is a nice area to describe in a self-contained way 
in a paper such as the present one. We will restrict ourselves to single 
perceptrons with various types of learning rules, since they form the most
transparent playground for explaining how the mathematical techniques work. 
For multi-layer perceptrons with a finite number of hidden neurons and
complete training sets the procedure to be followed is very
similar\footnote{The situation is different if we try to deal with
multi-layer perceptrons with a number of hidden neurons which scales
linearly with the 
number of input channels $N$. As far as we are aware, 
this still poses an unsolved problem, even in the case of complete training sets.}.

The picture changes dramatically if we move away from complete training sets
and consider those where the number of training examples is proportional 
to the number of degrees of freedom, i.e. in simple perceptrons and in 
two-layer perceptrons with a finite number of hidden neurons this implies 
$|\set|=\alpha N$ ($0<\alpha<\infty$). Now the dependence of the microscopic variables $\bJ$ on the realisation
of the training set $\set$ is non-negligible.  However, if the questions in the training set are drawn at random from the full question set $D$ one often finds 
that in the $N\to\infty$ limit the values of the {\em macroscopic} observables
only depend on the size $|\set|$ of the training set, not on its microscopic realisation. 
For those familiar with the statistical mechanical analysis  
of the operation of recurrent neural networks: learning  
dynamics with complete training sets is mathematically similar 
to the dynamics of attractor networks away from saturation,  
whereas 
learning dynamics with incomplete training sets is similar, if non equivalent, 
to the dynamics of attractor networks close to saturation (in turn equivalent to the
complex dynamics of spin-glasses). Here one needs much more powerful mathematical tools, which are as yet only partly available. This class of problems is therefore only beginning to be studied, and we cannot yet give a well rounded overview 
with a happy ending (as for the case of complete training sets). We will do the next best thing and try to explain as clearly as possible what the problem is. 

No review is unbiased and complete; and one always has to strike a balance between broadness and depth (equivalently: 
between being encyclopedic and being self-contained). Here we have
opted for the latter. As a result, the references we give 
are intended to serve as a guide only, not as a true
reflection of 
all the work that has been done; for each paper mentioned at least
fifty will have
been left out, and we wish to apologise beforehand to the authors of
the papers in the latter category. 
We aim to explain the ideas and techniques 
only for a subset of the field, in the hope that the text can then be sufficiently self-contained to serve not just the 
interested spectator but also those who wish to  become actively involved.  


\section{On-Line Learning: Complete Training Sets and Explicit Rules}

We will now derive explicitly macroscopic 
dynamical equations that describe the evolution in time for the error in large perceptrons, trained with several on-line learning rules to perform linearly separable tasks.  In this section we restrict ourselves to complete training sets $\set=D=\{-1,1\}^N$. There is consequently 
no difference 
between training and generalization error, 
and we can simply define $E=\lim_{N\to\infty} \bra E_{\rm g}\ket = \lim_{N\to\infty}\bra E_{\rm t}\ket$. 

\subsection{General On-Line Learning Rules}

\begin{figure}[h]
\setlength{\unitlength}{0.08mm}
\newcommand{\sq}{\thinlines\framebox(10,10)}
\newcommand{\here}{\makebox(0,0)}
\begin{picture}(1300,750)(-300,350)
\newcommand{\Smallsize}{130}
\newcommand{\Size}{270}
\put(50,1000){\here{$\xi_1$}}
\put(50,950){\here{$.$}}
\put(50,900){\here{$.$}}
\put(50,850){\here{$.$}}
\put(50,800){\here{$.$}}
\put(50,750){\here{$.$}}
\put(50,700){\here{$.$}}
\put(50,650){\here{$.$}}
\put(50,600){\here{$.$}}
\multiput(100,390)(0,50){13}{\sq}
\put(50,550){\here{$.$}}
\put(50,500){\here{$.$}}
\put(50,450){\here{$.$}}
\put(50,400){\here{$\xi_N$}}
\put(400,800){\framebox(600,200)}
\put(700,950){\here{\large Student Perceptron}}
\put(700,850){\here{$S(\bxi)=\sgn(\bJ\inn\bxi)$}}
\put(1010,900){\line(1,0){220}}
\put(1010,900){\vector(1,0){110}}
\put(1300,900){\here{$S(\bxi)$}}
\put(120,400){\line(3,1){\Size}}
\put(120,500){\line(3,4){\Size}}
\put(120,600){\line(1,1){\Size}}
\put(120,700){\line(3,2){\Size}}
\put(120,800){\line(3,1){\Size}}
\put(120,900){\line(1,0){\Size}}
\put(120,1000){\line(3,-1){\Size}}
\put(120,400){\line(3,5){\Size}}
\put(120,500){\line(1,0){\Size}}
\put(120,600){\line(3,-1){\Size}}
\put(120,700){\line(3,-2){\Size}}
\put(120,800){\line(1,-1){\Size}}
\put(120,900){\line(3,-4){\Size}}
\put(120,1000){\line(3,-5){\Size}}
\put(400,400){\framebox(600,200)}
\put(700,550){\here{\large Teacher Perceptron}}
\put(700,450){\here{$T(\bxi)=\sgn(\bB\inn\bxi)$}}
\put(1010,500){\line(1,0){220}}
\put(1010,500){\vector(1,0){110}}
\put(1300,500){\here{$T(\bxi)$}}
\put(120,400){\vector(3,1){\Smallsize}}
\put(120,500){\vector(3,4){\Smallsize}}
\put(120,600){\vector(1,1){\Smallsize}}
\put(120,700){\vector(3,2){\Smallsize}}
\put(120,800){\vector(3,1){\Smallsize}}
\put(120,900){\vector(1,0){\Smallsize}}
\put(120,1000){\vector(3,-1){\Smallsize}}
\put(120,500){\vector(1,0){\Smallsize}}
\put(120,600){\vector(3,-1){\Smallsize}}
\put(120,700){\vector(3,-2){\Smallsize}}
\put(120,800){\vector(1,-1){\Smallsize}}
\put(120,900){\vector(3,-4){\Smallsize}}
\end{picture}
\caption{A student perceptron 
$S$ is being trained according to on-line learning rules 
to perform a linearly separable operation, generated by some unknown  
teacher perceptron $T$.}  
\label{fig:percscenario}
\end{figure}
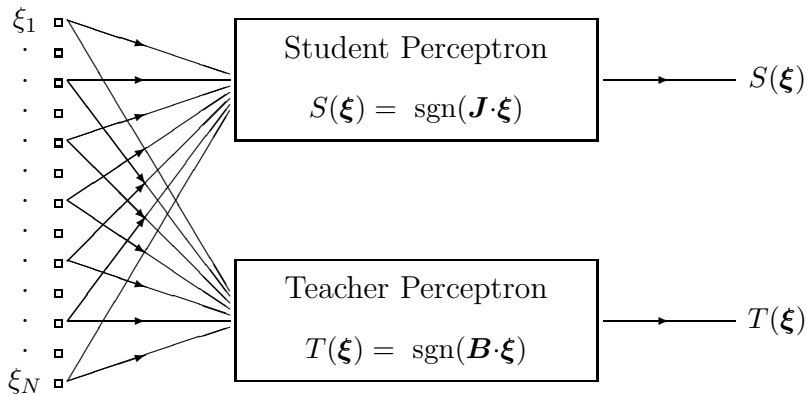

Consider a linearly separable 
binary classification task $T:\{-1,1\}^N\to\{-1,1\}$. It can be regarded as 
generated 
by a teacher 
perceptron with some unknown weight vector $\bB \in\Re^N$, i.e. $T(\bxi)=\sgn(\bB\inn\bxi)$, normalised according to 
$|\bB |=1$ (with the sign function  
$\sgn(z>0)=1$, $\sgn(z<0)=-1$).
A student perceptron with output $S(\bxi)=\sgn(\bJ\inn\bxi)$ (where
$\bJ\in\Re^N$) 
is being trained in an on-line fashion using  
randomly drawn examples of input vectors $\bxi\in\{-1,1\}^N$ with corresponding teacher answers $T(\bxi)$. 
The general picture of figure \ref{fig:scenario}  thus 
specialises to figure \ref{fig:percscenario}. 
We exploit our knowledge of the perceptron's scaling properties 
(see figure \ref{fig:standardperceptron}), and distinguish between the discrete time unit in terms of iteration steps, from now on to be denoted by $\mu=1,2,3,\ldots$, and the scale-invariant time unit $t_\mu=\mu/N$.  
Our goal is to derive well-behaved differential equations 
in the limit $N\to\infty$, so we require weight changes occurring in intervals $\Delta t=\frac{1}{N}$ to be of order $\order(\frac{1}{N})$ as well. In terms of equation (\ref{eq:weightdynamics}) this implies that $F[\dots]=\order(\frac{1}{N})$. 
If, finally, we restrict ourselves to those rules where weight changes are made in the direction of the example vectors (which includes most popular rules), we obtain the generic\footnote{One can obviously write down more general rules, and also write the present recipe (\ref{eq:genericrule}) in different ways.} recipe
\be
\bJ(t_\mu +\frac{1}{N})
=\bJ(t_\mu)+\frac{1}{N}\eta(t_\mu)\bxi^\mu\sgn(\bB\inn\bxi^\mu)\cF[|\bJ(t_\mu)|;
\bJ(t_\mu)\inn\bxi^\mu,
\bB\inn\bxi^\mu]
\label{eq:genericrule}
\ee
Here $\eta(t_\mu)$ denotes a (possibly time-dependent) learning rate and 
$\bxi^\mu$ is the input vector selected at iteration step $\mu$. 
$\cF[\ldots]$ is an as yet arbitrary
function of the length of the student weight vector and of the local
fields $u$ and $v$ of student and teacher (note: $\cF$ can depend on the sign of the teacher field only, not on its magnitude).
For example, for $\cF[J;u,v]=1$ we obtain a Hebbian rule, for
$\cF[J;u,v]=\theta[-uv]$
we obtain the perceptron learning rule, etc. 
\vsp

We now try to solve the dynamics of the learning process in terms of the two 
macroscopic 
observables that play a special role in the perceptron convergence proof:
\be
Q[\bJ]=\bJ^2
~~~~~~~~~~~~
R[\bJ]=\bJ\inn\bB
\label{eq:QandR}
\ee
(at this stage the selection of observables is still no more than intuition-driven guesswork). The formal approach would now
be to derive an expression for the (time-dependent) probability density $P(Q,R)=\bra
\delta[Q\minus Q[\bJ]]\delta[R\minus R[\bJ]]\ket$, however, it turns out that in the present case\footnote{This will be different in the case of incomplete training sets.} there is a short-cut.
Squaring (\ref{eq:genericrule}) and taking the inner product
of (\ref{eq:genericrule}) with the teacher vector $\bB$ gives, respectively
\bea
Q[\bJ(t_{\mu}\plus\frac{1}{N})] & \!
\!=\!\! &
Q[\bJ(t_\mu)]
+\frac{2}{N}\eta(t_\mu)(\bJ(t_\mu)\inn\bxi^\mu)
\sgn(\bB\inn\bxi^\mu)\cF[|\bJ(t_\mu)|;\bJ(t_\mu)\inn \bxi^\mu,\bB\inn \bxi^\mu]
\nonumber \\[2mm]
&& ~~~~~~~~~~~~~~~~~~~~
+\frac{1}{N}\eta^2(t_\mu)\cF^2[|\bJ(t_\mu)|;\bJ(t_\mu)\inn \bxi^\mu,\bB\inn \bxi^\mu]
\nonumber \\[2mm]
R[\bJ(t_{\mu}\plus\frac{1}{N})] & \!\!=\!\! &
R[\bJ(t_\mu)]+\frac{1}{N}\eta(t_\mu)
|\bB\inn\bxi^\mu|\cF[|\bJ(t_\mu)|;\bJ(t_\mu)\inn \bxi^\mu,\bB\inn \bxi^\mu]
\nonumber
\eea
(note: $\bxi^\mu\inn\bxi^\mu=N$). 
After $\ell$ discrete update steps we 
will have accumulated $\ell$ such  
modifications, and will thus arrive at:
\bd
\frac{Q[\bJ(t_{\mu}\plus \ell/N)]-
Q[\bJ(t_\mu)]}{\ell/N} =
~~~~~~~~~~~~~~~~~~~~~~~~~~~~~~~~~~~~~~~~~~~~~~~~~~~~~~~~~~~~~~~~~~~~~~~~~~~~~~~~~~~~~~~
\ed
\bd
\frac{1}{\ell}
\sum_{m=0}^{\ell-1}\left\{
2\eta(t_\mu\plus\frac{m}{N})(\bJ(t_\mu\plus\frac{m}{N})\inn\bxi^{\mu+m})
\sgn(\bB\inn\bxi^{\mu+m})\cF[|\bJ(t_\mu\plus\frac{m}{N})|;\bJ(t_\mu\plus\frac{m}{N})\inn \bxi^{\mu+m},\bB\inn \bxi^{\mu+m}]
\right.
\ed
\bd
\left.
+\eta^2(t_\mu\plus\frac{m}{N})\cF^2[|\bJ(t_\mu\plus\frac{m}{N})|;\bJ(t_\mu\plus\frac{m}{N})\inn \bxi^{\mu+m},\bB\inn \bxi^{\mu+m}]
\right\}
\ed
\bd
\frac{R[\bJ(t_{\mu}\plus \ell/N)]-
R[\bJ(t_\mu)]}{\ell/N} =
~~~~~~~~~~~~~~~~~~~~~~~~~~~~~~~~~~~~~~~~~~~~~~~~~~~~~~~~~~~~~~~~~~~~~~~~~~~~~~~~~~~~~~~
\ed
\bd
\frac{1}{\ell}\sum_{m=0}^{\ell-1}\left\{
\eta(t_\mu\plus\frac{m}{N})
|\bB\inn\bxi^{\mu+m}|\cF[|\bJ(t_\mu\plus\frac{m}{N})|;\bJ(t_\mu\plus\frac{m}{N})\inn \bxi^{\mu+m},\bB\inn \bxi^{\mu+m}]
\right\}
\ed
All is still exact, but at this stage we will have to make an assumption  
which is not entirely satisfactory\footnote{We will later find out that a more careful analysis gives the same results.}. We assume that $\bJ(t_\mu\plus\frac{m}{N})\inn\bxi^{\mu+m}\to \bJ(t_\mu)\inn\bxi^{\mu+m}$ if $N\to\infty$ for finite $m$. This is only true in a probabilistic sense, since, although $J_i(t_\mu\plus\frac{m}{N})=J_i(t_\mu)+\order(\frac{m}{N})$, the inner product is a sum of $N$ terms. 
If for now, however, we accept this step and also choose learning rates which vary sufficiently slowly over time to 
guarantee existence of the limit $\lim_{N\to\infty} \eta(t_\mu)$, 
we find that by taking the limit $N\to\infty$, followed by the limit $\ell\to\infty$, three pleasant  
simplifications occur: (i) the time unit $t_\mu=\mu/N$ becomes a continuous variable, 
(ii) the left-hand sides of the above equations for the evolution of the observables $Q$ and $R$ become temporal derivatives, and 
(iii) the summations in the right-hand sides of these equations become averages of the training set. Upon putting $Q(t)=Q[\bJ(t)]$ and $R(t)=R[\bJ(t)]$ the result can be written as:
\bd
\frac{d}{dt}Q(t)
=
2\eta(t)\bra (\bJ(t)\inn\bxi)
\sgn(\bB\inn\bxi)\cF[Q^{\frac{1}{2}}(t);\bJ(t)\inn \bxi,\bB\inn \bxi]
\ket_\set
+\eta^2(t)\bra\cF^2[Q^{\frac{1}{2}}(t);\bJ(t)\inn \bxi,\bB\inn \bxi]\ket_\set
\ed
\bd
\frac{d}{dt}R(t)=
\eta(t)\bra  
|\bB\inn\bxi|\cF[Q^{\frac{1}{2}}(t);\bJ(t)\inn \bxi,\bB\inn \bxi]
\ket_\set
\ed
The only dependence of the right-hand sides of these expressions on the microscopic variables $\bJ$ is via the student fields $\bJ(t)\inn\bxi=Q^{\frac{1}{2}}(t)
\hbJ(t)\inn\bxi$, with $\hbJ=\bJ/|\bJ|$\footnote{This property of
course depends crucially on our choice (\ref{eq:genericrule}) made for the
form of the learning rules.}. We therefore define the
stochastic variables $x=\hbJ\inn\bxi$ and $y=\bB\inn\bxi$ and their
joint probability distribution $P_t(x,y)$:
\be
P_t(x,y)=\bra \delta[x\minus\hbJ(t)\inn\bxi]\delta[y\minus \bB\inn\bxi] \ket_\set
~~~~~~~~~~~~
\bra f(x,y)\ket = \int\!dxdy~P_t(x,y)f(x,y)
\label{eq:fielddist}
\ee
Using brackets without subscripts for joint field averages cannot cause confusion, since such expressions always {\em replace} averages over $\bJ$, rather than occur simultaneously. 
Our previous result now takes the form
\be
\frac{d}{dt}Q(t)
=
2\eta(t)Q^{\frac{1}{2}}(t)\bra x 
\sgn(y)\cF[Q^{\frac{1}{2}}(t);Q^{\frac{1}{2}}(t)x,y]\ket 
+\eta^2(t)\bra\cF^2[Q^{\frac{1}{2}}(t);Q^{\frac{1}{2}}(t)x,y]
\ket
\label{eq:dQdt}
\ee
\be
\frac{d}{dt}R(t)=
\eta(t)\bra 
|y|\cF[Q^{\frac{1}{2}}(t);Q^{\frac{1}{2}}(t)x,y]
\ket
\label{eq:dRdt}
\ee
Since the operation performed by the student does not depend on the length $|\bJ|$ of its weight vector, and since both $Q$ and $R$ involve $|\bJ|$, 
it will be convenient at this stage to switch to another (equivalent) pair of observables:
\be
J(t)=|\bJ(t)|
~~~~~~~~~~~~
\omega(t)=\bB\cdot\hbJ(t)
\label{eq:newobservables}
\ee
Using the relations $\frac{d}{dt}Q=2J\frac{d}{dt}J$ and $\frac{d}{dt}R=J\frac{d}{dt}\omega+\omega\frac{d}{dt}J$, and upon 
dropping the various explicit time arguments (for
notational convenience) 
we then find the compact expressions
\begin{eqnarray}
\frac{d}{dt}J&\!\!=\!\!&
\eta\bra
x\sgn(y)\cF[J;Jx,y]\ket+\frac{\eta^2}{2J}\bra
\cF^2[J;Jx,y]\ket
\label{eq:genJinxy}\\
\frac{d}{dt}\omega &\!\!=\!\!&
\frac{\eta}{J}\bra\left[|y|\minus\omega
\,x\sgn(y)\right]\cF[J;Jx,y]\ket-\frac{\omega\eta^2}{2J^2}\bra\cF^2[J;Jx,y]\ket
\label{eq:genomegainxy}
\end{eqnarray}
Unless we manage to express $P(x,y)$ in terms of the
pair $(J,\omega)$, however, the equations (\ref{eq:genJinxy},\ref{eq:genomegainxy}) 
do not constitute a solution of our problem 
, since we would still be forced to solve
the original microsopic dynamical equations in order to find $P(x,y)$ as a
function of time and work out (\ref{eq:genJinxy},\ref{eq:genomegainxy}).

The final stage of the argument is to assume that the joint probability distribution (\ref{eq:fielddist}) has a Gaussian shape, since $\set=\{-1,1\}^N$ and since all $\bxi\in\set$ contribute equally to the average in (\ref{eq:fielddist}).  
This will be true in the vast majority of cases, e.g. it is true 
with probability one if the vectors $\bJ$ and $\bB$ are drawn at random from compact sets like $[-1,1]^N$, due to the central limit theorem\footnote{It is not true for all choices of {\bf J} and {\bf B}. A trivial counter-example is $J_k=\delta_{k1}$, less trivial counter-examples are e.g. $J_k=e^{-k}$ and 
$J_k=k^{-\gamma}$ with $\gamma>\frac{1}{2}$.}. Gaussian distributions are 
fully specified by their first and second order moments,  
which are here calculated trivially using $\bra \xi_i\ket=0$ and $\bra \xi_i\xi_j\ket=\delta_{ij}$:
\bd
\bra x\ket=\sum_i\hat{J}_i\bra \xi_i\ket =0~~~~~~~~~~~~
\bra y\ket=\sum_i B_i\bra \xi_i\ket=0
\ed
\bd
\bra x^2\ket=\sum_{ij}\hat{J}_i\hat{J}_j\bra \xi_i\xi_j\ket=1
~~~~~~~~~
\bra y^2\ket=\sum_{ij}B_i B_j\bra \xi_i\xi_j\ket=1
~~~~~~~~~
\bra xy\ket=\sum_{ij}\hat{J}_i B_j\bra \xi_i\xi_j\ket=\omega
\ed
giving
\be
P(x,y)=\frac{e^{-\frac{1}{2}[x^2+y^2-2xy\omega]/(1-\omega^2)}}
{2\pi\sqrt{1-\omega^2}}
\label{eq:gaussian}
\ee
Note that $P(x,y)=P(y,x)$. 
The simple fact that $P(x,y)$ depends on time only through $\omega$
ensures that the two equations
(\ref{eq:genJinxy},\ref{eq:genomegainxy})
are a {\em closed} set. 
Note also that now (\ref{eq:genJinxy},\ref{eq:genomegainxy})
 are deterministic equations; apparently the fluctuations in the macroscopic observables $Q[\bJ]$ and $R[\bJ]$ vanish in the $N\to\infty$ limit.  
\vsp

Finally, the generalization error $E_{\rm g}$ (here identical to the training error $E_{\rm t}$ due to $\set=D$) 
can be expressed in terms of our macroscopic observables. 
We define the error made in a single classification of an input $\bxi$ as $E[T(\bxi),S(\bxi)]=\theta[-(\bB\inn\bxi)(\bJ\inn\bxi)]\in\{0,1\}$. 
Averaged over $D$ this gives the probability of a misclassification for randomly drawn questions $\bxi\in D$:
\bd
\lim_{N\to\infty}E_{\rm g}(\bJ(t)) =\lim_{N\to\infty}
\bra[\theta[-(\bB\inn\bxi)(\bJ(t)\inn\bxi)]\ket_D
=\bra\theta[- xy]\ket 
\ed
\bd
=\int_0^\infty\!\!\int_0^\infty\!dxdy\left[
P(x,-y)+P(-x,y)\right]
\ed
The generalization error (from this stage onwards to be denoted  
simply by $E$) 
also evolves deterministically for $N\to\infty$, and can be expressed purely in terms of the observable $\omega$.  
The integral (with the distribution (\ref{eq:gaussian})) can even be done analytically (see appendix) and produces the simple result
\be
E=\frac{1}{\pi}\arccos(\omega)
\label{eq:Einomega}
\ee
The macrosopic equations (\ref{eq:genJinxy},\ref{eq:genomegainxy}) 
can now equivalently be written in terms of the pair $(J,E)$. 
We have hereby achieved our goal: we have derived a closed set of deterministic equations for a small number (two) of macroscopic observables, valid for $N\to\infty$, and we 
know the generalization error at any time. 

\subsection{Hebbian Learning with Constant Learning Rate}

We will now work out our general result (\ref{eq:genJinxy},\ref{eq:genomegainxy},\ref{eq:gaussian}) for specific members of the general class (\ref{eq:genericrule}) of on-line learning rules.
The simplest non-trivial choice to be made
is the Hebbian rule, obtained by choosing 
$\cF[J;Jx,y]=1$, with a constant learning rate $\eta$:
\be
\bJ(t_\mu+\frac{1}{N})=\bJ(t_\mu)+\frac{\eta}{N}\bxi^\mu \sgn(\bB\inn\bxi^\mu)
\label{eq:constetahebb}
\ee
Equations (\ref{eq:genJinxy}) and (\ref{eq:genomegainxy}), describing the macroscopic dynamics generated by (\ref{eq:constetahebb}) in the limit $N\to\infty$  now become
\bd
\frac{d}{dt}J=\eta\langle x\sgn(y)\rangle+\frac{\eta^2}{2J}
~~~~~~~~~~~~
\frac{d}{dt}\omega =\frac{\eta}{J}\langle|y|\minus \omega x\sgn(y)\rangle-\frac{\omega\eta^2}{2J^2}
\ed
or, in more explicit form with the function $P(x,y)$ (\ref{eq:gaussian}):
\begin{eqnarray*}
\frac{d}{dt}J &\!\!=\!\! &
\eta\int\!\!\int\!dxdy\,x\sgn(y)P(x,y)+\frac{\eta^2}{2J}\\
\frac{d}{dt}\omega &\!\!=\!\!&
\frac{\eta}{J}\int\!\!\int
\!dxdy\,|y|P(x,y)-\frac{\omega\eta}{J}\int\!\!\int\!dxdy\,x\sgn(y)P(x,y)-\frac{\omega\eta^2}{2J^2}
\end{eqnarray*}
\begin{figure}[t]
\centering
\vspace*{95mm}
\hbox to \hsize{\hspace*{10mm}\includegraphics{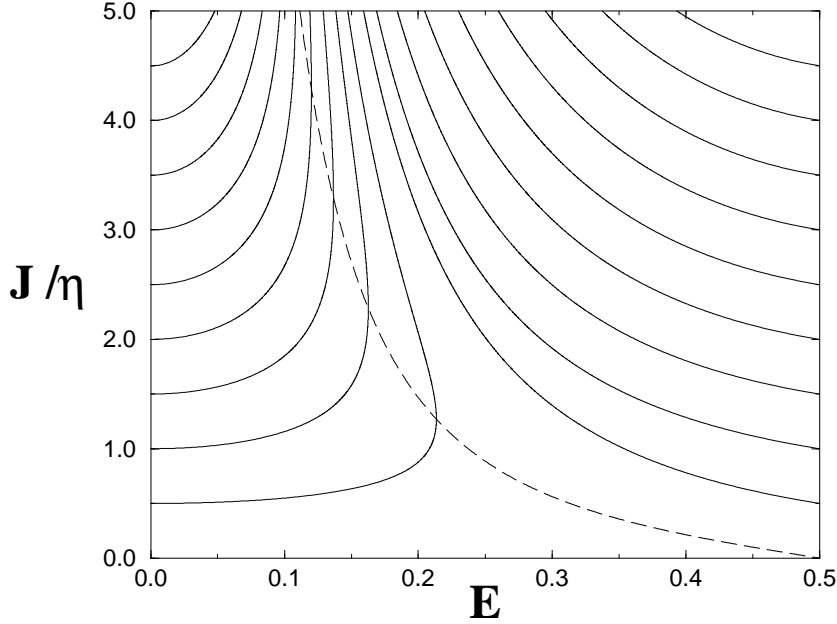}\hspace*{-10mm}}
\vspace*{-10mm}
\caption{Flow in the $(E,J)$ plane generated by the Hebbian 
learning rule with constant learning rate $\eta$, in the limit
$N\to\infty$. Dashed: the line where $dE/dt=0$ ($dJ/dt>0$
for any $(E,J)$). 
Note that the flow asymptotically gives $E\to 0$ and
$J\to\infty$.}
\label{fig:hebbflow}
\end{figure}
The integrals in these equations can be calculated analytically (see appendix) 
and we get 
\bd
\frac{d}{dt}J=\omega\eta\sqrt{\frac{2}{\pi}}+\frac{\eta^2}{2J}
~~~~~~~~~~~~
\frac{d}{dt}\omega =(1\minus \omega^2)\frac{\eta}{J}\sqrt{\frac{2}{\pi}}-\frac{\omega\eta^2}{2J^2}
\ed
Thus, upon elimination of the observable $\omega$  using equation (\ref{eq:Einomega}),  
we arrive at the following closed differential equations in terms of $J$ and $E$:
\begin{eqnarray}
\frac{d}{dt}J &\!\!=\!\!&
\eta\cos(\pi E)\sqrt{\frac{2}{\pi}}+\frac{\eta^2}{2J}
\label{eq:JHebb}\\
\frac{d}{dt}E&\!\!=\!\!&
-\frac{\eta\sin(\pi E)}{\pi J}\sqrt{\frac{2}{\pi}}
+\frac{\eta^2}{2\pi J^2\tan(\pi E)}
\label{eq:EHebb}
\end{eqnarray}
The flow in the $(E,J)$ plane described by these equations is drawn in figure \ref{fig:hebbflow} (which is obtained by numerical solution of (\ref{eq:JHebb},\ref{eq:EHebb})). From (\ref{eq:JHebb}) it follows that 
$\frac{d}{dt}J>0\ \forall t\geq 0$. 
From (\ref{eq:EHebb}) it follows that 
$\frac{d}{dt}E=0$ along the line
\bd
J_c(E)=\frac{\eta\cos(\pi E)}{2\sin^2(\pi E)}\sqrt{\frac{\pi}{2}}
\ed
(drawn as a dashed line in figure \ref{fig:hebbflow}). 

Let us now investigate the temporal properties of the solution (\ref{eq:JHebb},\ref{eq:EHebb}), and work out their predictions for the asymptotic 
decay of the generalization error. 
For small values of $E$ equations (\ref{eq:JHebb},\ref{eq:EHebb}) yield
\begin{eqnarray}
\frac{d}{dt}J &\!\!=\!\!&
\eta\sqrt{\frac{2}{\pi}}+\frac{\eta^2}{2J}+\order(E^2)\label{eq:smallEHebbJ}\\
\frac{d}{dt}E &\!\!=\!\!&
-\frac{\eta E}{J}\sqrt{\frac{2}{\pi}}+\frac{\eta^2}{2\pi^2 J^2 E}+\order(E^3/J,E/J^2)\label{eq:smallEHebbE}
\end{eqnarray}
From (\ref{eq:smallEHebbJ}) we infer 
that $J\sim\eta t\sqrt{\frac{2}{\pi}}$ for $t\rightarrow\infty$. Subsitution 
of this asymptotic solution into 
equation (\ref{eq:smallEHebbE}) gives
\begin{equation}
\frac{d}{dt}E=-\frac{E}{t}+\frac{1}{4\pi E t^2}+\order(E^3/t,E/t^2)
~~~~~~~~~~(t\to\infty)
\label{eq:HebbsmallE}
\end{equation}
We insert the ansatz $E=At^{-\alpha}$ into equation (\ref{eq:HebbsmallE}) and get
the solution $A=1/\sqrt{2\pi}$, $\alpha=1/2$. This implies that (in the $N\to\infty$ limit) on-line Hebbian learning with complete training sets 
produces an asymptotic decay of the generalization of the form
\be
E\sim\frac{1}{\sqrt{2\pi t}}
~~~~~~~~~~(t\to\infty)
\label{eq:Edecayhebb}
\ee
Figures 
\ref{fig:Ecompare1}, \ref{fig:Ecompare2} and \ref{fig:Ecomparelog} 
will show the theoretical results of this section together with the results of doing numerical simulations of the learning rule (\ref{eq:constetahebb}) and with similar results for other on-line learning rules with constant learning rates. The agreement between theory and simulations is quite convincing.  

\subsection{Perceptron Learning with Constant Learning Rate}

Our second application 
of (\ref{eq:genJinxy},\ref{eq:genomegainxy},\ref{eq:gaussian}) is making the choice 
$\cF[J;Jx,y]=\theta[-xy]$ in equation (\ref{eq:genericrule}), with constant learning rate $\eta$, which produces the perceptron learning algorithm:
\be
\bJ(t_\mu +\frac{1}{N})=\bJ(t_\mu)+\frac{\eta}{N}\bxi^\mu\theta\left[
-(\bB\inn\bxi^\mu)(\bJ(t_\mu)\inn\bxi^\mu)\right]
\label{eq:constetaperc}
\ee
In other words: the student weights are updated in accordance with the Hebbian rule only when 
$\sgn(\bB\cdot\bxi)= -\sgn(\bJ\cdot\bxi)$, i.e. when student and teacher are not in agreement.
Equations (\ref{eq:genJinxy},\ref{eq:genomegainxy}) now become
\begin{eqnarray*}
\frac{d}{dt}J&\!\!=\!\!&
\eta\langle x\sgn(y)\theta[-xy]\rangle+\frac{\eta^2}{2J}\langle\theta[-xy]\rangle\\
&\!\!=\!\!
&\eta\int\!\!\int\!dxdy~x\sgn(y)\theta[-xy]P(x,y)+\frac{\eta^2}{2J}\int\!\!\int
\!dxdy~\theta[-xy]P(x,y)
\end{eqnarray*}
\begin{eqnarray*}
\frac{d}{dt}\omega &\!\!=\!\!
&\frac{\eta}{J}\langle\left[|y|\minus \omega x\sgn(y)\right]\theta[-xy]\rangle
-\frac{\omega\eta^2}{2J^2}\langle\theta[-xy]\rangle\\
&\!\!=\!\!&
\frac{\eta}{J}
\int\!\!\int\!dxdy~|y|\theta[-xy]P(x,y)
-\frac{\omega\eta}{J}\int\!\!\int\!dxdy~
x\sgn(y)\theta[-xy]P(x,y)\\
& &~~~~~~~~~~
-\frac{\omega\eta^2}{2J^2}\int\!\!\int\!
dxdy~\theta[-xy]P(x,y)
\end{eqnarray*}
with $P(x,y)$ given by (\ref{eq:gaussian}). 
As before the various Gaussian integrals occurring in these expressions can be done analytically (see appendix), which results in 
\bd
\frac{d}{dt}J=
-\frac{\eta(1\minus\omega)}{\sqrt{2\pi}}+\frac{\eta^2 }{2\pi J}\arccos(\omega)
~~~~~~~~~~~~
\frac{d}{dt}\omega=
\frac{\eta(1\minus\omega^2)}{\sqrt{2\pi}J}-\frac{\omega\eta^2}{2\pi J^2}\arccos(\omega)
\ed
Elimination of $\omega$ using (\ref{eq:Einomega}) 
then gives us the dynamical equations in terms of the pair $(J,E)$:
\begin{eqnarray}
\frac{d}{dt}J&\!\!=\!\!&
- \frac{\eta(1\minus\cos(\pi E))}{\sqrt{2\pi}}+\frac{\eta^2 E}{2J}
\label{eq:Jperceptron}\\
\frac{d}{dt}E&\!\!=\!\!&
-\frac{\eta\sin(\pi E)}{\pi\sqrt{2\pi}J}+\frac{\eta^2E}{2\pi J^2\tan(\pi E)}
\label{eq:Eperceptron}
\end{eqnarray}
\begin{figure}[t]
\centering
\vspace*{95mm}
\hbox to \hsize{\hspace*{10mm}\includegraphics{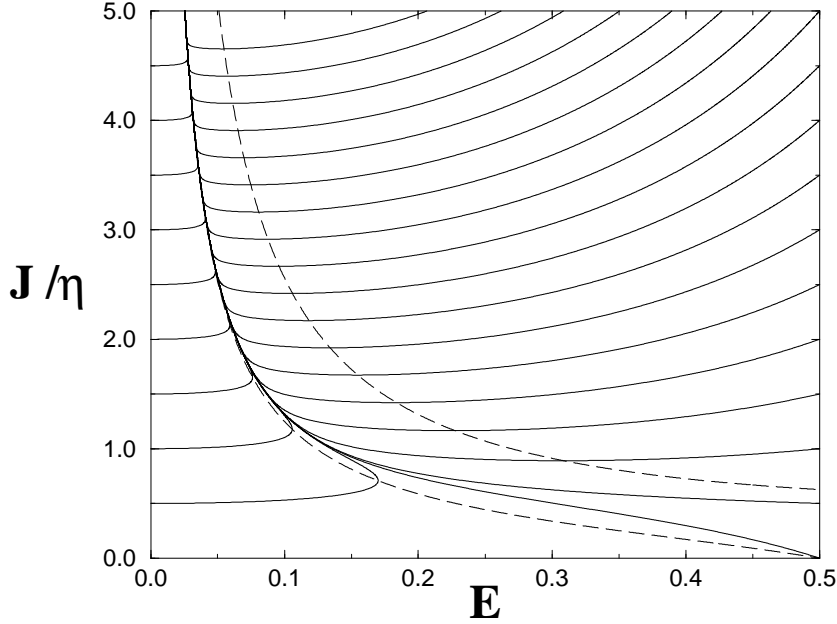}\hspace*{-10mm}}
\vspace*{-8mm}
\caption{Flow in the $(E,J)$ plane generated by the perceptron
learning rule with constant learning rate $\eta$, in the limit
$N\to\infty$. Dashed: the two lines where $dE/dt=0$ and $dJ/dt=0$,
respectively. Note that the flow is attracted into the gully between
these two dashed lines and asymptotically gives $E\to 0$ and
$J\to\infty$.}
\label{fig:percflow}
\end{figure}
Figure \ref{fig:percflow} shows the flow in the $(E,J)$ plane, obtained by numerical solution of (\ref{eq:Jperceptron},\ref{eq:Eperceptron}).   
The two lines where $\frac{d}{dt}J=0$ and where $\frac{d}{dt}E=0$ are 
found to be $J_{c,1}(E)$ and $J_{c,2}(E)$, respectively:
\bd
J_{c,1}(E)=\eta\sqrt{\frac{\pi}{2}}
\frac{E}{1\minus\cos(\pi E)}
~~~~~~~~~~
J_{c,2}(E)=\eta\sqrt{\frac{\pi}{2}}
\frac{E\cos(\pi E)}{1\minus\cos^2(\pi E)}
\ed
For $E\in [0,1/2]$ one always has  $J_{c,1}(E)\geq J_{c,2}(E)$, 
with equality only if $(J,E)=(\infty,0)$.
Figure \ref{fig:percflow} shows that the flow is drawn into the  
gully between the curves $J_{c,1}(E)$ and $J_{c,2}(E)$. 

As with the Hebbian rule we now wish to investigate the asymptotic behaviour of the generalization error. To do this we expand equations 
(\ref{eq:Jperceptron},\ref{eq:Eperceptron}) for small $E$:
\[
\frac{d}{dt}J=- \frac{\eta\pi^2 E^2}{2\sqrt{2\pi}}+\frac{\eta^2 E}{2J}+\order(E^4)
\]
\[
\frac{d}{dt}E =-\frac{\eta E}{\sqrt{2\pi}J}+\frac{\eta^2}{2\pi^2J^2}-\frac{\eta^2E^2}{6J^2}+\order(E^3)
\]
For small $E$ and large $t$ we know that $J\sim J_{c,1}(E)\sim 1/E$. 
Making the ansatz $J=A/E$ (and hence $\frac{d}{dt}E=-\frac{E^2}{A}\frac{d}{dt}J$) 
leads to a situation where we have two equivalent differential 
equations for $E$:
\bd
\frac{d}{dt}E=\frac{\eta\pi^2E^4}{2\sqrt{2\pi}A}-\frac{\eta^2E^4}{2A^2}+\order(E^6)
\ed
\bd
\frac{d}{dt}E=-\frac{\eta E^2}{\sqrt{2\pi}A}+\frac{\eta^2E^2}{2\pi^2A^2}+\order(E^4)
\ed
Since both describe the same dynamics, the leading term of the 
second expression 
should be identical to that of the first, i.e. 
$\order(E^4)$, giving us the condition
$A=\frac{\eta\sqrt{2\pi}}{2\pi^2}$. Substitution of this condition
into the first expression for $\frac{d}{dt}E$ then gives
\bd
\frac{d}{dt}E=-\frac{1}{2}\pi^3E^4+\order(E^5)
~~~~~~~~~~(t\to\infty)
\ed
which has the solution 
\be
E\sim\left(\frac{2}{3}\right)^{1/3}\pi^{-1} t^{-1/3}
~~~~~~~~~~(t\to\infty)
\label{eq:Edecayperc}
\ee
We find, somewhat surprisingly, that in large systems $(N\to\infty)$ the 
on-line perceptron learning rule is asymptotically much slower in converging towards the desired $E=0$ state than the simpler Hebbian rule. 
This will be different if we allow for time-dependent learning rates. 
Figures \ref{fig:Ecompare1}, \ref{fig:Ecompare2} and \ref{fig:Ecomparelog} will  show the theoretical results on the perceptron 
rule together with the results of doing numerical simulations and together 
with similar results for other on-line learning  rules. 
Again the agreement between theory and experiment is quite satisfactory. 

\subsection{AdaTron Learning with Constant Learning Rate}

As our third application we analyse the macroscopic dynamics of the 
AdaTron learning rule, corresponding to the choice $\cF[J;Jx,y]=|Jx|\theta[-xy]$ in the general recipe (\ref{eq:genericrule}). As in the perceptron rule, modifications are made only when student and teacher are in disagreement; however, here the modification made is proportional to the magnitude of the student's local field. Students are 
punished in proportion to their confidence in the wrong answer.  
The rationale is that wrong student answers 
$S(\bxi)=\sgn(\bJ\cdot\bxi)$ with large values of
$|\bJ\cdot\bxi|$ require more rigorous corrections to $\bJ$ to be
remedied than those with small values of $|\bJ\cdot\bxi|$.   
\be
\bJ(t_\mu+\frac{1}{N})=\bJ(t_\mu)+\frac{\eta}{N}\bxi^\mu\sgn(\bB\inn\bxi^\mu)
|\bJ(t_\mu)\inn\bxi^\mu|
\theta[-(\bB\inn\bxi^\mu)(\bJ(t_\mu)\inn\bxi^\mu)]
\label{eq:constetaadat}
\ee
\begin{figure}[t]
\centering
\vspace*{95mm}
\hbox to \hsize{\hspace*{10mm}\includegraphics{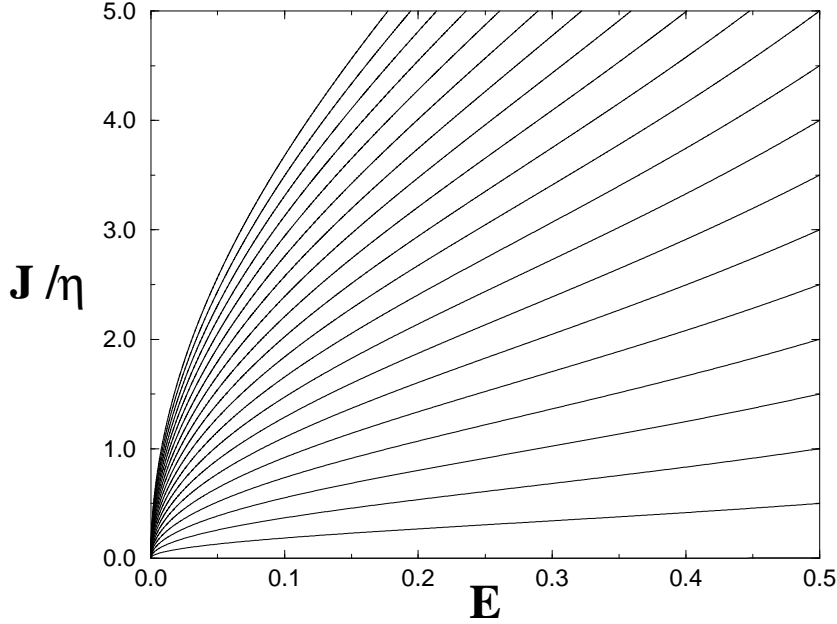}\hspace*{-10mm}}
\vspace*{-8mm}
\caption{Flow in the $(E,J)$ plane generated by the AdaTron learning rule 
with constant learning rate $\eta=1$, in the limit
$N\to\infty$ (in this case the influence of the value of the learning rate on the flow is more than just a rescaling of the length $J$).}
\label{fig:adatflow}
\end{figure}
Working out the general equations 
(\ref{eq:genJinxy},\ref{eq:genomegainxy}) 
for the learning rule (\ref{eq:constetaadat})   gives
\begin{eqnarray*}
\frac{d}{dt}J&\!\!=\!\!&
\eta J\int\!\!\int \!dxdy~x|x|\sgn(y)\theta[-xy]P(x,y)+\frac{1}{2}\eta^2J\int\!\!\int\! dxdy~x^2\theta[-xy]P(x,y)\\[1mm]
\frac{d}{dt}\omega &\!\!=\!\!&
\eta\int\!\!\int\! dxdy~|xy|\theta[-xy]P(x,y)-\eta\omega\int\!\!\int\! dxdy~x|x|\sgn(y)\theta[-xy]P(x,y)\nonumber\\
& &~~~~~~~~~~~~~~~~~~~~
-\frac{1}{2}\omega\eta^2\int\!\!\int\! dxdy
~x^2\theta[-xy]P(x,y)
\end{eqnarray*}
All integrals can again be done analytically (see appendix), so that we obtain  
explicit macroscopic flow equations:
\bd
\frac{d}{dt}J=\frac{J}{\omega}[\eta-\frac{\eta^2}{2}]I_2(\omega)
~~~~~~~~~~~~~
\frac{d}{dt}\omega
=\eta I_1(\omega)-[\eta-\frac{\eta^2}{2}]I_2(\omega)
\ed
with the short-hands
\bd
I_1(\omega)=\frac{(1-\omega^2)^{3/2}}{\pi}-\frac{\omega(1-\omega^2)}{\pi}\arccos(\omega)+\frac{\omega^2\sqrt{1-\omega^2}}{\pi}-\frac{\omega^3}{\pi}\arccos(\omega)
\ed
\bd
I_2(\omega)=-\frac{\omega(1-\omega^2)}{\pi}\arccos(\omega)+\frac{\omega^2 \sqrt{1-\omega^2}}{\pi}-\frac{\omega^3}{\pi}\arccos(\omega)
\ed
The usual translation from equations for the pair $(J,\omega)$ into one involving the pair $(J,E)$, following (\ref{eq:Einomega}), turns out to simplify matters considerably, since it gives
\begin{eqnarray}
\frac{d}{dt}J&\!\!=\!\!&
J[\frac{\eta^2}{2}-\eta]\left[E-\frac{\cos(\pi
E)\sin(\pi E)}{\pi}\right]\label{eq:adatdJ2}\\
\frac{d}{dt}E&\!\!=\!\!&
-\frac{\eta\sin^2(\pi E)}{\pi^2}+\frac{\eta^2E}{2\pi\tan(\pi E)}-\frac{\eta^2\cos^2(\pi E)}{2\pi^2}
\label{eq:adatdE}
\end{eqnarray}
The flow described by the equations (\ref{eq:adatdJ2},\ref{eq:adatdE}) 
is shown in figure \ref{fig:adatflow}, for the case $\eta=1$. 
In contrast with the 
Hebbian and the perceptron learning rules we here observe from the equations (\ref{eq:adatdJ2},\ref{eq:adatdE}) that the learning 
rate $\eta$ cannot be eliminated from the macroscopic laws by a rescaling of the weight vector length $J$. Moreover, 
the state $E=0$ is stable only 
for $\eta<3$, in which case $\frac{d}{dt}E<0$ for all $t$.  
For $\eta<2$ one has $\frac{d}{dt}J<0$ for all $t$, for $\eta=2$ one has $J(t)=J(0)$ for all $t$, and for $2<\eta<3$ we have $\frac{d}{dt}J>0$ for all $t$. 

For small $E$ equation (\ref{eq:adatdE}) reduces to
\bd
\frac{d}{dt}E=[\frac{\eta^2}{3}-\eta]E^2 +\order(E^4)
\ed
giving
\be
E\sim \frac{3t^{-1}}{\eta(3\minus\eta)}
~~~~~~~~~~(t\to\infty)
\label{eq:Edecayadat}
\ee
For $\eta=1$, which gives the standard representation of the AdaTron alrorithm, we 
find $E\sim\frac{3}{2}t^{-1}$. 
Note from equation (\ref{eq:adatdJ2}) that 
for the AdaTron rule 
there is a value for $\eta$ which 
normalises the length $J$ of the student's weight vector, 
$\eta=2$, which again gives 
$E\sim\frac{3}{2}t^{-1}$. 
The optimal value for $\eta$, however, is 
$\eta=\frac{3}{2}$ in which case we find 
$E\sim\frac{4}{3}t^{-1}$ (see (\ref{eq:Edecayadat})).

\subsection{Theory Versus Simulations}

\begin{figure}[t]
\centering
\vspace*{235mm}
\hbox to \hsize{\hspace*{-40mm}\includegraphics{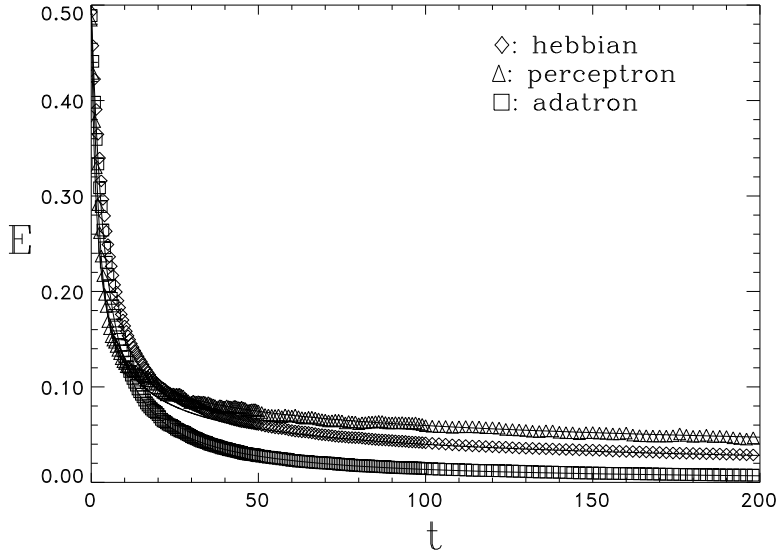}\hspace*{40mm}}
\vspace*{-155mm}
\caption{Evolution in time of the generalization error $E$  
as measured during numerical simulations 
(with $N=1000$ neurons)  
of three different learning rules:
Hebbian (diamonds), perceptron (triangles) and
AdaTron (squares). 
Initial state: $E(0)=\frac{1}{2}$ 
(random guessing) and $J(0)=1$. Learning rate: $\eta=1$. The solid lines give for each 
learning rule the
prediction of the $N=\infty$ theory, obtained by numerical solution of the flow equations for $(E,J)$. 
} 
\label{fig:Ecompare1}
\end{figure}
\begin{figure}[t]
\centering
\vspace*{235mm}
\hbox to \hsize{\hspace*{-40mm}\includegraphics{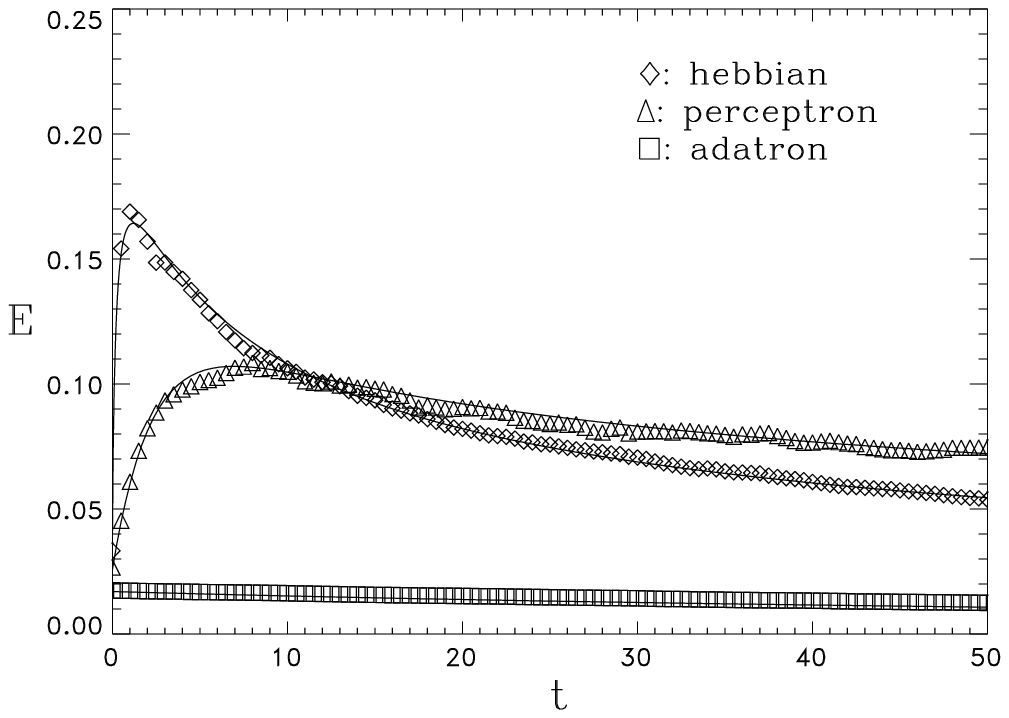}\hspace*{40mm}}
\vspace*{-155mm}
\caption{Evolution in time of the generalization error $E$  
as measured during numerical simulations 
(with $N=1000$ neurons)  
of three different learning rules:
Hebbian (diamonds), perceptron (triangles) and
AdaTron (squares). 
Initial state: $E(0)\approx 0.025$  
and $J(0)=1$. Learning rate: $\eta=1$. The solid lines give for each 
learning rule the
prediction of the $N=\infty$ theory, obtained by numerical solution of the flow equations for $(E,J)$. 
} 
\label{fig:Ecompare2}
\end{figure}
\begin{figure}[t]
\centering
\vspace*{235mm}
\hbox to \hsize{\hspace*{-40mm}\includegraphics{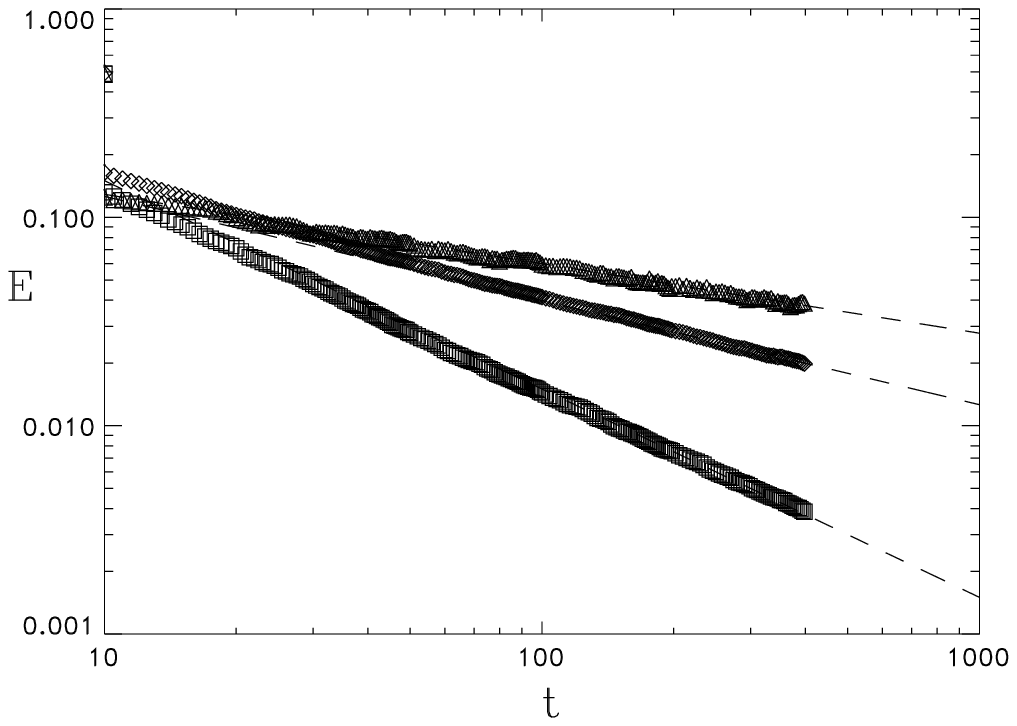}\hspace*{40mm}}
\vspace*{-155mm}
\caption{Asymptotic behaviour of the generalization error $E$  
measured during numerical simulations 
(with $N=1000$)  
of three different learning rules:
Hebbian (diamonds, middle curve), perceptron (triangles, upper curve) and
AdaTron (squares, lower curve). 
Initial state: $E(0)=\frac{1}{2}$ 
and $J(0)=1$. Learning rate: $\eta=1$. The dashed lines 
give for each 
learning rule the corresponding power law 
predicted  
by the $N=\infty$ theory (equations (\ref{eq:Edecayhebb},\ref{eq:Edecayperc},\ref{eq:Edecayadat}), respectively). 
} 
\label{fig:Ecomparelog}
\end{figure}
We close this section with results of comparing the dynamics described by the various macroscopic flow equations with the results of measuring 
the error $E$ during numerical simulations of the various (microscopic) learning rules discussed so far. 
This will serve to support the analysis and its implicit and explicit assumptions, but also illustrates how the three learning rules compare among one another. 
Figures \ref{fig:Ecompare1} and \ref{fig:Ecompare2} show the initial stage of the learning processes, 
for initialisations corresponding to random guessing ($E=0.5$) and
almost correct classification ($E$ small), 
respectively (note that for the perceptron and adatron rules starting at precisely $E=0$ produces in finite systems a stationary state). 
Here the solutions of the flow equations (solid lines) were obtained by numerical iteration. The initial increase in the error $E$, as observed for the Hebbian and perceptron rule, 
following initialisation with small values of $E$ can be understood as follows.
The error depends only on the angle of the weight vector $\bJ$, not on its length $J$, this means that the modifications generated by the Hebbian and Perceptron learning rules (which are of uniform magnitude) generate large changes in $E$ when $J$ is small, but small changes in $E$ when $J$ is large, with corresponding effects on the stability of low $E$ states. 
The AdaTron rule, in contrast, involves weight changes which scale with the length $J$, so that the stability of the $E=0$ state does not depend on the value of $J$. 
Figure \ref{fig:Ecomparelog} shows the asymptotic relaxation of the error $E$, in a log-log plot, together with the three corresponding asymptotic (power law) predictions (\ref{eq:Edecayhebb},\ref{eq:Edecayperc},\ref{eq:Edecayadat}).  
All simulations were carried out with networks of $N=1000$ neurons, which apparently is already
sufficiently large for the $N=\infty$ theory to apply. The teacher weight vectors $\bB$ were in all cases drawn at random from $[-1,1]^N$. We conclude that the theory describes the simulations essentially perfectly.


\pagebreak
\section{On-Line Learning: Complete Training Sets and Optimised Rules}

We now set out to use our macroscopic equations in `reverse mode'. Rather 
than calculate the macroscopic dynamics for a given choice of learning 
rule, we will try to find learning rules that optimise the macroscopic 
dynamical laws in the sense that they produce the fastest decay towards the
desired $E=0$ state. As a bonus it will turn out that in many cases we
can even solve the corresponding macroscopic differential  equations analytically, and find explicit expressions for $E(t)$, or rather its inverse $t(E)$. 

\subsection{Time-Dependent Learning Rates}

First we illustrate how modifying existing learning rules in a 
simple way, by just
allowing for suitably chosen time-dependent learning rates $\eta(t)$, 
can already lead to a drastic improvement in the asymptotic behaviour of the error $E$. 

We will 
inspect two specific choices of time-dependent learning rates for the
perceptron rule. Without loss of generality we can always 
put 
$\eta(t)=K(t)J(t)$ in our dynamic equations (for notational
convenience we will drop the explicit time argument of $K$). This
choice will enable us to decouple the dynamics of $J$ from that of the
generalization 
error $E$. 
For the perceptron rule  
we subsequently find equation (\ref{eq:Eperceptron}) being replaced 
by  
\begin{displaymath}
\frac{d}{dt}E=-\frac{K\sin(\pi E)}{\pi\sqrt{2\pi}}+\frac{K^2E}{2\pi\tan(\pi E)}
\end{displaymath}
giving for small $E$
\begin{displaymath}
\frac{d}{dt}E=-\frac{KE}{\sqrt{2\pi}}+\frac{K^2}{2\pi^2}+\order(K^2 E^2)
\end{displaymath} 
In order to obtain $E\rightarrow0$ for $t\rightarrow\infty$ it is clear that we need $K\rightarrow0$. Applying the ansatz $E=A/t^\alpha$, $K=B/t^\beta$ for the asymptotic forms in the previous equation produces
\begin{displaymath}
-At^{-\alpha-1}=\frac{-ABt^{-\alpha-\beta}}{\sqrt{2\pi}}+\frac{B^2 t^{-2\beta}}{2\pi^2}+\order(t^{-2\alpha-2\beta})
\end{displaymath}
and so: $\alpha=\beta=1$ and $A=\frac{1}{\pi\sqrt{2\pi}}\frac{B^2}{(B-\sqrt{2\pi})}$. 
Our aim is to obtain the fastest approach of the $E=0$ state,
i.e. we wish to maximise $\alpha$ (for which we found $\alpha=1$) 
and subsequently minimise $A$. 
Apparently the value of $B$ for which   $A$ is minimized is $B=2\sqrt{2\pi}$, 
in which case we obtain the error decay given by
\be
\eta\sim \frac{2J\sqrt{2\pi}}{t}:
~~~~~~~~~~E\sim\frac{4}{\pi t }~~~~~~~~~~(t\to\infty)
\label{eq:percvaryingeta1}
\ee
This is clearly a great improvement upon the result for the perceptron rule with 
constant $\eta$, i.e. equation (\ref{eq:Edecayperc}); in fact it is the fastest relaxation we have derived so far. 
\vsp

Let us now move to an alternative choice for the time-dependent learning rate for the perceptron. 
According to equation (\ref{eq:Jperceptron}) there is one specific recipe for $\eta(t)$ such that the length $J$ of the student's weight vector will remain constant, given by 
\be
\eta=\sqrt{\frac{2}{\pi}}\frac{J}{E}(1\minus \cos(\pi E))
\label{eq:expressionforeta}
\ee
Making this choice converts equation (\ref{eq:Eperceptron}) for the evolution of $E$ 
into
\begin{equation}
\frac{d}{dt}E=-\frac{(1\minus \cos(\pi E))^2}{\pi^2E\sin(\pi E)}
\label{eq:etanormE}
\end{equation}
Equation  (\ref{eq:etanormE}) can be written in the form $\frac{d}{dE}t=g(E)$, so that $t(E)$ becomes a simple integral which can be done analytically, with the result
\begin{equation}
t(E)=\frac{\pi E+\sin(\pi E)}{1-\cos(\pi E)}-\frac{\pi E_0+\sin(\pi E_0)}{1-\cos(\pi E_0)}
\label{eq:exact1}
\end{equation}
(which can also be verified directly by substitution into (\ref{eq:etanormE})). 
Expansion of (\ref{eq:exact1}) and (\ref{eq:expressionforeta}) for small $E$ gives the asymptotic behaviour also 
encountered in (\ref{eq:percvaryingeta1}): 
\be
\eta\sim \frac{2J\sqrt{2\pi}}{t},
~~~~~~E\sim\frac{4}{\pi t }~~~~~~~~~~(t\to\infty)
\label{eq:percvaryingeta2}
\ee
It might appear that implementation of the recipe (\ref{eq:expressionforeta}) 
is in practice impossible, since it involves information which is not available to the student perceptron 
(namely the instantaneous error $E$).  However, since we know  
(\ref{eq:exact1}) we can simply calculate the required 
$\eta(t)$ explicitly as a function of time. 
\vsp

One has to be somewhat careful in 
extrapolating results such as those obtained in this section. For instance, choosing the time-dependent learning rate (\ref{eq:expressionforeta}) enforces the constraint $\bJ^2(t)=1$ in the macroscopic equations for $N\to \infty$. 
This is not identical to choosing $\eta(t_\mu)$ in the original equation (\ref{eq:genericrule}) such as to enforce $\bJ^2(t_\mu+\frac{1}{N})=\bJ^2(t_\mu)$ at the level of individual iteration steps, as can be seen by working out the dynamical laws. The latter case would correspond to the {\em microscopically fluctuating} choice 
\bd
\eta(t_\mu)=-2\frac{(\bJ(t_\mu)\inn\bxi^\mu)
\sgn(\bB\inn\bxi^\mu)}{\cF[|\bJ(t_\mu)|;\bJ(t_\mu)\inn\bxi^\mu,\bB\inn\bxi^\mu]}~~~~~~~~~~{\rm if}~~~~\cF[|\bJ(t_\mu)|;\bJ(t_\mu)\inn\bxi^\mu,\bB\inn\bxi^\mu]\neq 0
\ed
If we now choose for example $\cF[J;Jx,y]=\theta[-xy]$, implying $\eta(t_\mu)=2|\bJ(t_\mu)\inn\bxi^\mu|$, we find by insertion into 
(\ref{eq:genericrule}) that the perceptron rule with `hard' 
weight normalisation at each iteration step via adaptation of the learning rate is identical to the AdaTron rule with constant learning rate $\eta=2$. 
We know therefore that in this case one obtains $E\sim 3/2t$, whereas 
for the Perceptron rule with `soft' weight normalisation via (\ref{eq:expressionforeta}) (see the analysis above) one obtains $E\sim 4/\pi t$. 
Apparently the two procedures are not equivalent.

\subsection{Spherical On-Line Learning Rules}

We arrive in a natural way at the question of how to find the optimal time-dependent learning rate for any given learning rule, or more generally: of how to find the optimal learning rule. 
This involves variational calculations in two-dimensional flows (since our macroscopic equations are defined in terms of the evolving pair $(J,E)$). 
Such calculations would be much simpler if our macroscopic equations were just one-dimensional, 
e.g. describing only the evolution of the error $E$ with a stationary (or simply irrelevant) value of the length $J$. Often it will turn out that for finding the optimal learning rate or the optimal learning rule the problem can indeed be reduced to a one-dimensional one. To be able to obtain results also 
for those cases where this reduction does
not happen
we will now construct so-called spherical learning rules, where $\bJ^2(t)=1$ for all $t$. 
This can be arranged in several equivalent ways. 

The first method is to add to the general rule (\ref{eq:genericrule}) a term proportional to the instantaneous weight vector $\bJ$, whose sole purpose is to achieve the constraint $\bJ^2=1$:
\be
\bJ(t_\mu +\frac{1}{N})
=\bJ(t_\mu)+\frac{1}{N}\left\{
\eta(t_\mu)\bxi^\mu\sgn(\bB\inn\bxi^\mu)\cF[|\bJ(t_\mu)|;
\bJ(t_\mu)\inn\bxi^\mu,
\bB\inn\bxi^\mu]-\lambda(t_\mu)\bJ(t_\mu)\right\}
\label{eq:sphericalrule1}
\ee
The evolution of the two observables $Q[\bJ]$ and $R[\bJ]$ (\ref{eq:QandR}) is now given by
\bea
Q[\bJ(t_{\mu}\plus\frac{1}{N})] & \!
\!=\!\! &
Q[\bJ(t_\mu)](1\minus\frac{2\lambda(t_\mu)}{N})
+\frac{2}{N}\eta(t_\mu)(\bJ(t_\mu)\inn\bxi^\mu)
\sgn(\bB\inn\bxi^\mu)\cF[|\bJ(t_\mu)|;\bJ(t_\mu)\inn \bxi^\mu,\bB\inn \bxi^\mu]
\nonumber \\[2mm]
&& ~~~~~~~~~~~~~~~~~~~~
+\frac{1}{N}\eta^2(t_\mu)\cF^2[|\bJ(t_\mu)|;\bJ(t_\mu)\inn \bxi^\mu,\bB\inn \bxi^\mu]
+\order(N^{-2})
\nonumber \\[2mm]
R[\bJ(t_{\mu}\plus\frac{1}{N})] & \!\!=\!\! &
R[\bJ(t_\mu)](1\minus \frac{\lambda(t_\mu)}{N})+\frac{1}{N}\eta(t_\mu)
|\bB\inn\bxi^\mu|\cF[|\bJ(t_\mu)|;\bJ(t_\mu)\inn \bxi^\mu,\bB\inn \bxi^\mu]
\nonumber
\eea
Following the procedure of section 1.2 to arrive at the $N\to\infty$ limit of the dynamical equations for $Q$ and $R$ 
then leads to (we drop explicit time arguments for notational convenience):
\bd
\frac{d}{dt}Q=2\eta Q^{\frac{1}{2}}
\bra x\sgn(y)\cF[Q^{\frac{1}{2}};Q^{\frac{1}{2}}x,y]\ket +\eta^2 \bra \cF^2[Q^{\frac{1}{2}};Q^{\frac{1}{2}}x,y]\ket 
-2\lambda Q
\ed
\bd
\frac{d}{dt}R=\eta \bra |y|\cF[Q^{\frac{1}{2}};Q^{\frac{1}{2}}x,y] \ket -\lambda R
\ed
We now choose the function $\lambda(t)$ such that $Q(t)=1$ for all $t\geq 0$. This ensures that $R(t)=\omega(t)=\hat{\bJ}(t)\inn\bB$, and gives (via $\frac{d}{dt}Q=0$) 
a recipe for $\lambda(t)$
\bd
\lambda=
\eta 
\bra x\sgn(y)\cF[1;x,y]\ket +\frac{1}{2}\eta^2 \bra \cF^2[1;x,y]\ket 
\ed
which 
can then be substituted into our equation for $\frac{d}{dt}\omega$:
\be
\frac{d}{dt}\omega=\eta \bra \left[
|y|-\omega x\sgn(y)\right]\cF[1;x,y]-\frac{1}{2}\omega\eta^2\bra \cF^2[1;x,y]\ket
\label{eq:sphericalomega1}
\ee
with averages as usual defined with respect to the Gaussian joint field distribution (\ref{eq:gaussian}), which depends only on $\omega$, so that equation (\ref{eq:sphericalomega1}) is indeed autonomous. 

The second method to arrange the constraint $\bJ^2=1$ is to explicitly normalise the weight vector $\bJ$ after each modification step, i.e.
\be
\bJ(t_\mu +\frac{1}{N})
=
\frac{\bJ(t_\mu)+\frac{1}{N}
\eta(t_\mu)\bxi^\mu\sgn(\bB\inn\bxi^\mu)\cF[
1;\bJ(t_\mu)\inn\bxi^\mu,
\bB\inn\bxi^\mu]}
{|\bJ(t_\mu)+
\frac{1}{N}
\eta(t_\mu)\bxi^\mu\sgn(\bB\inn\bxi^\mu)\cF[
1;\bJ(t_\mu)\inn\bxi^\mu,
\bB\inn\bxi^\mu]|}
\label{eq:sphericalrule2}
\ee
\bd
=
\hat{\bJ}(t_\mu)+\frac{1}{N}\eta(t_\mu)
\left\{\left[
\bxi^\mu
-\hat{\bJ}(t_\mu)
(\hat{\bJ}(t_\mu)\inn\bxi^\mu)
\right]
\sgn(\bB\inn\bxi^\mu)\cF[
1;\bJ(t_\mu)\inn\bxi^\mu,
\bB\inn\bxi^\mu]
\right.
\ed
\bd
\left.
-\frac{1}{2}\eta(t_\mu)\bJ(t_\mu)\cF^2[
1;\bJ(t_\mu)\inn\bxi^\mu,
\bB\inn\bxi^\mu]
\right\}
+\order(N^{-2})
\ed
The evolution of the observable $\omega[\bJ]=\hat{\bJ}\inn\bB$ is thus given by
\bd
\omega[\bJ(t_{\mu}\plus\frac{1}{N})] =
\omega[\bJ(t_\mu)]
+\frac{1}{N}\eta(t_\mu)\left\{
\left[
|\bB\inn\bxi^\mu|
-\omega[\bJ(t_\mu)]
(\hat{\bJ}(t_\mu)\inn\bxi^\mu)\sgn(\bB\inn\bxi^\mu)\right]
\cF[1;
\bJ(t_\mu)\inn\bxi^\mu,
\bB\inn\bxi^\mu]
\right.
\ed
\bd
\left.
-\frac{1}{2}\omega\eta(t_\mu)
\cF^2[1;
\bJ(t_\mu)\inn\bxi^\mu,
\bB\inn\bxi^\mu]
\right\}
+\order(N^{-2})
\ed
Following the procedure of section 1.2 
then leads to
\be
\frac{d}{dt}\omega=\eta \bra \left[
|y|-\omega x\sgn(y)\right]\cF[1;x,y]-\frac{1}{2}\omega\eta^2\bra \cF^2[1;x,y]\ket
\label{eq:sphericalomega2}
\ee
which is identical to equation (\ref{eq:sphericalomega1}). 

Finally we convert equation (\ref{eq:sphericalomega1}) into a dynamical  equation for the error $E$, using (\ref{eq:Einomega}), which gives the final result  
\begin{equation}
\frac{d}{dt}E=-\frac{\eta}{\pi\sin(\pi E)}\bra
\left[|y|-\cos(\pi E) x\sgn(y)\right]
\cF[1;x,y]\ket
+\frac{\eta^2}{2\pi\tan(\pi E)}\bra\cF^2[1;x,y]\ket
\label{eq:genE} 
\end{equation}
with averages defined with respect to the distribution (\ref{eq:gaussian}), in which $\omega=\cos(\pi E)$. 

For spherical models described by either 
of the equivalent classes of on-line rules (\ref{eq:sphericalrule1}) or (\ref{eq:sphericalrule2}) the evolution of the error is described by a single first-order non-linear differential equation, rather 
than a pair of coupled non-linear differential equations. This will allow us to push the analysis further, but the price we pay is that of a loss in generality. 

\subsection{Optimal Time-Dependent Learning Rates}

We wish to optimise the approach to the $E=0$ state of our macroscopic equations, by choosing a suitable time-dependent learning rate. Let us distinguish between the possible situations we can find ourselves in.
If our learning rule is of the general form (\ref{eq:genericrule}), without spherical normalisation, we have two coupled macroscopic equations:
\be
\frac{d}{dt}J=
\eta\bra
x\sgn(y)\cF[J;Jx,y]\ket+\frac{\eta^2}{2J}\bra
\cF^2[J;Jx,y]\ket
\label{eq:dJdtgeneral}
\ee
\be
\frac{d}{dt}E =
-\frac{\eta}{J\pi\sin(\pi E)}\bra\left[|y|\minus\cos(\pi E)
x\sgn(y)\right]\cF[J;Jx,y]\ket
+\frac{\eta^2}{2\pi J^2\tan(\pi E)}\bra\cF^2[J;Jx,y]\ket
\label{eq:dEdtgeneral}
\ee
which are obtained by combining (\ref{eq:genJinxy},\ref{eq:genomegainxy}) with (\ref{eq:Einomega}). The probability distribution (\ref{eq:gaussian}) 
with which the averages are computed depends on $E$ only, not on $J$. 
If, on the other hand, we complement the rule (\ref{eq:genericrule}) with weight vector normalisation as in (\ref{eq:sphericalrule1}) or (\ref{eq:sphericalrule2}) (the spherical rules), we obtain a single equation for $E$ only:
\begin{equation}
\frac{d}{dt}E=-\frac{\eta}{\pi\sin(\pi E)}\bra
\left[|y|-\cos(\pi E) x\sgn(y)\right]
\cF[1;x,y]\ket
+\frac{\eta^2}{2\pi\tan(\pi E)}\bra\cF^2[1;x,y]\ket
\label{eq:genEagain} 
\end{equation}
Since equation (\ref{eq:genEagain}) is autonomous (there are no
dynamical variables other than $E$),  
the optimal choice of the function 
$\tilde{\eta}(t)$ (i.e. the one that generates the fastest decay of the error $E$) 
is obtained by simply minimising the temporal derivative of the error {\em at each time-step}:
\be
\forall t\geq 0:~~~~~~~~\frac{\partial}{\partial\tilde{\eta}(t)}\left[\frac{d}{dt}E\right]=0
\label{eq:greedy}
\ee
which is called the `greedy' recipe. Note, however, that the same is true for equation 
(\ref{eq:dEdtgeneral}) if we restrict ourselves to rules with the property
that 
$\cF[J;Jx,y]=\gamma(J)\cF[1;x,y]$ for some function $\gamma(J)$, 
such as the Hebbian ($\gamma(J)=1$), perceptron ($\gamma(J)=1$) and
AdaTron ($\gamma(J)=J$) rules.  
This property can also be written as
\be
\frac{\partial}{\partial x}\frac{\cF[J;Jx,y]}{\cF[1;x,y]}
=\frac{\partial}{\partial y}\frac{\cF[J;Jx,y]}{\cF[1;x,y]}=0
\label{eq:Jirrelevant}
\ee
For rules which obey (\ref{eq:Jirrelevant}) we can simply  
write the time-dependent learning rate as
$\eta=\tilde{\eta}J/\gamma(J)$, such that equations
(\ref{eq:dJdtgeneral},\ref{eq:dEdtgeneral})  acquire the form:
\be
\frac{d}{dt}\log J=
\tilde{\eta}\bra
x\sgn(y)\cF[1;x,y]\ket+\frac{1}{2}\tilde{\eta}^2\bra
\cF^2[1;x,y]\ket
\label{eq:dJvaryeta}
\ee
\be
\frac{d}{dt}E =
-\frac{\tilde{\eta}}{\pi\sin(\pi E)}\bra\left[|y|\minus\cos(\pi E)
x\sgn(y)\right]\cF[1;x,y]\ket
+\frac{\tilde{\eta}^2}{2\pi \tan(\pi E)}\bra\cF^2[1;x,y]\ket
\label{eq:dEvaryeta}
\ee
In these cases, precisely since we are free to choose the function $\tilde{\eta}(t)$ as we wish, the evolution of $J$ decouples from our problem of optimising the evolution of $E$.
For learning rules where $\cF[J;Jx,y]$ truly depends on $J$, on the other
hand (i.e. where (\ref{eq:Jirrelevant}) does not hold), optimisation of the error relaxation is considerably more difficult, and 
is likely to depend on the particular time $t$ for which one wants to minimise $E(t)$.
We will not deal with such cases here. 

If the `greedy' recipe applies (for spherical rules and for ordinary
ones with the property 
(\ref{eq:Jirrelevant})) 
working out the derivative in (\ref{eq:greedy}) 
immediately gives us
\begin{equation}
\tilde{\eta}(t)_{\opt}=\frac{\langle\{|y|\minus\cos(\pi E) x\sgn(y)\}\cF[1;x,y]\rangle}{\cos(\pi E)\langle\cF^2[1;x,y]\rangle}
\label{eq:etaopt}  
\end{equation}
Insertion of this choice into 
equation (\ref{eq:genE}) subsequently leads to
\begin{equation}
\left.\frac{dE}{dt}\right|_\opt=-\frac{\langle\{|y|\minus \cos(\pi E) x\sgn(y)\}\cF[1;x,y]\rangle^2}{2\pi\sin(\pi E)\cos(\pi E)\langle\cF^2[1;x,y]\rangle}
\label{eq:dEopt}   
\end{equation}
These and subsequent expressions we will write in terms of $\tilde{\eta}$, 
defined as $\tilde{\eta}(t)=\eta(t)$ for the spherical learning
rules and as $\tilde{\eta}(t)=\eta(t)J(t)/\gamma(J(t))$ for
the  non-spherical learning rules. 
We will now work out the details of the results
(\ref{eq:etaopt},\ref{eq:dEopt}) upon making the familiar choices for the function $\cF[\ldots]$:
the Hebbian, perceptron and AdaTron rules.  
\vsp

For the (ordinary and spherical)  
Hebbian rules, corresponding to  $\cF[J;Jx,y]=1$, 
the various Gaussian integrals in (\ref{eq:etaopt},\ref{eq:dEopt}) are
the same as those we already did (analytically) in the case of constant learning rate $\eta$.
Substitution of the outcomes of the integrals (see appendix) into the equations (\ref{eq:etaopt},\ref{eq:dEopt}) gives  
\bd
\tilde{\eta}_\opt
=\sqrt{\frac{2}{\pi}}~\frac{\sin^2(\pi E)}{\cos(\pi E)}
~~~~~~~~~~~~~~
\left.
\frac{dE}{dt}\right|_\opt =
-
\frac{\sin^3(\pi E)}{\pi^2\cos(\pi E)}
\ed
The equation for the error $E$ can be solved explicitly, giving (to be verified by substitution):
\begin{equation}
t(E)=\frac{1}{2}\pi \sin^{-2}(\pi E)-\frac{1}{2}\pi \sin^{-2}(\pi E_0)
\label{eq:Hebbopt}
\end{equation}
The asymptotic behaviour of the process follows from expansion of (\ref{eq:Hebbopt}) for small $E$, and gives
\bd
E_\opt\sim\frac{1}{\sqrt{2\pi t}}
~~~~~~~~~~~~~~
\tilde{\eta}_\opt\sim\sqrt\frac{\pi}{2}\,\frac{1}{t}
~~~~~~~~~~~~~~(t\to\infty)
\ed
Asymptotically there is  
nothing to be gained by choosing the optimal time-dependent learning
rate, since the same asymptotic form for $E$ was also obtained for
constant $\eta$ (see (\ref{eq:Edecayhebb})). 
 Note that 
the property $\cF[J;Jx,y]=\cF[1;x,y]$ of the Hebbian recipe guarantees
that the 
result (\ref{eq:Hebbopt})  applies to both the ordinary and the
spherical Hebbian rule. 
The only difference between the two cases is
in the definition of $\tilde{\eta}$: 
for the ordinary (non-spherical) version
$\tilde{\eta}(t)=\eta(t)/J(t)$, whereas for the spherical version
$\tilde{\eta}(t)=\eta(t)$.

We move on to the (ordinary and spherical) perceptron learning rules, where
$\cF[J;Jx,y]=\theta[-xy]$, with time-dependent learning rates $\eta(t)$
which we aim  to optimise. As in the Hebbian case
all integrals occurring in (\ref{eq:etaopt},\ref{eq:dEopt}) upon
substitution of the present choice $\cF[J;Jx,y]=\theta[-xy]$ have been
done already (see the appendix) . Insertion of the outcomes of these
integrals  into (\ref{eq:etaopt},\ref{eq:dEopt}) gives
\begin{displaymath}
\tilde{\eta}_\opt=\frac{\sin^2(\pi E)}{\sqrt{2\pi}E\cos(\pi E)}
~~~~~~~~~~~~~~
\left.\frac{dE}{dt}\right|_\opt =- \frac{\sin^3(\pi E)}{4\pi^2 E\cos(\pi E)}
\end{displaymath}
Again the non-linear differential equation describing the evolution of the error $E$ can be solved
exactly:
\be
t(E)=\frac{2[\pi E+\sin(\pi E)\cos(\pi E)]}{\sin^2(\pi E)}-\frac{2[\pi E_0+\sin(\pi E_0)\cos(\pi E_0)]}{\sin^2(\pi E_0)}
\label{eq:exact2}
\ee
Expansion of (\ref{eq:exact2}) for small $E$ gives the asymptotic
behaviour
\bd
E_\opt \sim\frac{4}{\pi t} 
~~~~~~~~~~~~
\tilde{\eta}_\opt\sim\frac{2\sqrt{2\pi}}{t}
~~~~~~~~~~~~(t\rightarrow\infty)
\ed
which is identical to that found in the beginning of this section,
i.e. equations (\ref{eq:percvaryingeta1},\ref{eq:percvaryingeta2}), 
upon exploring the consequences of making two simple ad-hoc  choices
for the time-dependent
learning rate 
(since $\tilde{\eta}=\eta/J$). 
As with the Hebbian rule 
the property $\cF[J;Jx,y]=\cF[1;x,y]$ of the perceptron recipe guarantees
that the 
result (\ref{eq:exact2})  applies to both the ordinary and the
spherical version. 

Finally we try to optimise the learning rate for the spherical AdaTron
learning rule, corresponding to the choice
$\cF[J;Jx,y]=|Jx|\theta[-xy]$. Working out the averages in
(\ref{eq:etaopt},\ref{eq:dEopt}) again does not require doing any new
integrals. Using those already encountered in analysing the AdaTron rule
with constant learning rate (to be found in the appendix), we obtain  
\bd
\tilde{\eta}_\opt=\frac{\sin^3(\pi E)}{\pi}\left[E\cos(\pi
E)-\frac{\cos^2(\pi E)\sin(\pi E)}{\pi}\right]^{-1}
\ed
\bd
\left.\frac{dE}{dt}\right|_\opt
=-\frac{\sin^5(\pi E)}{2\pi^2\cos(\pi E)}\left[\frac{1}{\pi E-\cos(\pi
E)\sin(\pi E)}\right]
\ed
(note that in both versions, ordinary and spherical, of the AdaTron
rule we simply have $\tilde{\eta}(t)=\eta(t)$). 
It will no longer come as a surprise that also this equation for the
evolution of the error allows for analytical solution:
\begin{equation}
t(E)=\frac{\pi}{8}\left[\frac{4\pi E-\sin(4\pi E)}{\sin^4(\pi
E)}-\frac{4\pi E_0-\sin(4\pi E_0)}{\sin^4(\pi E_0)}\right]
\label{eq:adatexact}
\end{equation}
Asymptotically we find, upon expanding (\ref{eq:adatexact}) for small
$E$, a relaxation of the form 
\bd
E_\opt\sim \frac{4}{3t}
~~~~~~~~~~~~~~
\tilde{\eta}_\opt \sim \frac{3}{2}
~~~~~~~~~~~~~~(t\rightarrow\infty)
\ed
So for the AdaTron rule the asymptotic behaviour for optimal
time-dependent learning rate $\eta$ is identical to that 
found for optimal {\em constant} learning rate $\eta$ (which is indeed
$\eta=\frac{3}{2}$, see (\ref{eq:Edecayadat})).  
As with the previous two rules, 
the property $\cF[J;Jx,y]=J\cF[1;x,y]$ of the AdaTron recipe guarantees
that the 
result (\ref{eq:Hebbopt})  applies to both the ordinary and the
spherical version. 
\vsp

It is quite remarkable that the simple perceptron learning rule, which
came out at the bottom of the league among the three learning rules 
considered so far in
the case of having constant learning rates, all of a sudden 
comes out with `douze
points' as soon as we allow for optimised time-dependent learning
rates. 
It is in addition quite satisfactory that in a number
of cases one can actually find an explicit expression for the relation
$t(E)$ between the duration of the learning stage and the
generalization error achieved, i.e. equations
(\ref{eq:exact1},\ref{eq:Hebbopt},\ref{eq:exact2},\ref{eq:adatexact}).

\subsection{Optimal On-line Learning Rules}

We need not restrict our optimisation attempts to varying
the learning rate $\eta$ only, but we can also vary the full form 
$\eta\cF[J;Jx,y]$ of the learning rule. 
The aim, as always,
is to minimise the generalisation error, but there will be limits to
what is achievable.  
So far all examples of on-line learning rules we have
studied gave an asymptotic relaxation of the error of the form $E\sim
t^{-q}$ with $q\leq 1$. 
It can be shown using general probabilistic
arguments that if one only has $p=\alpha N$
examples of randomly drawn 
question/answer pairs $\{\bxi^\mu,T(\bxi^\mu)\}$ with
which to calculate the weight vector $\bJ$ of an
$N$-neuron binary student perceptron (whether in an on-line or a batch
fashion), the generalisation error $E_g(\bJ)$ obeys the inequality
$E_g(\bJ)\geq 0.44\ldots/\alpha$ for $N\to\infty$ (this is the one result we will mention
without derivation). For on-line learning rules of the class
(\ref{eq:genericrule}) or
(\ref{eq:sphericalrule1},\ref{eq:sphericalrule2})
we have used at time $t$ a number of examples
$p\leq tN$, so this inequality translates into 
\be
\lim_{t\to\infty} ~tE(t)~\geq~ 0.44\ldots
\label{eq:opper}
\ee
No on-line learning rule can 
violate (\ref{eq:opper})\footnote{This will be different for
graded-response perceptrons.}. On the other hand: we have already
encountered several rules with at least the optimal power $E\sim t^{-1}$. 
The optimal on-line
learning rule is thus one which gives asymptotically $E\sim A/t$, but
with the smallest value of $A$ possible.  

The function $\cF[J;Jx,y]$ in the learning rules
is allowed to depend only on the {\em sign} of 
the teacher field
$y=\bB\inn\bxi$, not on its magnitude, since otherwise it would
describe a situation where considerably more than  just the answers $T(\bxi)=\sgn[\bB\inn\bxi]$ of the
teacher are used for updating the parameters of the student. One can
easily see that using unavailable information indeed violates
(\ref{eq:opper}). Suppose, for instance, we would consider spherical
on-line rules, i.e. 
(\ref{eq:sphericalrule1}) or (\ref{eq:sphericalrule2}), 
and  make the forbidden choice 
\bd
\eta\cF[1;x,y]=\frac{|y|\minus \cos(\pi E) x\sgn(y)}{\cos(\pi E)}
\ed
We would then find for the corresponding equation (\ref{eq:genEagain})
describing the evolution of the error $E$ for $N\to\infty$:
\bd
\frac{d}{dt}E=-
\frac{\bra\left[ |y|\minus \cos(\pi E) x\sgn(y)\right]^2\ket}{2\pi\sin(\pi E)\cos(\pi E)}
\ed
(with averages as always calculated with the distribution
(\ref{eq:gaussian})) 
from which it follows, upon using the Gaussian intregrals done in the appendix:
\bd
\frac{d}{dt}E = -\frac{\tan(\pi E)}{2\pi}
\ed
This produces exponential decay of the error, and thus indeed violates
(\ref{eq:opper}).  
\vsp

Taking into account the restrictions on available information, 
and
anticipating the form subsequent expressions will take, we write 
the function $\cF[J;Jx,y]$ (which we will be varying, and which we
will also allow to have an explicit time-dependence\footnote{By
allowing for an explit time-dependence, we can drop the 
dependence on $J$ in $\cF[J;Jx,y]$ if we wish, 
without loss of generality, since
$J$ is itself just some function of time.}  
) in the following form
\be
\eta\cF[J;Jx,y]=\left\{
\begin{array}{lll}
J\cF_+(x,t) && {\rm if}~ y>0\\[1mm]
J\cF_-(x,t) && {\rm if}~ y<0
\end{array}
\right.
\label{eq:optimalrule}
\ee
If our learning rule is of the general form (\ref{eq:genericrule}),
without spherical normalisation, the coupled equations
(\ref{eq:dJdtgeneral},\ref{eq:dEdtgeneral}) describe the macroscopic
dynamics. For the spherical rules
(\ref{eq:sphericalrule1},\ref{eq:sphericalrule2})  we have the single
macroscopic equation (\ref{eq:genEagain}). 
Both (\ref{eq:dEdtgeneral}) and (\ref{eq:genEagain}) now acquire the
form 
\bd
\frac{d}{dt}E =
-\frac{1}{\pi\sin(\pi E)}\left\{\room
\bra(y\minus\omega x)\theta[y]\cF_+(x,t)\ket
-\bra(y\minus\omega x)\theta[\minus y]\cF_-(x,t)\ket
~~~~~~
\right.
\ed
\be
~~~~~~
\left.\room
-\frac{1}{2}\omega \bra
\theta[y]\cF^2_+(x,t)
\ket
-\frac{1}{2}\omega \bra
\theta[\minus y]\cF^2_-(x,t)
\ket
\right\}
\label{eq:dEdtavailable}
\ee
with the usual short-hand $\omega=\cos(\pi E)$ and with averages
calculated with the (time-dependent) distribution (\ref{eq:gaussian}).
To simplify notation we now introduce the two functions   
\bd
\int\! dy~\theta[y] P(x,y) = \Omega(x,t)
~~~~~~~~~~~~
\int\! dy ~\theta[y](y\minus \omega x)P(x,y) = \Delta(x,t)
\ed
and hence, using the symmetry $P_t(x,y)=P_t(\minus x,\minus y)$, equation
(\ref{eq:dEdtavailable}) acquires the compact form
\bd
\frac{d}{dt}E =
-\frac{1}{\pi\sin(\pi E)}
\int\!dx\left\{
\Delta(x,t)\cF_+(x,t)-\frac{1}{2}\omega
\Omega(x,t)\cF^2_+(x,t)\right\}~~~~~~
\ed
\be
~~~~~~
-\frac{1}{\pi\sin(\pi E)}
\int\!dx\left\{
\Delta(\minus x,t)\cF_-(x,t)\ket
-\frac{1}{2}\omega \Omega(\minus x,t)\cF^2_-(x,t)
\right\}
\label{eq:dEdtavailablenew}
\ee
Since there is only one dynamical variable, the error $E$, our
optimisation problem is solved by the `greedy' recipe which here
involves functional derivatives:
\bd
\forall x,~\forall t:~~~~~~
\frac{\delta}{\delta \cF_+(x,t)}\left[\frac{dE}{dt}\right]=
\frac{\delta}{\delta \cF_-(x,t)}\left[\frac{dE}{dt}\right]=0
\ed
with the solution
\bd
\cF_+(x,t)
=\frac{\Delta(x,t)}{\omega\Omega(x,t)}
~~~~~~~~~~
\cF_-(x,t)=
\frac{\Delta(\minus x,t)}{\omega \Omega(\minus x,t)}=\cF_+(\minus x,t)
\ed
Substitution of this solution into
(\ref{eq:dEdtavailablenew}) gives the corresponding law describing the
optimal error evolution of (ordinary and spherical) 
on-line rules:
\bd
\left.
\frac{dE}{dt}\right|_\opt =
-\frac{1}{\pi\sin(\pi E)\cos(\pi E)}
\int\!dx
\frac{\Delta^2(x,t)}{\Omega(x,t)}
\ed
Explicit calculation of the integrals  $\Delta(x,t)$ and $\Omega(x,t)$ (see
appendix) gives: 
\bd
\Delta(x,t)=\frac{\sin(\pi E)}{2\pi}e^{-\frac{1}{2}x^2/\sin^2(\pi E)}
~~~~~~~~~~
\Omega(x,t)=
\frac{e^{-\frac{1}{2}x^2}}{2\sqrt{2\pi}}\left[1\plus \erf\left(x/\sqrt2 \tan(\pi E)\right)\right]
\ed
\begin{figure}[t]
\centering
\vspace*{235mm}
\hbox to \hsize{\hspace*{-40mm}\includegraphics{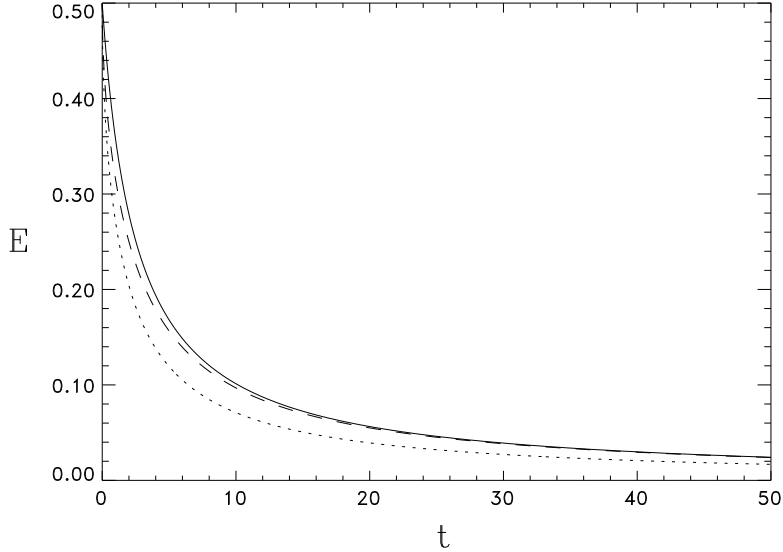}\hspace*{40mm}}
\vspace*{-155mm}
\caption{Evolution of the error $E$ 
for three on-line learning 
rules: Perceptron rule
with a learning rate such that $J(t)=1$ for all $t\geq 0$ (solid line), Perceptron rule with optimal learning rate (dashed line) and the optimal spherical learning rule (dotted line). 
Initial state: $E(0)=\frac{1}{2}$ and $J(0)=1$. 
The curves for the Perceptron rules are given by (\ref{eq:exact1}) and (\ref{eq:exact2}). The curve
for the optimal spherical rule was obtained by numerical solution of equation ((\ref{eq:optimalonlinedEdt}). 
}
\label{fig:exact}
\end{figure}
\begin{figure}[t]
\centering
\vspace*{235mm}
\hbox to \hsize{\hspace*{-40mm}\includegraphics{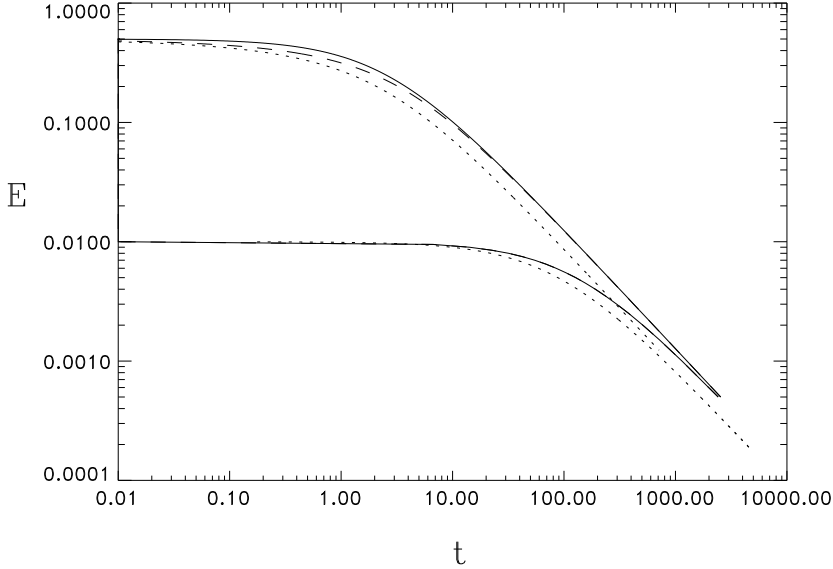}\hspace*{40mm}}
\vspace*{-155mm}
\caption{Evolution of the error $E$ 
for the  
on-line Perceptron rule
with a learning rate such that $J(t)=1$ for all $t\geq 0$ (solid line), the on-line Perceptron rule with optimal learning rate (dashed line) and the optimal spherical on-line learning rule (dotted line). 
Initial states: $(J,E)=(1,\frac{1}{2})$ (upper curves), and $(J,E)=(1,\frac{1}{100})$ (lower curves).
The curves for the Perceptron rules are given by (\ref{eq:exact1}) and (\ref{eq:exact2}). The curves 
for the optimal spherical rule were obtained by numerical solution of equation (\ref{eq:optimalonlinedEdt}). 
}
\label{fig:exactlog}
\end{figure}
\noindent
with which we finally obtain an explicit expression for the optimal
form of the learning rule, via (\ref{eq:optimalrule}), as well as for the dynamical law
describing the corresponding error evolution:
\be
\eta\cF[J;Jx,y]_\opt 
=\sqrt{\frac{2}{\pi}}~
\frac{J\tan(\pi E) e^{-\frac{1}{2}x^2/\tan^2(\pi E)}}
{1\plus \sgn(xy)\erf\left(|x|/\sqrt2\tan(\pi E)\right)}
\label{eq:optimalonlinerule}
\ee
\be
\left.\frac{dE}{dt}\right|_\opt=-\frac{\tan^2(\pi
E)}{\pi^2\sqrt{2\pi}}
\int\!dx~\frac{e^{-\frac{1}{2}x^2[1+\cos^2(\pi E)]/\cos^2(\pi E)}}
{1+\erf(x/\sqrt2)}
\label{eq:optimalonlinedEdt}
\ee
The asymptotic form of the error relaxation towards the  $E=0$ state
follows from expansion of equation (\ref{eq:optimalonlinedEdt}) for small
$E$, which gives
\begin{displaymath}
\frac{dE}{dt}=-E^2\int\!dy\frac{e^{-y^2}}{\sqrt{2\pi}[1+\erf(y/\sqrt2)] }+\order(E^4)
\end{displaymath}
so that we can conclude that the optimum asymptotic decay for on-line
learning rules (whether spherical or non-spherical) is given by
$E\sim\frac{A}{t}$ for $t\to\infty$, with
\bd
A^{-1}=\int\!dx~\frac{e^{-x^2}}{\sqrt{2\pi}[1+\erf(x/\sqrt2)]}
\ed
Numerical evaluation of this integral (which is somewhat delicate due
to the behaviour of the integrand for $y\to\minus \infty$) 
finally gives
\begin{displaymath}
E~\sim~\frac{0.883\ldots}{t}~~~~~~~~~~(t\to\infty)
\end{displaymath} 
It is instructive to investigate briefly the form of the optimal
learning rule (\ref{eq:optimalonlinerule}) for large values of $E$ (as
in the initial stages of learning processes) and for small values of
$E$ (as in the final stages of learning processes). 
Initially we find 
\bd
\lim_{E\uparrow \frac{1}{2}}~ 
\frac{\eta\cF[J;Jx,y]_\opt}{\tan(\pi E)}
=J\sqrt{\frac{2}{\pi}}
\ed
which describes a Hebbian-type learning rule with diverging learning
rate (note that $\tan(\pi E)\to\infty$ for $E\uparrow \frac{1}{2}$). In contrast, in the final stages the optimal learning rule 
(\ref{eq:optimalonlinerule}) acquires the form  
\bd
\lim_{E\downarrow 0}~ \eta\cF[J;Jx,y]_\opt
=\frac{J|x|}{\sqrt{\pi}}~\theta[-xy]~
\lim_{z\to \infty}
\frac{e^{-z^2}}
{z\left[1\minus \erf(z)\right]}
=J|x|\theta[-xy]
\ed
which is the AdaTron learning rule with learning rate $\eta=1$\footnote{The reason that, in spite
of the asymptotic equivalence of the two rules,  the
optimal rule does not asymptotically give the same relaxation of the
error $E$ as the AdaTron rule is that in order to determine the
asymptotics one has to take the limit $E\to 0$ in the full 
macroscopic differential equation for $E$, which, in addition to the 
function $\cF[\ldots]$ defining the learning rule, involves
the Gaussian probability distribution (\ref{eq:gaussian}) which
depends on $E$ in a non-trivial way, especially near $E=0$.}. 
\vsp

In figures \ref{fig:exact} (short times and ordinary axes) and
\ref{fig:exactlog} (large times and log-log axes) we finally compare the
evolution of the error for the optimal on-line learning rule
(\ref{eq:optimalonlinerule}) with the two on-line learning rules which so far were
found to give the fastest relaxation: the perceptron rule with
normalising time-dependent learning rate (giving the error of (\ref{eq:exact1})), and the
perceptron rule with
optimal time-dependent learning rate (giving the error of
(\ref{eq:exact2})). This in order to assess whether choosing the
optimal on-line learning rule (\ref{eq:optimalonlinerule}) rather than its simpler 
competitors is actually worth the effort. 
The curves for the optimal on-line rule were obtained by numerical
solution of equation (\ref{eq:optimalonlinedEdt}).   

\subsection{Summary in a Table}

We close this section with an overview of some of the results on
on-line learning in perceptrons 
described/derived so far.
The upper part of this table contains results for specific learning
rules with either arbitrary constant learning rates $\eta$ (first 
column), optimal constant learning rate $\eta$ (second column), and
where possible, a time-dependent learning rate $\eta(t)$ chosen such as to
realise the normalisation $J(t)=1$ for all $t$.  The lower part of the
table gives results for specific learning rules with optimised time dependent
learning rates $\eta(t)$, as well as lower bounds on the asymptotic 
generalization error.  
\vspace*{5mm}

\noindent
\hspace*{-3mm} 
\renewcommand{\arraystretch}{2}
\begin{tabular}{|l|p{2.3in}|p{2in}|p{1in}|}
\hline
\multicolumn{4}{|c|}{GENERALIZATION ERROR IN PERCEPTRONS WITH ON-LINE
LEARNING RULES} \\
\hline
\hline
&\multicolumn{2}{|c|}{Constant learning rate $\eta$}& Variable $\eta$\\
\hline
Rule & Asymptotic decay for constant $\eta$ & Optimal asymptotic decay for constant $\eta$ & $\eta$ chosen to normalise $J$ \\
\hline
\hline
Hebbian & $E\sim\frac{1}{\sqrt{2\pi}} t^{-1/2}$ \hfill for $\eta>0$ & $E\sim\frac{1}{\sqrt{2\pi}} t^{-1/2}$ \hfill for $\eta>0$ & N/A \\
\hline
Perceptron & $E\sim(\frac{2}{3})^{1/3} \pi^{-1} t^{-1/3}$ \hfill for $\eta>0$ & $E\sim(\frac{2}{3})^{1/3} \pi^{-1} t^{-1/3}$ \hfill for $\eta>0$ &   $E\sim\frac{4}{\pi}t^{-1}$ \\
\hline
AdaTron & $E\sim(\frac{3}{3\eta-\eta^2})t^{-1}$ \hfill for $0<\eta<3$ & $E\sim\frac{4}{3}t^{-1}$ \hfill for $\eta=\frac{3}{2}$ & $E\sim\frac{3}{2}t^{-1}$ \\
\hline
\hline
\multicolumn{4}{|c|}{OPTIMAL GENERALIZATION}\\
\hline
\hline
&\multicolumn{3}{|c|}{Optimal time-dependent learning rate $\eta$} \\
\hline
Rule & \multicolumn{2}{|l|}{Generalization error for optimal
time-dependent $\eta$} &Asymptotics   \\
\hline
\hline
Hebbian & \multicolumn{2}{|l|}{$t=\frac{\pi}{2}[\frac{1}{\sin^2(\pi E)}-\frac{1}{\sin^2(\pi E_0)}]$} & $E\sim\frac{1}{\sqrt{2\pi}}t^{-1/2}$ \\
\hline
Perceptron & \multicolumn{2}{|l|}{$t=2[\frac{\pi E+\sin(\pi E)\cos(\pi E)}{\sin^2(\pi E)}-\frac{\pi E_0+\sin(\pi E_0)\cos(\pi E_0)}{\sin^2(\pi E_0)}]$} & $E\sim\frac{4}{\pi}t^{-1}$ \\
\hline
AdaTron & \multicolumn{2}{|l|}{$t=\frac{\pi}{8}[\frac{4\pi E-\sin(4\pi E)}{\sin^4(\pi E)}-\frac{4\pi E_0-\sin(4\pi E_0)}{\sin^4(\pi E_0)}]$} & $E\sim\frac{4}{3}t^{-1}$  \\
\hline
\multicolumn{3}{|l|}{Lower bound for on-line learning (asymptotics of the optimal
learning rule)} & $E\sim 0.88 t^{-1}$\\
\hline
\multicolumn{3}{|l|}{Lower bound for any learning rule} & $E\sim 0.44 t^{-1}$\\
\hline
\end{tabular}


\pagebreak
\section{The Formal Approach}

The main reason for developing a more formal
approach to learning dynamics is that in the complicated cases 
 of incomplete
training sets  or layered systems with large numbers of hidden neurons 
we can no longer get
away with the relatively simple methods used so far. 
For perceptrons with $N$ inputs the situation of incomplete training
sets arises when  
the number of `questions' scales as  $|\tilde{D}|=\alpha N$.  
We show how in the limit $N\to\infty$ 
the dynamics of any finite set of mean-field observables 
will be described by a (macroscopic) Fokker-Planck equation, of which the flow- and
diffusion terms can be calculated explicitly.  
In addition the more formal analysis  will allow us to recover the 
previous results on on-line learning 
in a more rigorous way, and will clarify the relation between the macroscopic
laws for the 
on-line and batch scenarios.

\subsection{From Discrete to Continuous Times}

We will describe the formal procedure for calculating macroscopic 
dynamical laws from the microscopic ones 
for on-line and batch learning processes in simple perceptrons. 
It involves several distinct stages. 
Our starting point is the formulation (\ref{eq:markovprocess}) in terms of a Markov process: 
\be
\hat{p}_{m+1}(\bJ)=
\int\!d\bJ^\prime~W[\bJ;\bJ^\prime]\hat{p}_m(\bJ^\prime)
\label{eq:markovchain}
\ee
with transition probability densities   corresponding to the class
(\ref{eq:genericrule}) of
generic  learning rules\footnote{As before this is just one choice of
many.  We could e.g. easily add a term of the form
$\frac{\eta}{N}\bJ^\prime {\cal
K}[|\bJ^\prime|;\bJ^\prime\cdot\bxi,\bB\cdot\bxi]$ to account for
weight 
decay (constant, `hard' spherical, `soft' spherical, or otherwise), 
without making the analysis significantly more difficult.}  
\be
\begin{array}{ll}
{\rm On\!\!-\!Line:} & 
W[\bJ;\bJ^\prime]=\bra \delta\left\{\room
\bJ\minus \bJ^\prime\minus \frac{\eta}{N}\bxi\sgn(\bB\cdot\bxi) 
\cF\left[|\bJ^\prime|;\bJ^\prime\inn\bxi,\bB\inn\bxi\right]\right\}\ket_\set
\\[3mm]
{\rm Batch:} & 
W[\bJ;\bJ^\prime]=\delta\left\{\room
\bJ\minus \bJ^\prime\minus \frac{\eta}{N}
\bra \bxi\sgn(\bB\cdot\bxi)\cF\left[
|\bJ^\prime|;\bJ^\prime\inn\bxi,\bB\inn\bxi\right]\ket_\set\right\}
\end{array}
\label{eq:transitions}
\ee
Note that in the previous approach the limit $N\to\infty$ realised 
several simplifications at once 
(continuous versus discrete time, stochastic versus deterministic
macroscopic evolution) which for technical reasons we 
would prefer to control independently.

We will first describe a method to 
make the transition from the discrete-time process
(\ref{eq:markovchain}) to a description 
involving real-valued times in a more transparent and exact way. The
idea is to choose 
the duration of each discrete iteration step in the
process (\ref{eq:markovchain}) to be a real-valued random
number, such that the probability that at time $t$ precisely $m$ steps
have been made is given by the Poisson expression
\bd
\pi_m(t)=\frac{1}{m!}(Nt)^me^{-Nt}
\vspace*{-2mm} 
\ed
with the properties
\be
\frac{d}{dt}\pi_{m>0}(t)=N[\pi_{m-1}(t)\minus \pi_m(t)]
~~~~~~~~~~~~
\frac{d}{dt}\pi_{0}(t)=-N\pi_0(t)
\label{eq:poisson1}
\ee
\be
\bra m\ket=Nt~~~~~~~~~~~~\bra m^2\ket\minus \bra m\ket^2=Nt
\label{eq:poisson2}
\ee
This move at first sight appears to make the problem more complicated,
but will turn out to do precisely the opposite.  
From (\ref{eq:poisson2}) it follows that for times $t\ll N$ one has $t=m/N+\order(N^{-\frac{1}{2}})$,
the usual time unit.   
Due to the random durations of the iteration steps we also have to
replace 
the microscopic probability distribution $\hat{p}_m(\bsigma)$ in
(\ref{eq:markovchain}) by one that takes the variations in numbers of iteration
steps performed at a given time $t$ into account:
\be
p_t(\bJ)=
\sum_{m\geq 0}\pi_m(t)\hat{p}_m(\bJ)
\label{eq:conttime}
\ee  
This distribution obeys a simple differential equation, which
follows from combining the equations 
(\ref{eq:markovchain},\ref{eq:poisson1},\ref{eq:poisson2},\ref{eq:conttime}):
\be
\frac{d}{dt}p_t(\bJ)
= 
N\int\!d\bJ^\prime~\left\{
W[\bJ;\bJ^\prime]-\delta[\bJ\minus \bJ^\prime]\right\}
p_t(\bJ^\prime)
\label{eq:conttimemarkov}
\ee
So far no approximations have been made, equation
(\ref{eq:conttimemarkov}) which replaces (\ref{eq:markovchain}) is 
exact for any $N$. 
We have made the transition from discrete-time iterations to
differential equations (which are usually much easier to handle) without invoking the limit $N\to \infty$, but at
the price of an uncertainty in where we are on the time axis. This
uncertainty, however, is guaranteed to vanish in the limit $N\to\infty$.   

\subsection{From Microscopic to Macroscopic Laws}

We next wish to investigate the dynamics of a number of as yet
arbitrary {\em macroscopic} observables
$\bOmega[\bJ]=(\Omega_1[\bJ],\ldots,\Omega_k[\bJ])$. 
They are assumed to be $\order(1)$ each for $N\to\infty$, and
finite in number. 
To do so we
introduce the associated macroscopic probability distribution
\be
P_t(\bOmega)=
\int\!d\bJ~p_t(\bJ)\delta\left[\bOmega-\bOmega[\bJ]\right]
\ee 
Its time derivative immediately follows from that in (\ref{eq:conttimemarkov}):
\bd
\frac{d}{dt}P_t(\bOmega)=N\int\!d\bJ
d\bJ^\prime~\delta\left[\bOmega\minus \bOmega[\bJ]\right]
\left\{
W[\bJ;\bJ^\prime]\minus \delta[\bJ\minus \bJ^\prime]\right\}
p_t(\bJ^\prime)
\ed 
This equation can be written in the standard form 
\be
\frac{d}{dt}P_t(\bOmega)
=
\int\!d\bOmega^\prime~ {\cal W}_t[\bOmega;\bOmega^\prime]
P_t(\bOmega^\prime)
\label{eq:macrodynamics}
\ee
where
\bd
{\cal W}_t[\bOmega;\bOmega^\prime]=
\frac{\int\!d\bJ^\prime~p_t(\bJ^\prime)
\delta\left[\bOmega^\prime\minus \bOmega[\bJ^\prime]\right]
\int\!d\bJ 
\delta\left[\bOmega\minus \bOmega[\bJ]\right]
N\left\{
W[\bJ;\bJ^\prime]\minus \delta[\bJ\minus \bJ^\prime]\right\}
}
{\int\!d\bJ^\prime~p_t(\bJ^\prime)\delta\left[\bOmega^\prime\minus
\bOmega[\bJ^\prime]\right]}
\ed
(this statement can be verified by substitution of ${\cal
W}_t[\bOmega;\bOmega^\prime]$ into (\ref{eq:macrodynamics})). 
Note that the macroscopic process (\ref{eq:macrodynamics}) need not be
Markovian, due to the explicit time-dependence of the macroscopic
transition density ${\cal W}_t[\bOmega;\bOmega^\prime]$. 
If we now insert the relevant expressions
(\ref{eq:transitions}) for $W[\bJ;\bJ^\prime]$, we can perform the
$\bJ$-integrations, and obtain an expression in
terms of so-called sub-shell averages (or conditional averages) $\bra
f(\bJ)\ket_{\bOmega;t}$, 
which
are defined as
\bd 
\bra f(\bJ)\ket_{\bOmega;t}=
\frac{\int\!d\bJ~p_t(\bJ)\delta\left[\bOmega\minus
\bOmega[\bJ]\right]f(\bJ)}
{\int\!d\bJ~p_t(\bJ)\delta\left[\bOmega\minus
\bOmega[\bJ]\right]}
\ed
For the two types of learning rules at hand (on-line and batch) we obtain:
\bd
{\cal W}^{\rm onl}_t[\bOmega;\bOmega^\prime]=N\bra\bra
\delta\left[
\bOmega\minus
\bOmega[\bJ\plus\frac{\eta}{N}\bxi\sgn(\bB\inn\bxi)\cF[|\bJ|;\bJ\inn\bxi,\bB\inn\bxi]]
\right]\ket_\set
\minus 
\delta\left[\bOmega\minus \bOmega[\bJ]\right]
\ket_{\bOmega^\prime;t}
\ed
\bd
{\cal W}^{\rm bat}_t[\bOmega;\bOmega^\prime]=
N\bra 
\delta\left[\bOmega\minus 
\bOmega[\bJ
\plus\frac{\eta}{N}\bra\bxi\sgn(\bB\inn\bxi)\cF[|\bJ|;\bJ\inn\bxi,\bB\inn\bxi]\ket_\set
]
\right]
\minus 
\delta\left[\bOmega\minus \bOmega[\bJ]\right]
\ket_{\bOmega^\prime;t}
\ed
We now insert integral representations for the $\delta$-distributions
\bd
\delta[\bOmega\minus
\bQ]=\int\!\frac{d\hat{\bOmega}}{(2\pi)^k}e^{i\hat{\bOmega}\inn[\bOmega-\bQ]}
\ed
which gives for our two learning scenario's:
\be
{\cal W}^{\rm onl}_t[\bOmega;\bOmega^\prime]=
\int\!\frac{d\hat{\bOmega}}{(2\pi)^k}e^{i\hat{\bOmega}\inn\bOmega}~
N\bra\bra
e^{-i\hat{\bOmega}\inn
\bOmega[\bJ
\plus\frac{\eta}{N}\bxi\sgn(\bB\inn\bxi)\cF[|\bJ|;\bJ\inn\bxi,\bB\inn\bxi]
]}
\ket_\set
\minus 
e^{-i\hat{\bOmega}\inn\bOmega[\bJ]}
\ket_{\bOmega^\prime;t}
\label{eq:macromap1}
\ee
\be
{\cal W}^{\rm bat}_t[\bOmega;\bOmega^\prime]=
\int\!\frac{d\hat{\bOmega}}{(2\pi)^k}e^{i\hat{\bOmega}\inn\bOmega}~
N\bra 
e^{-i\hat{\bOmega}\inn\bOmega
[\bJ
\plus\frac{\eta}{N}\bra\bxi\sgn(\bB\inn\bxi)\cF[|\bJ|;\bJ\inn\bxi,\bB\inn\bxi]\ket_\set
]}
\minus 
e^{-i\hat{\bOmega}\inn\bOmega[\bJ]}
\ket_{\bOmega^\prime;t}
\label{eq:macromap2}
\ee
Still no approximations have been made. The above two expressions
differ only in at which stage the averaging over the training set
$\set$ 
occurs. \vsp

Our aim is to obtain from (\ref{eq:macrodynamics}) an {\em autonomous} set of macroscopic dynamic
equations, i.e. we want to choose the observables $\bOmega[\bJ]$ such
that for $N\to \infty$    
the explicit
time-dependence in ${\cal W}_t[\bOmega;\bOmega^\prime]$, induced by the appearance of
the microscopic distribution $p_t(\bJ)$ will 
vanish. This can happen either because $p_t(\bJ)$ drops out, or 
because $p_t(\bJ)$
depends on $\bJ$ only via $\bOmega[\bJ]$, or 
even through combinations of these mechanisms.
In expanding equations (\ref{eq:macromap1},\ref{eq:macromap2}) for
large $N$ we have to be somewhat careful, since the
system size $N$ enters both as a small parameter to control the
magnitude of the modification of individual components of the weight
vector $\bJ$, but also determines the dimensions and lengths of various
vectors. Upon inspection of 
the general Taylor expansion
\bd
\Omega[\bJ\plus\bk]
=\sum_{\ell\geq 0}\frac{1}{\ell
!}\sum_{i_1=1}^N\cdots\sum_{i_\ell=1}^N k_{i_1}\cdots k_{i_\ell}
\frac{\partial^\ell \Omega[\bJ]}{\partial
J_{i_1}\cdots\partial J_{i_\ell}}
\ed
we see that if all derivatives were to be treated as $\order(1)$ (i.e.
if we only take into account the dependence of the components of $\bk$
on $N$, we end up in trouble, since in the
cases of interest (where $\bk^2=\order(N^{-1})$) 
this series could give $\Omega[\bJ\plus\bk]=
\sum_{\ell\geq 0}(\sum_i k_i)^\ell =\sum_{\ell\geq 0}\order(1)$. 
We need to restrict ourselves to observables $\Omega_\mu[\bJ]$ of the mean-field type, 
where all compoments $J_i$ play an equivalent role in determining
the overall scaling with respect to $N$ (which makes sense). For instance:
\bd
\begin{array}{ll}
\Omega[\bJ]=\sum_k B_kJ_k: &  \order(\partial_i
\Omega[\bJ])=\order(B_i)=N^{-1}\order(\Omega[\bJ])/\order(J_i)\\[-2mm]
\Omega[\bJ]=\sum_k J^2_k: &  \order(\partial_i
\Omega[\bJ])=\order(J_i)=N^{-1}\order(\Omega[\bJ])/\order(J_i)\\[-2mm]
\Omega[\bJ]=\sum_{kl} J_kA_{kl}J_l: &  \order(\partial_i
\Omega[\bJ])=\order(\sum_k A_{ik}J_k)=N^{-1}\order(\Omega
[\bJ])/\order(J_i)
\end{array}
\ed
The pattern is clear. 
The only additional point to be taken into account is that
in the case of multiple derivatives with respect to the {\em
same}
component $J_i$, our scaling requirement will be less
severe due to the fact that such terms occur less frequently than
multiple derivatives with respect to different components (i.e. in
$\sum_{ij}J_iA_{ij}J_j$ we have $N(N\minus 1)$ terms
with $i\neq j$, but just $N$ with $i=j$). We thus define mean-field
observables $\Omega[\bJ]$ as
\be
{\rm mean\!-\!field~observables}:~~~~~~
\frac{\partial^\ell \Omega[\bJ]}{\partial
J_{i_1}\cdots\partial J_{i_\ell}}=\order\left(\!
N^{-\frac{1}{2}\ell}\frac{\Omega[\bJ]}{|\bJ|^\ell}.N^{\ell-d}
\!\right)
~~~~~~(N\to\infty)
\label{eq:meanfield}
\ee
in which $d$ is the number of {\em different} elements in the set
$\{i_1,\ldots,i_\ell\}$.   
For mean-field observables we can estimate the scaling
of the various terms in the Taylor expansion:
\be
\Omega[\bJ\plus\bk]
=
\Omega[\bJ]+\sum_i k_i\frac{\partial \Omega[\bJ]}{\partial J_i}
+\frac{1}{2}\sum_{ij}k_i k_j\frac{\partial^2 \Omega[\bJ]}{\partial J_i \partial J_j}
+\sum_{\ell\geq 3} \order\left(\!
\Omega[\bJ]\left[\frac{|\bk|}{|\bJ|}\right]^\ell\!\right)
\label{eq:taylor}
\ee
(where we have used $\sum_i k_i=\order(\sqrt{N}|\bk|)$).

We now apply (\ref{eq:taylor}) to our equations 
(\ref{eq:macromap1},\ref{eq:macromap2}), restricting ourselves
henceforth 
to mean-field observables $\Omega_\mu[\bJ]$ in the sense of
(\ref{eq:meanfield}). 
The shifts $\bk$,  
being either 
$\frac{\eta}{N}\bxi\sgn(\bB\inn\bxi)\cF[|\bJ|;\bJ\inn\bxi,\bB\inn\bxi]$ 
or 
$\frac{\eta}{N}\bra
\bxi\sgn(\bB\inn\bxi)\cF[|\bJ|;\bJ\inn\bxi,\bB\inn\bxi]\ket_\set$, 
scale as $|\bk|=\order(N^{-\frac{1}{2}})$. Furthermore, if we choose
one of our observables to be $\Omega_1[\bJ]=\bJ^2$, 
the subshells in (\ref{eq:macromap1},\ref{eq:macromap2}) 
will ensure $\bJ^2=\order(1)$, so that 
the $\ell$-th order term in the expansions (\ref{eq:taylor}) will be of
order $N^{-\frac{1}{2}\ell}$ in both cases. This allows us to expand:
\bd
e^{-i\hat{\bOmega}\inn
\bOmega[\bJ\plus\bk]}-e^{-i\hat{\bOmega}\inn
\bOmega[\bJ]}
=
e^{-i\hat{\bOmega}\inn\left\{
\bOmega[\bJ]+\sum_i k_i\frac{\partial}{\partial J_i}\bOmega[\bJ]
+\frac{1}{2}\sum_{ij}k_i k_j\frac{\partial^2}{\partial J_i \partial
J_j}
\bOmega[\bJ]
+\order(N^{-\frac{3}{2}})\right\}}-e^{-i\hat{\bOmega}\inn
\bOmega[\bJ]}
\ed
\bd
=-
e^{-i\hat{\bOmega}\inn\bOmega[\bJ]}\left\{
i\sum_i k_i\frac{\partial}{\partial J_i}(\hat{\bOmega}\inn\bOmega[\bJ])
\plus \frac{i}{2}\sum_{ij}k_i k_j\frac{\partial^2}{\partial J_i \partial
J_j}
(\hat{\bOmega}\inn\bOmega[\bJ])
\plus \frac{1}{2}\left[
\sum_i k_i\frac{\partial}{\partial J_i}(\hat{\bOmega}\inn\bOmega[\bJ])
\right]^2\right\}
\ed
\bd
~~~~~~~~~~~~~~~~~~~~~~~~~~~~~~~~~~~~~~~~~~~~~~~~~~~~~~~~~~~~~~~~~~~~~~~~~~~~~~~~~~~~~~~~~~~~~~~~~~~~~~~~~~~
+\order(N^{-\frac{3}{2}})
\vspace*{-3mm}
\ed
so that 
\bd
N\int\!\frac{d\hat{\bOmega}}{(2\pi)^k}e^{i\hat{\bOmega}\inn\bOmega}~
\left[e^{-i\hat{\bOmega}\inn
\bOmega[\bJ\plus\bk]}
\minus e^{-i\hat{\bOmega}\inn
\bOmega[\bJ]}\right]
=
-N\int\!\frac{d\hat{\bOmega}}{(2\pi)^k}e^{i\hat{\bOmega}\inn[\bOmega-\bOmega[\bJ]]}~\times
~~~~~~~~~~~~~~~~~~~~~~~~~~~~~~~~~~~~~~~~~~~~~~~~~~~~~~~~~~~~~~~~~~~
\ed
\bd
\left\{\room
i\sum_{\mu}\hat{\Omega}_\mu \sum_i k_i\frac{\partial \Omega_\mu[\bJ]}{\partial J_i}
\plus \frac{i}{2}\sum_\mu\hat{\Omega}_\mu\sum_{ij}k_i
k_j\frac{\partial^2\Omega_\mu[\bJ]}{\partial J_i \partial J_j}
\plus \frac{1}{2}\sum_{\mu\nu}\hat{\Omega}_\mu \hat{\Omega}_\nu
\sum_{ij} k_ik_j\frac{\partial\Omega_\mu[\bJ]}{\partial J_i}
\frac{\partial\Omega_\nu[\bJ]}{\partial J_j}\right\}
+\order(\frac{1}{\sqrt{N}})
\ed
\bd
=
\minus N\int\!\frac{d\hat{\bOmega}}{(2\pi)^k}
\left\{
\sum_{\mu}\frac{\partial}{\partial\Omega_\mu} 
\left[
\sum_i k_i\frac{\partial \Omega_\mu[\bJ]}{\partial J_i}
\plus \frac{1}{2}\sum_{ij}k_i
k_j\frac{\partial^2\Omega_\mu[\bJ]}{\partial J_i \partial J_j}\right]
\minus \frac{1}{2}\sum_{\mu\nu}\frac{\partial^2}{\partial\Omega_\mu
\partial\Omega_\nu}
\sum_{ij} k_ik_j\frac{\partial\Omega_\mu[\bJ]}{\partial J_i}
\frac{\partial\Omega_\nu[\bJ]}{\partial J_j}\right\}
\ed
\bd
~~~~~~~~~~~~~~~~~~~~~~~~~~~~~~~~~~~~~~~~~~~~~~~~~~~~~~~~~~~~~
\times e^{i\hat{\bOmega}\inn[\bOmega-\bOmega[\bJ]]}
+\order(N^{-\frac{1}{2}})
\ed
\bd
=
\minus N \left\{
\sum_{\mu}\frac{\partial}{\partial\Omega_\mu} 
\left[
\sum_i k_i\frac{\partial \Omega_\mu[\bJ]}{\partial J_i}
\plus \frac{1}{2}\sum_{ij}k_i
k_j\frac{\partial^2\Omega_\mu[\bJ]}{\partial J_i \partial J_j}\right]
\minus \frac{1}{2}\sum_{\mu\nu}\frac{\partial^2}{\partial\Omega_\mu
\partial\Omega_\nu}
\sum_{ij} k_ik_j\frac{\partial\Omega_\mu[\bJ]}{\partial J_i}
\frac{\partial\Omega_\nu[\bJ]}{\partial J_j}\right\}
\ed
\bd
~~~~~~~~~~~~~~~~~~~~~~~~~~~~~~~~~~~~~~~~~~~~~~~~~~~~~~~~~~~~~
\times\delta\left[\bOmega-\bOmega[\bJ]\right]
+\order(N^{-\frac{1}{2}})
\ed
We now find, upon insertion of this expansion into the expressions
(\ref{eq:macromap1}) and (\ref{eq:macromap2}), 
that both types of learning dynamics (on-line and batch) are described
by 
macroscopic laws with
transition probabilities of the
general form 
\bd
{\cal W}^{\rm\star\star\star}_t[\bOmega;\bOmega^\prime]=
\left\{
-\sum_{\mu}F_\mu[\bOmega^\prime;t]\frac{\partial}{\partial\Omega_\mu} 
+
\frac{1}{2}\sum_{\mu\nu}G_{\mu\nu}[\bOmega^\prime;t]
\frac{\partial^2}{\partial\Omega_\mu\partial\Omega_\nu}
\right\}\delta\left[\bOmega-\bOmega^\prime\right]
\ed
which, in combination with the dynamic equation
(\ref{eq:macrodynamics}), 
leads to convenient  and transparent description of the  
macroscopic dynamics in the form of a Fokker-Planck equation:
\be
\frac{d}{dt}P_t(\bOmega)
=
-\sum_{\mu=1}^k\frac{\partial}{\partial\Omega_\mu} 
\left\{F_\mu[\bOmega;t]P_t(\bOmega)\right\}
+\frac{1}{2}\sum_{\mu\nu=1}^k
\frac{\partial^2}{\partial\Omega_\mu\partial\Omega_\nu}
\left\{ G_{\mu\nu}[\bOmega;t]P_t(\bOmega)\right\}
\label{eq:fokkerplanck}
\ee
(modulo contributions which vanish for $N\to\infty$). 
The differences between on-line and batch learning are in the explicit
expressions for the functions $F_\mu[\bOmega;t]$ and
$G_{\mu\nu}[\bOmega;t]$ in the flow- and diffusion terms. Upon
introducing the short hand $\cF[\ldots]$ for 
$\cF[|\bJ|;\bJ\inn\bxi,\bB\inn\bxi]$ these can be written as: 
\be
F^{\rm onl}_\mu[\bOmega;t]=
\eta \bra\bra\sum_i \xi_i\sgn(\bB\inn\bxi)\cF[\ldots] 
\frac{\partial \Omega_\mu[\bJ]}{\partial
J_i}\ket_\set\ket_{\bOmega;t}
+ \frac{\eta^2}{2N}\bra\bra\sum_{ij}
\xi_i\xi_j \cF^2[\ldots]
\frac{\partial^2\Omega_\mu[\bJ]}{\partial J_i \partial J_j}
\ket_\set\ket_{\bOmega;t}~~~
\label{eq:Fonline}
\ee
\be
G^{\rm onl}_{\mu\nu}[\bOmega;t]=
\frac{\eta^2}{N} \bra\bra \sum_{ij} 
\xi_i\xi_j\cF^2[\ldots]
\left[\frac{\partial\Omega_\mu[\bJ]}{\partial J_i}\right]
\left[\frac{\partial\Omega_\nu[\bJ]}{\partial J_j}\right]
\ket_\set\ket_{\bOmega;t}
~~~~~~~~~~~~~~~~~~~~~~~~~~~~~~~~~~~~
\label{eq:Gonline}
\ee
\bd
F^{\rm bat}_\mu[\bOmega;t]=
\eta \bra\sum_i 
\bra\xi_i\sgn(\bB\inn\bxi)\cF[\ldots]\ket_\set
\frac{\partial \Omega_\mu[\bJ]}{\partial J_i}\ket_{\bOmega;t}
~~~~~~~~~~~~~~~~~~~~~~~~~~~~~~~~~~~~~~~~~~~~~~~~~~~~
\ed
\be
+ \frac{\eta^2}{2N}\bra\sum_{ij}
\bra\xi_i\sgn(\bB\inn\bxi)\cF[\ldots]
\ket_\set
\bra\xi_j\sgn(\bB\inn\bxi)\cF[\ldots] 
\ket_\set
\frac{\partial^2\Omega_\mu[\bJ]}{\partial J_i \partial J_j}
\ket_{\bOmega;t}
\label{eq:Fbatch}
\ee
\be
G^{\rm bat}_{\mu\nu}[\bOmega;t]=
\frac{\eta^2}{N}\bra \sum_{ij} 
\bra\xi_i\sgn(\bB\inn\bxi)\cF[\ldots]
\ket_\set
\bra\xi_j\sgn(\bB\inn\bxi)\cF[\ldots]
\ket_\set
\left[\frac{\partial\Omega_\mu[\bJ]}{\partial J_i}\right]
\left[\frac{\partial\Omega_\nu[\bJ]}{\partial J_j}\right]
\ket_{\bOmega;t}
\label{eq:Gbatch}
\ee
The result (\ref{eq:fokkerplanck}) is still fairly general. The only
conditions on the observables $\Omega_\mu[\bJ]$ needed for
(\ref{eq:fokkerplanck})  to hold
are $(i)$ all are of order unity for $N\to\infty$, $(ii)$ all are of
the mean-field type (\ref{eq:meanfield}), and $(iii)$ one of them is
the squared length $\bJ^2$ of the student's weight vector. 
\vsp

The Fokker-Planck equation (\ref{eq:fokkerplanck}) subsequently quantifies the properties of the
ideal choice(s) for our macroscopic observables $\Omega_\mu[\bJ]$, if
our aim is to find closed deterministic equations. Firstly:
\be
{\rm deterministic~ laws:}~~~~~~
\lim_{N\to\infty}G_{\mu\nu}[\bOmega;t]=0
\label{eq:deterministic}
\ee
If (\ref{eq:deterministic}) holds, equation (\ref{eq:fokkerplanck}) 
reduces to a Liouville equation, with solutions of the desired form  
$P_t(\bOmega)=\delta[\bOmega\minus\bOmega^*(t)]$ in which the trajectory
$\bOmega^*(t)$, in turn, is the solution of the deterministic equation
\be
\frac{d}{dt}\bOmega=\bF[\bOmega;t]
\label{eq:flow}
\ee
with the flow field $\bF$ given either by (\ref{eq:Fonline}) (for
on-line learning) or by (\ref{eq:Fbatch}) (for batch learning).
Note that condition (\ref{eq:deterministic}) is not only sufficient to
guarantee deterministic evolution, but also necessary. 
Secondly, we want the deterministic laws to be closed:
\be
{\rm closed~ laws:}~~~~~~
\lim_{N\to\infty}\frac{\partial}{\partial
t}F_{\mu}[\bOmega;t]=0
\label{eq:closed}
\ee
(again this condition is sufficient and necessary). 
A set of mean-field observables $\Omega_\mu[\bJ]$ meeting
the criteria (\ref{eq:deterministic},\ref{eq:closed}) constitutes for
$N\to\infty$ an 
exact autonomous macroscopic level of description of the learning process,
in the form of the coupled deterministic 
differential equations (\ref{eq:flow}). 
However, in general  there will be no a priori guarantee 
that such a set of observables 
actually exists. 

\subsection{Application to $(Q,R)$ Evolution}

We now apply the general results of this section to the specific duo of
observables that we considered in the previous sections to describe
on-line learning with complete training sets:
\be
\Omega_1[\bJ]=Q[\bJ]=\bJ^2~~~~~~~~~
\Omega_2[\bJ]=R[\bJ]=\bJ\cdot \bB
\label{eq:learningobservables}
\ee
These observables are indeed of the mean-field type
(\ref{eq:meanfield}) if all $B_i=\order(N^{-\frac{1}{2}})$, and are defined to be of order unity.  
However, the training set $\set$ is now chosen to consist of $|\set|=\alpha N$ randomly drawn
questions $\bxi^\mu\in\{-1,1\}^N$. We will 
show that $Q$ and $R$ obey deterministic macroscopic
equations 
for any $\alpha$. These equations, however, fail to close as soon as the training set is
incomplete (for $\alpha<\infty$).  In contrast, our previous results 
are recovered for
the case of complete training sets (for $\alpha\to\infty$). 
In addition we will derive for the case of
complete training sets the
 macroscopic equations for the batch version of some of the most
popular learning rules.   

As could have been expected, we will also need the joint input distribution
\be
P(x,y)=
\bra\bra \delta[x\minus
\hbJ\inn\bxi]\delta[y\minus\bB\inn\bxi]\ket_\set\ket_{Q,R;t}
\label{eq:generalP}
\ee
Note that we cannot simply assume the distribution (\ref{eq:generalP}) to be of
a Gaussian form; it will 
depend on $\alpha$. We will now first show that the second order
moments of $P(x,y)$ remain finite for any $\alpha$ in the limit
$N\to\infty$.  
For arbitrary vectors $\bx$ and $\by$ 
we find
\bd
\bra(\bx\inn\bxi)(\by\inn\bxi)\ket_\set=\bx\inn\by+
L(\bx,\by)~~~~~~~~~~
L(\bx,\by)=
\sum_{i\neq j}x_i y_j
\left[\frac{1}{\alpha N}\sum_{\mu=1}^{\alpha N}\xi_i^\mu \xi_j^\mu
\right]
\ed
The second term is bounded according to 
$\lambda_{\rm min}|\bx||\by| \leq L(\bx,\by)\leq \lambda_{\rm
max}|\bx||\by|$, in which the $\lambda's$ denote the (real)
eigenvalues of the matrix   
$M_{ij}=\frac{1\minus\delta_{ij}}
{\alpha N}\sum_{\mu=1}^{\alpha N}\xi_i^\mu \xi_j^\mu$. 
For large $N$ the spectrum of the matrix $M$ can be calculated using
random matrix theory (see e.g. 
\cite{hertz}) and the eigenvalues will be bounded:
\be
{\rm eigenvalues~of}~~ M_{ij}=\frac{1\minus\delta_{ij}}
{\alpha N}\sum_{\mu=1}^{\alpha N}\xi_i^\mu \xi_j^\mu:~~~~~
\left\{\begin{array}{lll}
\alpha\leq 1: & \lambda_{\rm min}=\minus 1 & 
\lambda_{\rm max}=\frac{1}{\alpha}+\frac{2}{\sqrt{\alpha}}\\[-2mm]
\alpha> 1: & \lambda_{\rm
min}=\frac{1}{\alpha}-\frac{2}{\sqrt{\alpha}} &
\lambda_{\rm max}=\frac{1}{\alpha}+\frac{2}{\sqrt{\alpha}}
\end{array}
\right.
\label{eq:spectra}
\ee
From this it follows that all second order moments (and therefore also
all first order moments) of the
distribution $P(x,y)$ are finite, whatever the value of $\alpha$, but
also 
that only for $\alpha\to\infty$ we recover the familiar previous
expressions (derived for complete training sets in section 2) for 
the second-order moments in terms of $Q$ and $R$: 
\be
\alpha\to\infty:~~~~\int\!dxdy~x^2P(x,y)=\int\!dxdy~y^2P(x,y)=1,~~
\int\!dxdy~xy P(x,y)=R/Q
\label{eq:simplemoments}
\ee  
\vsp

The next stage is to assess the scaling of the various diffusion terms
$G_{\mu\nu}^{\star\star\star}$ in the Fokker-Planck equation
(\ref{eq:fokkerplanck}).  These should vanish for $N\to\infty$
if our observables are to behave determistically in the 
limit $N\to\infty$. 
For the present observables the diffusion terms
(\ref{eq:Gonline},\ref{eq:Gbatch}) 
become 
\bd
\begin{array}{l}
G^{\rm onl}_{QQ}[Q,R;t]=\frac{4\eta^2}{N} Q
\int\!dxdy~P(x,y) x^2\cF^2[Q^{\frac{1}{2}};Q^{\frac{1}{2}}x,y] 
\\[-2mm]
G^{\rm onl}_{QR}[Q,R;t]=\frac{2\eta^2}{N} Q^{\frac{1}{2}}
\int\!dxdy~P(x,y) xy\cF^2[Q^{\frac{1}{2}};Q^{\frac{1}{2}}x,y] 
\\[-2mm]
G^{\rm onl}_{RR}[Q,R;t]=\frac{\eta^2}{N} 
\int\!dxdy~P(x,y) y^2\cF^2[Q^{\frac{1}{2}};Q^{\frac{1}{2}}x,y] 
\\[2mm]
G^{\rm bat}_{QQ}[Q,R;t]= \frac{4\eta^2}{N}Q
\left\{ \int\!dxdy~P(x,y) x \sgn(y)\cF[Q^{\frac{1}{2}};Q^{\frac{1}{2}}x,y] \right\}^2\\[-2mm]
G^{\rm bat}_{QR}[Q,R;t]= 
\frac{2\eta^2}{N}Q^{\frac{1}{2}}
\left\{ \int\!dxdy~P(x,y) x\sgn(y)\cF[Q^{\frac{1}{2}};Q^{\frac{1}{2}}x,y] \right\}
\left\{ \int\!dxdy~P(x,y) |y|\cF[Q^{\frac{1}{2}};Q^{\frac{1}{2}}x,y]
\right\}
\\[-2mm]
G^{\rm bat}_{RR}[Q,R;t]=
\frac{\eta^2}{N}
\left\{ \int\!dxdy~P(x,y)|y|\cF[Q^{\frac{1}{2}};Q^{\frac{1}{2}}x,y] \right\}^2
\end{array}
\ed
We conclude 
that all diffusion terms
$G_{\star\star}^{\star\star\star}$ are of order $\order(\frac{1}{N})$
provided $\cF[\ldots]$ is bounded (which we assumed from the start). 
This implies that for $N\to\infty$ 
our macroscopic  observables $Q$ and $R$ indeed 
evolve deterministically for any $\alpha>0$. 
\vsp

The resulting deterministic equations for the duo $(Q,R)$ for on-line
and batch learning are given by
combining (\ref{eq:flow}) with the flow terms  (\ref{eq:Fonline}) and
(\ref{eq:Fbatch}), respectively.  
These equations we now work out explicitly, starting with the
on-line scenario. Insertion of (\ref{eq:Fonline}) into (\ref{eq:flow}) gives
\be
\frac{d}{dt}Q 
= \lim_{N\to\infty}\left\{
2\eta Q^{\frac{1}{2}}\int\!dxdy~P(x,y) x\sgn(y)\cF[Q^{\frac{1}{2}};Q^{\frac{1}{2}}x,y] 
+ \eta^2\int\!dxdy~
P(x,y)\cF^2[Q^{\frac{1}{2}};Q^{\frac{1}{2}}x,y]\right\}
\label{eq:onlinedQdt}
\ee
\be
\frac{d}{dt}R
=\lim_{N\to \infty}
\eta  \int\!dxdy~ P(x,y)
|y|\cF[Q^{\frac{1}{2}};Q^{\frac{1}{2}}x,y] 
\label{eq:onlinedRdt}
\ee
Note that these equations are of the same form as those derived
earlier for complete training sets, i.e.
(\ref{eq:dQdt},\ref{eq:dRdt}). The differences between complete and
incomplete training sets are purely in the joint
distribution $P(x,y)$, i.e. equation (\ref{eq:generalP}). 
\vsp 

Working out the macroscopic equations for the case of batch learning is somewhat less straightforward, although
the final result will be simpler. Insertion of (\ref{eq:Fbatch}) into 
(\ref{eq:flow}) gives, with the usual short-hand $\cF[\ldots]=
\cF[|\bJ|;\bJ\inn\bxi,\bB\inn\bxi]$:
\bd
\frac{d}{dt}Q
= \lim_{N\to\infty}\left\{
2\eta Q^{\frac{1}{2}}\int\!dxdy~P(x,y) x\sgn(y)
\cF[Q^{\frac{1}{2}};Q^{\frac{1}{2}}x,y] 
+ \frac{\eta^2}{N}\bra\sum_{i}
\bra\xi_i\sgn(\bB\inn\bxi)\cF[\ldots]
\ket_\set^2
\ket_{\bOmega;t}
\right\}
\ed
\bd
\frac{d}{dt}R
=\lim_{N\to \infty}
\eta  \int\!dxdy~ P(x,y)
|y|\cF[Q^{\frac{1}{2}};Q^{\frac{1}{2}}x,y] 
\ed
The second term in the temporal derivative of $Q$ can be written as
the subshell average of a quantity of 
the form 
\bd
\frac{1}{\alpha N^2}\sum_{i}\sum_{\mu\nu=1}^{\alpha N}
x_\mu \xi^\mu_i \xi_i^\nu x_\nu 
=\frac{\bx^2}{\alpha N}+K(\bx)~~~~~~~~~~
K(\bx)=
\frac{1}{\alpha N}\sum_{\mu\neq \nu=1}^{\alpha N}x_\mu  x_\nu 
\left[\frac{1}{N}\sum_{i}
\xi^\mu_i \xi_i^\nu  \right]
\ed 
with $\bx^2=\order(1)$ for $N\to\infty$. 
The second term in this expression is bounded according to 
$\tilde{\lambda}_{\rm min}\bx^2/\alpha N \leq K(\bx)\leq \tilde{\lambda}_{\rm
max}\bx^2/\alpha N$, in which the $\tilde{\lambda}'s$ denote the (real)
eigenvalues of the matrix   
$\tilde{M}_{\mu\nu}=\frac{1\minus\delta_{\mu\nu}}
{N}\sum_{i=1}^{N}\xi_i^\mu \xi_i^\nu$. Note that for $N\to\infty$ 
the eigenvalues of the 
matrix $\tilde{M}$ are related to those of the matrix $M$ in
(\ref{eq:spectra}) by simply replacing $\alpha\to 1/\alpha$ (since
the relation between the two cases is interchanging $\alpha N$ and
$N$). 
From this it follows that $\lim_{N\to\infty} K(\bx)=0$ and that
in the temporal derivative of $Q$ only the first term survives the
limit $N\to\infty$. This leaves the final result:
\be
\frac{d}{dt}Q
= \lim_{N\to\infty}
2\eta Q^{\frac{1}{2}}\int\!dxdy~P(x,y) x\sgn(y)
\cF[Q^{\frac{1}{2}};Q^{\frac{1}{2}}x,y] 
\label{eq:batchdQdt}
\ee
\be
\frac{d}{dt}R
=\lim_{N\to \infty}
\eta  \int\!dxdy~ P(x,y)
|y|\cF[Q^{\frac{1}{2}};Q^{\frac{1}{2}}x,y]
\label{eq:batchdRdt} 
\ee
For any value of $\alpha$, the difference between the macroscopic  
equations for 
on-line learning (\ref{eq:onlinedQdt},\ref{eq:onlinedRdt}) 
and batch learning (\ref{eq:batchdQdt},\ref{eq:batchdRdt}) 
(apart from a possible difference in the expressions one might find for the
distribution $P(x,y)$) is simply the presence/absence of
terms which are quadratic in the learning rate $\eta$.
\vsp
 
For finite $\alpha$, the case of incomplete training sets, 
we observe that the macroscopic equations for the
pair $(Q,R)$ (i.e. (\ref{eq:onlinedQdt},\ref{eq:onlinedRdt}) and
(\ref{eq:batchdQdt},\ref{eq:batchdRdt})) 
do not close, since the distribution $P(x,y)$  (\ref{eq:generalP}) 
need not be of a Gaussian form, and its moments need not (and almost
certainly will not) be expressible in terms of the quantities $Q$ and $R$.

For $\alpha\to\infty$, the case of complete training sets, we can
express the second order moments of $P(x,y)$ (\ref{eq:generalP}) 
in terms of the observables $(Q,R)$
via (\ref{eq:simplemoments}). Moreover, we can show that the first
order moments of $P(x,y)$ are zero, since for any normalised vector
$\bx\in\Re^N$:
\bd
\bra \bx\inn\bxi\ket^2_\set=\left[\sum_{i=1}^N x_i 
\left(\frac{1}{\alpha N}\sum_{\mu=1}^{\alpha N}\xi_i^\mu\right)
\right]^2 \leq ~
\sum_{i=1}^N \left(\frac{1}{\alpha N}\sum_{\mu=1}^{\alpha
N}\xi_i^\mu\right)^2
\equiv \gamma(\bxi)
\ed
in which $\gamma(\bxi)$ obeys (with brackets denoting averages over the
possible training sets):  
\bd
\bra   \gamma^2(\bxi)\ket = 
\frac{1}{\alpha^4 N^4}
\sum_{ij=1}^N \sum_{\mu\nu\rho\lambda=1}^{\alpha
N} \bra x_i^\mu x_i^\nu x_j^\rho x_j^\lambda\ket
=\frac{1}{\alpha^2}+\order(\frac{1}{N})
\ed
This shows that $\lim_{\alpha\to\infty}\gamma(\bxi)=0$ and that the
first order moments of $P(x,y)$ will be zero. 
What cannot be demonstrated rigorously, however, is that for
$\alpha\to\infty$ the
distribution $P(x,y)$ is of a Gaussian form. This is impossible in
principle, even for $\alpha\to\infty$. We could,
for instance, choose an initial state $\bJ(0)$ for the student weight
vector of the form $J_i(0)\sim e^{-i}$, in which case the Gaussian
assumption would be violated for short times. If we choose our
teacher vector of the form $B_i\sim e^{-i}$ the situation is even
worse: now the system will be forced to evolve into a macroscopic state with a
non-Gaussian distribution $P(x,y)$. It will be clear that all we can
hope for is that for non-pathological initial conditions $\bJ(0)$ and
non-pathological teacher vectors $\bB$ one can derive a dynamic
equation for $P(x,y)$ with Gaussian solutions. 

\subsection{Complete Training Sets: Batch Learning versus On-Line Learning} 

Here we will work out the macroscopic equations for the batch versions
of the Hebbian, perceptron and AdaTron learning rules, and compare the
results to those of the on-line scenarios. It turns out that 
for these cases one can solve the macroscopic dynamical laws
explicitly. 
We restrict ourselves to
complete training sets. For $\alpha \to\infty$ and $N\to\infty$ 
the (exact) results of the 
previous subsection can be written as
\be
\frac{d}{dt}Q 
= 
2\eta Q^{\frac{1}{2}}\int\!dxdy~P(x,y) x\sgn(y)\cF[Q^{\frac{1}{2}};Q^{\frac{1}{2}}x,y] 
+ \Delta\eta^2\int\!dxdy~
P(x,y)\cF^2[Q^{\frac{1}{2}};Q^{\frac{1}{2}}x,y]
\label{eq:bothdQdt}
\ee
\be
\frac{d}{dt}R
=
\eta  \int\!dxdy~ P(x,y)
|y|\cF[Q^{\frac{1}{2}};Q^{\frac{1}{2}}x,y] 
\label{eq:bothdRdt}
\ee
in which $\Delta=1$ for the on-line scenario and $\Delta=0$ for the
batch scenario. Of the distribution $P(x,y)$ we know, without
additional assumptions:
\bd
P(x,y)=\lim_{N\to\infty}
\bra\bra \delta[x\minus
\hbJ\inn\bxi]\delta[y\minus\bB\inn\bxi]\ket_\set\ket_{Q,R;t}
~~~~~~~~~~
\bra x \ket=\bra y\ket =0,~~~
\bra x^2\ket =\bra y^2\ket =1,~~~
\bra xy \ket =R/Q
\ed  
If we now assume $\bJ(0)$ and $\bB$ to be such that $P(x,y)$ has 
a Gaussian shape, the above
expressions for the moments immediately dictate that for both
scenario's $P(x,y)$  will be
identical to (\ref{eq:gaussian}).  
We now firstly recover our
previous macroscopic equations (\ref{eq:dQdt},\ref{eq:dRdt}) for the
case of on-line learning ($\Delta=1$), and secondly find that the
macroscopic  equations for the case of batch learning can, for any
choice $\cF[\ldots]$ of the details of the learning rule, be obtained
from the on-line equations by simply removing from the latter all
terms which are quadratic in the learning rate $\eta$. This also holds 
if we write the macroscopic equations in terms of the observables
$(E,J)$, since the transformation $(Q,R)\to(E,J)$ does not involve the
learning rate $\eta$.
\vsp

For the Hebbian rule $\cF[|\bJ|;\bJ\inn\bxi,\bB\inn\bxi]=1$ 
we obtain the macroscopic equations describing
batch learning by elimination of the $\eta^2$ terms from the on-line
equations (\ref{eq:JHebb},\ref{eq:EHebb}), giving
\be
\frac{d}{dt}J=\eta \cos(\pi E)\sqrt{\frac{2}{\pi}}
~~~~~~~~~~~~~~
\frac{d}{dt}E=-\frac{\eta\sin(\pi E)}{\pi J}\sqrt{\frac{2}{\pi}}
\label{eq:Hebbbatch}
\ee
We can solve these equations by exploiting the existence of a
conserved quantity. If we define $D(J,E)=J\sin(\pi E)$ we find, using
(\ref{eq:Hebbbatch}), that $\frac{d}{dt}D=0$, which allows us to
express the length $J(t)$ at any time as
\[
J=J_0~\frac{\sin(\pi E_0)}{\sin(\pi E)}
\]
Substitution into the differential equation for the generalization
error $E$ then leads to a single non-linear differential equation
involving $E$ only:
\bd
\frac{d}{dt}E=-\sqrt{\frac{2}{\pi}}\frac{\eta\sin^2(\pi E)}{\pi J_0 \sin(\pi E_0)}
\ed
This equation is easily solved:
\be
t(E)=\frac{1}{\eta}\sqrt{\frac{\pi}{2}}J_0\sin(\pi E_0)
\left[\frac{1}{\tan(\pi E)}-\frac{1}{\tan(\pi E_0)}\right]
\label{eq:Hebbbatchsolution}
\ee
Asymptotically this gives
\[
E\sim\frac{J_0\sin(\pi E_0)}{\eta t\sqrt{2\pi}}
\]
Asymptotically the gain in using the batch scenario rather than the
on-line scenario is having a power law error relaxation of the form
$t^{-1}$ rather than $t^{-\frac{1}{2}}$. 
\vsp

For the perceptron rule $\cF[|\bJ|;\bJ\inn\bxi,\bB\inn\bxi]=\theta[\minus(\bJ\inn\bxi)(\bB\inn\bxi)]$ we obtain the macroscopic equations describing
batch learning by elimination of the $\eta^2$ terms from the on-line
equations (\ref{eq:Jperceptron},\ref{eq:Eperceptron}), giving
\be
\frac{d}{dt}J =- \frac{\eta[1\minus \cos(\pi E)]}{\sqrt{2\pi}}
~~~~~~~~~~~~
\frac{d}{dt}E = - \frac{\eta\sin(\pi E)}{\pi \sqrt{2\pi}J}
\label{eq:percbatch}
\ee
Here we find that the quantity $D(J,E)=J[1\plus \cos(\pi E)]$ is
conserved,  which leads to
\[
J=J_0~\frac{1\plus\cos(\pi E_0)}{1\plus \cos(\pi E)}
\]
Substitution into the differential equation for the generalization
error $E$ then again leads to a single non-linear differential equation
involving $E$ only:
\bd
\frac{d}{dt}E =-\frac{\eta\sin(\pi E)[1\plus\cos(\pi E)]}{\pi\sqrt{2\pi}J_0[1\plus \cos(\pi E_0)]}
\ed
which can be solved by writing $t$ as an integral over
$\frac{dt}{dE}$, and by using 
\bd
\int\frac{dx}{\sin(x)[1\plus \cos(x)]}=\frac{1}{2}
\left\{\log\tan(\frac{x}{2})+\frac{1}{1\plus\cos(x)}\right\}
\ed
(see \cite{GR}). This results in 
\be
t(E)=\frac{J_0}{\eta}\sqrt{\frac{\pi}{2}}[1\plus \cos(\pi
E_0)]\left[\log\tan(\frac{\pi E_0}{2})+\frac{1}{1\plus\cos(\pi E_0)}-\log\tan(\frac{\pi E}{2})-\frac{1}{1\plus\cos(\pi E)}\right]
\label{eq:percbatchsolution}
\ee
Asymptotically we now find an exponential decay of the generalization error:
\bd
E\sim e^{-\sqrt{\frac{2}{\pi}}\frac{\eta t}{J_0(1+\cos(\pi E_0))}}
\ed
The gain in using the batch scenario rather than the
on-line scenario for the perceptron learning rule is quite
significant. 
The batch scenario gives an
exponentially fast decay of the generalization error, compared to a
power law relation of the form
$t^{-1/3}$ for on-line learning. 
\vsp

Finally we turn to the AdaTron rule
$\cF[|\bJ|;\bJ\inn\bxi,\bB\inn\bxi]=|\bJ\inn\bxi|\theta[\minus(\bJ\inn\bxi)(\bB\inn\bxi)]$.
Here  we obtain the macroscopic equations describing
batch learning by elimination of the $\eta^2$ terms from the on-line
equations (\ref{eq:adatdJ2},\ref{eq:adatdE}), giving
\bd
\frac{d}{dt}J =-\eta JE+\frac{\eta J}{\pi}\cos(\pi E)\sin(\pi E)
~~~~~~~~~~~~
\frac{d}{dt}E=-\frac{\eta \sin^2(\pi E)}{\pi^2}
\ed
The equation for the generalization error is already decoupled from
the equation giving the evolution of the length $J$, and can be solved
directly:
\be
t(E)=\frac{\pi}{\eta \tan(\pi E)}-\frac{\pi}{\eta\tan(\pi E_0)}
\label{eq:adatbatchsolution}
\ee
Asymptotically this behaves as
\[
E\sim\frac{1}{\eta t}
\]
For the AdaTron rule there is only little to be gained in switching
from on-line learning to batch learning. Both scenario's give a power
law error relaxation of the form $t^{-1}$ (albeit with
different prefactors). 
\begin{figure}[t]
\centering
\vspace*{235mm}
\hbox to \hsize{\hspace*{-40mm}\includegraphics{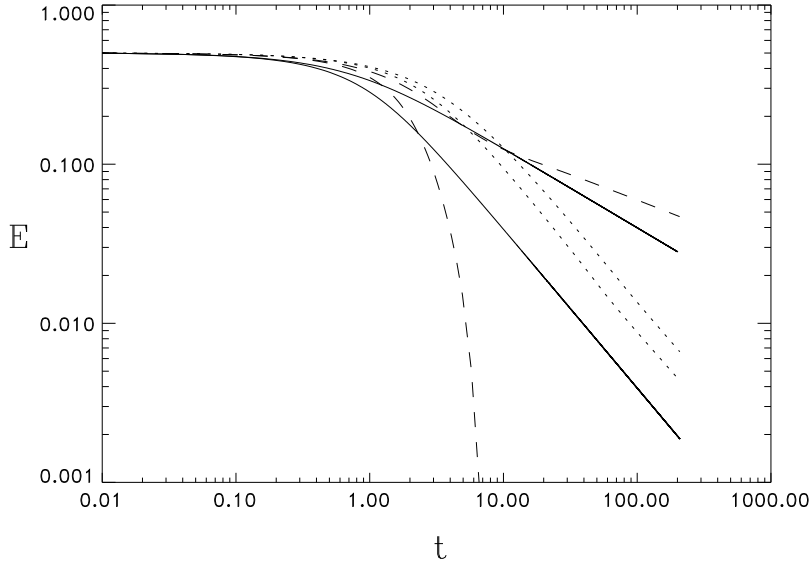}\hspace*{40mm}}
\vspace*{-155mm}
\caption{Qualitative comparison of the evolution of the error 
for batch versus on-line learning rules with constant learning rates $\eta=1$.
Solid lines: Hebbian rule (upper solid: on-line learning, lower solid: batch learning). Dashed lines: Peceptron rule (upper dashed: on-line learning, lower dashed: batch learning). Dotted lines: AdaTron rule (upper dotted: on-line learning, lower dotted: batch learning). 
} 
\label{fig:batchlog}
\end{figure}
\vsp

We summarise the results of this section on batch learning with
complete training sets in the table below, and also illustrate the
differences between the batch results and the on-line results in 
figure \ref{fig:batchlog}. 
Whereas the error evolution for the batch versions of the Hebbian and AdaTron
rules is almost identical, 
there is clearly a remarkable difference between
the perceptron learning rule on the one hand and the Hebbian and
AdaTron rules on the other, in the degree to which they benefit
from being executed in a batch scenario rather than an on-line
scenario. 
Only the perceptron rule manages to significantly 
capitalise on the advantage of batch
learning (where all
question/answer pairs in the training set $\set$ are available at each
iteration step, rather than just a single question/answer pair) and realise an exponential decay of the generalization
error.  

\pagebreak
\noindent
\renewcommand{\arraystretch}{2}
\begin{tabular}{|l|p{2in}|p{1in}|p{1.5in}|}
\hline
\multicolumn{4}{|c|}{GENERALIZATION ERROR IN PERCEPTRONS WITH BATCH LEARNING RULES} \\
\hline
\hline
Rule & \multicolumn{2}{|l|}{Generalization error} &Asymptotics   \\
\hline
\hline
Hebbian & \multicolumn{2}{|l|}{$t=\frac{J_0\sin(\pi E_0)}{\eta}\sqrt{\frac{\pi}{2}}[\frac{1}{\tan(\pi E)}-\frac{1}{\tan(\pi E_0)}]$} & $E\sim\frac{J_0\sin(\pi E_0)}{\eta\sqrt{2\pi}}t^{-1}$ \\
\hline
Perceptron &
\multicolumn{2}{|l|}{$t=\frac{J_0}{\eta}\sqrt{\frac{\pi}{2}}(1+\cos(\pi
E_0))[\ln\tan(\frac{\pi E_0}{2})+\frac{1}{1+\cos(\pi E_0)}$} & $E\sim e^{-\sqrt{\frac{2}{\pi}}\frac{\eta t}{J_0[1+\cos(\pi E_0)]}}$ \\
& \multicolumn{2}{|r|}{$-\ln\tan(\frac{\pi E}{2})-\frac{1}{1+\cos(\pi E)}]$} & \\
\hline
AdaTron & \multicolumn{2}{|l|}{$t=\frac{\pi}{\eta}\left[\frac{1}{\tan(\pi E)}-\frac{1}{\tan(\pi E_0)}\right]$} & $E\sim\frac{1}{\eta}t^{-1}$  \\
\hline
\end{tabular}


\pagebreak    
\section{Incomplete Training Sets}

\subsection{The Problem and Our Options}

We have seen in the previous section that 
in the case of incomplete training set (where $|\set|=\alpha N$) 
the equations for our familiar
observables $Q[\bJ]=\bJ^2$ and $R[\bJ]=\bJ\inn\bB$ (or, equivalently,
for $|\bJ|$ and the generalization error
$E_g[\bJ]=\frac{1}{\pi}\arccos(R[\bJ]/\sqrt{Q[\bJ]})$) no longer close, since the
distribution $P(x,y)$ (\ref{eq:generalP}) will no longer be Gaussian
and cannot be written in such a way that its dependence on the weight
vector $\bJ$ is  only through the observables $Q[\bJ]$ and $R[\bJ]$. One can in
fact show that for $\alpha<\infty$ no finite set of observables will ever
obey a closed set of dynamic equations. 

Closely related to this problem  
is the fact that our macroscopic equations always involve averages over the training
set $\set$, which for $\alpha<\infty$  will generally depend on
the details of the choice made for the $\alpha N$ questions $\bxi^\mu$
in $\set$. Since we cannot expect to  be able so solve the dynamics   
for any given microscopic realisation $\{\bxi^1,\ldots,\bxi^{\alpha
N}\}$ of the set $\set$, we will be forced to restrict ourselves to calculating
{\em averages} of observables {\em over all possible realisations of the
training set}. 
In order to avoid thereby ending up with irrelevant statements
(since we really aim to arrive at predictions for actual simulation
experiments, rather than averages over many such predictions), it is of
vital 
importance to focus on those observables which in the limit
$N\to\infty$ tend towards their averages over all possible training
sets anyway. Numerical simulations show that macroscopic observables 
such as the generalisation- and training errors have this property: 
if e.g. one chooses the questions $\bxi^\mu$ in the training set
$\set$ at
random from $D=\{\minus 1,1\}^N$, one will simply observe that 
for large $N$ the curves for $E_g$ and $E_t$ as
functions of time are reproducible, and depend only on the relative
size $\alpha$ of $\set$, not
on its detailed  composition  $\{\bxi^1,\ldots,\bxi^{\alpha
N}\}$:\footnote{This property is called `self-averaging'.}    
\be
\lim_{N\to\infty}\bra E_{\rm t}\ket =
\lim_{N\to\infty}\bra \bra E_{\rm t}\ket\ket_{\rm sets} =
\lim_{N\to\infty}\bigbra
\int\!d\bJ~p_t(\bJ|\bxi^1,\ldots,\bxi^{\alpha N})~
\frac{1}{\alpha N}\sum_{\mu=1}^{\alpha N}\theta[-(\bJ\inn\bxi^\mu)(\bB\inn\bxi^\mu)]
\bigket_{\rm sets}
\label{eq:averageEt}
\ee
\be
\lim_{N\to\infty}\bra E_{\rm g}\ket =
\lim_{N\to\infty}\bra \bra E_{\rm g}\ket\ket_{\rm sets} =
\lim_{N\to\infty}\bigbra
\int\!d\bJ~p_t(\bJ|\bxi^1,\ldots,\bxi^{\alpha N})~
\bra\theta[-(\bJ\inn\bxi)(\bB\inn\bxi)]\ket_D 
\bigket_{\rm sets}
\label{eq:averageEg}
\ee
(with the microscopic probability density $p_t(\bJ|\set)$ for the student weight
vector, given a realisation of the training set $\set$).

In equilibrium calculations the problem is often less severe, since 
in many cases one at least knows the stationary microscopic
probability density $p_\infty(\bJ|\set)$, 
so that one can write down 
the (exact) expressions for the equilibrium expectation
values of the training- and generalization errors
(\ref{eq:expectederrors})
and their averages over the realisations of the training set
(\ref{eq:averageEt},\ref{eq:averageEg}).  
One can then work out these expressions and obtain transparent results
in the $N\to\infty$ limit upon exchanging the order of the  
various summations and integrations.  The remaining problem is
of a technical nature. 
In dynamical studies away from
equilibrium, on the other
hand, we usually do not have an expression for $p_t(\bJ|\set)$ at our
disposal, and our problem is of a conceptual rather than a technical
nature.  In order to proceed we need to average over the realisations 
of the training sets, but we have as yet no object to average ...

The toolbox of non-equilibrium statistical mechanics at present offers
two (in a way complementary) techniques to deal with this situation, 
which is a familiar one in
the field of disordered magnetic systems, namely the technique of
generating 
functionals (involving 
path integrals) and dynamical replica theory. Following the generating
functional route one performs the average over the realisations of the
training sets on an object from which one
can derive all relevant observables by differentiation. In the limit
$N\to\infty$ this procedure leads to
exact equations for two-time correlation- and response functions,
which, however, are highly complicated and can be solved in practice
only near equilibrium. 
In dynamical replica theory one derives deterministic macroscopic
equations for an observable {\rm function} (equivalent to an infinite
number of ordinary scalar observables), which are averaged over the
realisations of the training set using the so-called replica method. 
Here one assumes that the chosen function obeys closed deterministic 
equations in the $N\to\infty$ limit; the exactness of the resulting
theory depends on the degree to which this assumption is correct. 
Solving the resulting equations numerically is feasible for
transients, but as yet too CPU-intensive to allow for solution close to
equilibrium. 
 
\subsection{Route 1: Generating Functionals and Path Integrals}

This rather elegant approach, which to our knowledge has so far only
been applied to learning rules with binary weights, is based on calculating a
generating functional $Z[\bpsi]$  
which is an average
over all possible `paths'  $\{\bJ(t)\}$ $(t\geq 0)$ of the student's weight
vectors through the state space $\Re^N$, given the dynamics
(\ref{eq:conttimemarkov}), 
\be
Z\left[\bpsi\right]=
\bra e^{-i\sum_i\int_0^t ds~ \psi_i(s)J_i(s)}\ket \room
\label{eq:generatingfunction}
\ee
in which time is a continuous variable. 
As with all path
integrals, 
averages such as (\ref{eq:generatingfunction}) are understood to be 
defined in the following way:
$(i)$ one discretises time in the dynamic equation
(\ref{eq:conttimemarkov}), $(ii)$ one calculates the desired average, and
subsequently $(iii)$ one takes the  continuum limit in the resulting
expression. 
From (\ref{eq:generatingfunction}) one can calculate all relevant
single- and multiple time 
observables by functional differentiation. 
Averaging the generating functional over the possible realisations of
the training set $\set$ gives relations such as
\be
\bra J_i(t)\ket_{\rm sets}= i\lim_{\bpsi\to \bnul}
\frac{\delta}{\delta \psi_i(t)}\bra Z[\bpsi]\ket_{\rm sets}
\ee
\be
\bra J_i(t)J_j(t^\prime)\ket_{\rm sets}= -\lim_{\bpsi\to \bnul}
\frac{\delta^2}{\delta \psi_i(t)\delta\psi_j(t^\prime)}\bra Z[\bpsi]\ket_{\rm sets}
\ee
etc. 
Overall constant prefactors in $Z[\bpsi]$ can always be recovered {\em
a posteriori} with the identity $Z[\bnul]=1$. 

The discretised version of our equation (\ref{eq:conttimemarkov}) 
and the corresponding discretised expression for the generating functional
(\ref{eq:generatingfunction}),
with time-steps of duration 
$\Delta$, 
 would be 
\be
p_{t+\Delta}(\bJ)=\int\!d\bJ^\prime 
\left\{\room
\delta[\bJ\minus\bJ^\prime]
+ \Delta N\left[W[\bJ;\bJ^\prime]\minus \delta[\bJ\minus\bJ^\prime]\right]
\right\}
p_t(\bJ^\prime)
~~~~~(0<\Delta\ll 1)
\label{eq:discretised1}
\ee
\be
Z\left[\bpsi\right]=
\bra e^{-i\sum_i\sum_{\ell=0}^{L}\Delta 
\psi_i(\ell.\Delta)J_i(\ell.\Delta)}\ket 
\label{eq:discretised2}
\ee
At the end of our calculation the dependence of any physical
observable on $\Delta$, other than via $t=\ell\Delta$, ought to
disappear.  
Note that, although (\ref{eq:discretised1}) appears to be almost
identical to (\ref{eq:markovchain}) (equation (\ref{eq:markovchain})
can be obtained from (\ref{eq:discretised1}) by choosing $\Delta=N^{-1}$), there is a crucial technical
difference.  In (\ref{eq:discretised1}), in contrast to
(\ref{eq:markovchain}), 
we can control the parameter
that converts time into a continuous variable ($\Delta$) independently
of the parameter that controls the fluctuations ($N$). This allows us
to take the limit $N\to\infty$ before the limit $\Delta\to 0$. 
The discretised process (\ref{eq:discretised1}) gives for the
probability density $P[\bJ(t_0),\ldots,\bJ(t_\ell)]$ of a
temporally discretised path (with $t_n=n\Delta$):
\bd
P[\bJ(t_0),\ldots,\bJ(t_\ell)]
=
\prod_{n=0}^{\ell-1}\left\{\room
\delta[\bJ(t_{n+1}\minus\bJ(t_n)]
+ \Delta N\left[W[\bJ(t_{n+1};\bJ(t_n)]\minus \delta[\bJ(t_{n+1}\minus\bJ(t_n)]\right]
\right\}
\ed
so that we find for (\ref{eq:discretised2}) after averaging over all possible training sets $\set$:
\bd
\bra Z\left[\bpsi\right]\ket_{\rm sets}=
\int\!\cdots\int\!
\prod_{n=0}^{t/\Delta}\left[d\bJ(t_n) e^{-i\sum_i\Delta 
\psi_i(t_n)J_i(t_n)}\right]
\times
~~~~~~~~~~~~~~~~
\ed
\be
\bigbra
\prod_{n=0}^{t/\Delta-1}\left\{\room
\delta[\bJ(t_{n+1}\minus\bJ(t_n)]
+ \Delta N\left[W[\bJ(t_{n+1};\bJ(t_n)]\minus \delta[\bJ(t_{n+1}\minus\bJ(t_n)]\right]
\right\}
\bigket_{\rm sets}
\label{eq:discretised3}
\ee
The problem has hereby again turned into a technical one, albeit of a
highly non-trivial nature. The strategy
would now be to $(i)$ insert into (\ref{eq:discretised3}) the recipe
(\ref{eq:transitions}) for the learning rule to be studied, 
$(ii)$ introduce appropriate $\delta$-distributions
that will isolate all occurrences of the vectors $\bxi^\mu\in\set$ in
(\ref{eq:discretised3}) in 
such a way that the average over all training sets can be performed,
$(iii)$ take the limit $N\to\infty$ for finite $\Delta$ (this will lead
to a saddle-point integral, involving integration variables with
two time-arguments), $(iv)$ take the limit $\Delta\to 0$ which
restores the original dynamics and converts all integrals into path
integrals, 
and finally $(v)$ solve the
saddle-point equations. 

The saddle-point equations will describe a non-Markovian stochastic 
dynamical problem for an effective single weight variable; it will involve a retarded
self-interaction and a stochastic noise which is not local in time
(i.e. with an auto-correlation function of finite width). This causes 
these saddle-point equations to be extremely hard to solve, especially in the
transient stages of the learning dynamics. 
Here we will not follow this procedure further, mainly because for the
types of rules we have been considering in this review such
calculations 
have not yet been performed (this program has so far only been carried
out for learning rules involving binary weight vectors $\bJ\in\{-1,1\}^N$).

\subsection{Route 2: Dynamical Replica Theory}

The second procedure to deal with incomplete training sets is closer to 
the methods used so far for dealing with complete training sets than
the above formalism, 
since it involves macroscopic differential equations for 
single-time observables. The ground work has already been done in
section four, where we found that for learning rules of the usual type
(\ref{eq:genericrule}), and under certain conditions, 
the evolution of macroscopic 
observables $\bOmega[\bJ]=(\Omega_1[\bJ],\ldots,\Omega_\ell[\bJ])$ 
is in the limit $N\to\infty$ described by deterministic laws. 
With  the short hand $\cF[\ldots]$ for 
$\cF[|\bJ|;\bJ\inn\bxi,\bB\inn\bxi]$, and with the definition of
sub-shell averages introduced in section four
\be
\bra f(\bJ)\ket_{\bOmega;t}=
\frac{\int\!d\bJ~p_t(\bJ)f(\bJ)\delta[\bOmega\minus\bOmega[\bJ]]}
{\int\!d\bJ~p_t(\bJ)\delta[\bOmega\minus\bOmega[\bJ]]}
\label{eq:subshells}
\ee
these deterministic laws can be written as: 
\be
{\rm On\!\!-\!\!Line:}~~~~~
\frac{d}{dt}\bOmega=
\eta \bra\bra\sum_i \xi_i\sgn(\bB\inn\bxi)\cF[\ldots] 
\frac{\partial \bOmega[\bJ]}{\partial
J_i}\ket_\set\ket_{\bOmega;t}
+ \frac{\eta^2}{2N}\bra\bra\sum_{ij}
\xi_i\xi_j \cF^2[\ldots]
\frac{\partial^2\bOmega[\bJ]}{\partial J_i \partial J_j}
\ket_\set\ket_{\bOmega;t}~~~
\label{eq:macrolawsonline}
\ee
\bd
{\rm Batch:}~~~~~~~
\frac{d}{dt}\bOmega=
\eta \bra\sum_i 
\bra\xi_i\sgn(\bB\inn\bxi)\cF[\ldots]\ket_\set
\frac{\partial \bOmega[\bJ]}{\partial J_i}\ket_{\bOmega;t}
~~~~~~~~~~~~~~~~~~~~~~~~~~~~~~~~~~~~~~~~~~~~~~~~~~~~
\ed
\be
+ \frac{\eta^2}{2N}\bra\sum_{ij}
\bra\xi_i\sgn(\bB\inn\bxi)\cF[\ldots]
\ket_\set
\bra\xi_j\sgn(\bB\inn\bxi)\cF[\ldots] 
\ket_\set
\frac{\partial^2\bOmega[\bJ]}{\partial J_i \partial J_j}
\ket_{\bOmega;t}
\label{eq:macrolawsbatch}
\ee
Sufficient  conditions for (\ref{eq:macrolawsonline},\ref{eq:macrolawsbatch})
to hold for $N\to\infty$ were found to be:
\begin{enumerate}
\item All $\Omega_\mu[\bJ]$ are of order unity for $N\to\infty$
\item All $\Omega_\mu[\bJ]$ are mean-field observables in the sense of
(\ref{eq:meanfield})
\item $\Omega_1[\bJ]=\bJ^2$
\item For all $\mu,\nu\leq \ell:$~
$\lim_{N\to\infty}G_{\mu\nu}[\bOmega;t]=0$
\end{enumerate}
in which the diffusion coefficients (for on-line and batch learning, 
respectively) are given by
\bd
G^{\rm onl}_{\mu\nu}[\bOmega;t]=
\frac{\eta^2}{N} \bra\bra \sum_{ij} 
\xi_i\xi_j\cF^2[\ldots]
\left[\frac{\partial\Omega_\mu[\bJ]}{\partial J_i}\right]
\left[\frac{\partial\Omega_\nu[\bJ]}{\partial J_j}\right]
\ket_\set\ket_{\bOmega;t}
\ed
\bd
G^{\rm bat}_{\mu\nu}[\bOmega;t]=
\frac{\eta^2}{N}\bra \sum_{ij} 
\bra\xi_i\sgn(\bB\inn\bxi)\cF[\ldots]
\ket_\set
\bra\xi_j\sgn(\bB\inn\bxi)\cF[\ldots]
\ket_\set
\left[\frac{\partial\Omega_\mu[\bJ]}{\partial J_i}\right]
\left[\frac{\partial\Omega_\nu[\bJ]}{\partial J_j}\right]
\ket_{\bOmega;t}
\ed
The basic idea of the formalism is to note that for those observables
$\bOmega[\bJ]$ which obey closed
deterministic dynamical laws which are self-averaging in the limit
$N\to\infty$, we can use
(\ref{eq:macrolawsonline},\ref{eq:macrolawsbatch}) to fully 
determine these laws.
If $\bOmega$ obeys closed equations we know that, at least for
$N\to\infty$, 
 the right-hand 
sides of (\ref{eq:macrolawsonline},\ref{eq:macrolawsbatch}) by
definition {\em
cannot} 
depend on the distribution of the microscopic 
probabilities $p_t(\bJ)$ within the
$\bOmega$-sub-shells of (\ref{eq:subshells}). As a consequence we can
simplify the evaluation of 
(\ref{eq:macrolawsonline},\ref{eq:macrolawsbatch}) 
by making a convenient
choice for $p_t(\bJ)$: one that describes probability
equipartitioning within the $\bOmega$-sub-shells, i.e.
\be
\bra f(\bJ)\ket_{\bOmega;t}~\rightarrow~
\bra f(\bJ)\ket_{\bOmega}=
\frac{\int\!d\bJ~f(\bJ)\delta[\bOmega\minus\bOmega[\bJ]]}
{\int\!d\bJ~\delta[\bOmega\minus\bOmega[\bJ]]}
\label{eq:equipartitioning}
\ee
Combination of (\ref{eq:equipartitioning}) with
(\ref{eq:macrolawsonline},\ref{eq:macrolawsbatch}), and usage of the 
self-averaging property, then leads to the following closed and
deterministic laws:
\bd
{\rm On\!\!-\!\!Line:}~~~~~~~
\frac{d}{dt}\bOmega=
\eta \lim_{N\to\infty}
\bigbra\bra\bra\sum_i \xi_i\sgn(\bB\inn\bxi)\cF[\ldots] 
\frac{\partial \bOmega[\bJ]}{\partial
J_i}\ket_\set\ket_{\bOmega}\bigket_{\rm sets}
~~~~~~~~~~~~~~~~~~~~~~~~~~
\ed
\be
+~\eta^2 \lim_{N\to\infty}\bigbra\frac{1}{2N}\bra\bra\sum_{ij}
\xi_i\xi_j \cF^2[\ldots]
\frac{\partial^2\bOmega[\bJ]}{\partial J_i \partial J_j}
\ket_\set\ket_{\bOmega}\bigket_{\rm sets}
\label{eq:closedonline}
\ee
\bd
{\rm Batch:}~~~~~~~~
\frac{d}{dt}\bOmega=
\eta \lim_{N\to\infty}\bigbra \bra\sum_i 
\bra\xi_i\sgn(\bB\inn\bxi)\cF[\ldots]\ket_\set
\frac{\partial \bOmega[\bJ]}{\partial J_i}\ket_{\bOmega}\bigket_{\rm sets}
~~~~~~~~~~~~~~~~~~~~~~~~~~~~~~~~~~~~~~~~~~~~
\ed
\be
+~ \eta^2\lim_{N\to\infty}\bigbra \frac{1}{2N}\bra\sum_{ij}
\bra\xi_i\sgn(\bB\inn\bxi)\cF[\ldots]
\ket_\set
\bra\xi_j\sgn(\bB\inn\bxi)\cF[\ldots] 
\ket_\set
\frac{\partial^2\bOmega[\bJ]}{\partial J_i \partial J_j}
\ket_{\bOmega}\bigket_{\rm sets}
\label{eq:closedbatch}
\ee
Given the choice for the observables $\bOmega[\bJ]$, our problem has
now again been converted into a technical one. One performs the
average over all training sets using the replica identity
\bd
\bigbra \frac{\int\!d\bJ~f(\bJ|\set)W(\bJ|\set)}
{\int\!d\bJ~W(\bJ|\set)} \bigket_{\rm sets}
=\lim_{n\to 0}\bigbra 
\int\cdots\int\prod_{\alpha=1}^n \left[d\bJ^\alpha
W(\bJ^\alpha|\set)\right] f(\bJ^1|\set)\bigket_{\rm sets}
\ed
The key question that remains is how to select the observables
$\bOmega[\bJ]$, since
although the
theory is guaranteed to generate the exact dynamic equations for observables
which indeed obey closed, deterministic and self-averaging laws, it does not
tell us which observables will have these properties beforehand. 
If the chosen observables $\bOmega[\bJ]$ do not 
obey closed deterministic laws, the
method will generate an approximate theory in which one simply 
has made the closure approximation that all microscopic states $\bJ$
with identical values for the macroscopic observables $\bOmega[\bJ]$ 
are assumed to be equally probable. The
available contraints to guide us in finding the appropriate
$\bOmega[\bJ]$ are the four properties listed below equation 
(\ref{eq:macrolawsbatch}) and the knowledge 
that we will need an infinite number
(i.e. $\ell\to\infty$), or. equivalently, an observable function.  
In addition, for those systems where the equilibrium microscopic
probability 
density $p_\infty(\bJ|\set)$ is know and is of a Boltzmann form, i.e.
$p_\infty(\bJ|\set)\sim e^{-\beta H(\bJ|\set)}$, one can guarantee 
exactness of the theory in equilibrium by choosing one of the
observables to be $H(\bJ|\set)/N$ (or equivalently a set of
observables that determine  $H(\bJ|\set)/N$ uniquely), since in that
case the equipartitioning assumption (\ref{eq:equipartitioning}) 
is exact in equilibrium. 
\vsp

For the learning dynamics of the type (\ref{eq:genericrule}), the
results of section 4.3 automatically lead us to the
following choice:
\be
\Omega_1[\bJ]=Q[\bJ]=\bJ^2~~~~~~~~~
\Omega_2[\bJ]=R[\bJ]=\bJ\cdot \bB~~~~~~~~~
\Omega_{xy}[\bJ]=P[x,y;\bJ]=\bra\delta[x\minus \bJ\cdot\bxi] 
\delta[x\minus\bB\cdot\bxi]\ket_\set
\label{eq:drtobservables}
\ee
(with $x,y\in\Re$). Note that here we have defined the distribution 
$P[x,y;\bJ]$
without explicit normalisation of $\bJ$, i.e. with $x=\bJ\inn\bxi$
rather than $x=\hat{\bJ}\inn\bxi$, which will make the subsequent equations 
somewhat simpler. 
The procedure for dealing with the distribution $P[x,y;\bJ]$ is to
first represent it by a finite number of
$\ell$ 
values $P[x_\mu,y_\mu;\bJ]$ (e.g. as a histogram), and take the
limit $\ell\to\infty$ after the limit
$N\to\infty$ has been taken.  
It can be shown that the observables (\ref{eq:drtobservables}) 
satisfy the four conditions for obeying deterministic laws in the
$N\to\infty$ limit if all
$B_i=\order(N^{-\frac{1}{2}})$ 
(demonstrating this is not entirely trivial in the case of the
distribution $P[x,y;\bJ]$). 
Working out the closed equations
(\ref{eq:closedonline},\ref{eq:closedbatch})  for the observables
(\ref{eq:drtobservables}) 
gives the following result:
\be
\frac{d}{dt}Q =
2\eta \int\!dxdy~P[x,y]x\sgn(y)\cF[\sqrt{Q};x,y]
+ \Delta\eta^2\int\!dxdy~ P[x,y]
\cF^2[\sqrt{Q};x,y]
\label{eq:dQdtfurther}
\ee
\be
\frac{d}{dt}R
=\eta  \int\!dxdy~ P[x,y]|y|\cF[\sqrt{Q};x,y] 
\label{eq:dRdtfurther}
\ee
\bd
\frac{\partial }{\partial t}P[x,y] =
-\frac{\eta}{\alpha}\frac{\partial}{\partial x}\left[\sgn(y)
\cF[\sqrt{Q};x,y] P[x,y]\right]
-\eta\frac{\partial}{\partial x}\int\!dx^\prime dy^\prime 
\sgn(y^\prime)\cF[\sqrt{Q};x^\prime,y^\prime] \cA[x,y;x^\prime,y^\prime]
\ed
\be
+\frac{1}{2}\Delta\eta^2\frac{\partial^2}{\partial x^2}\left[P[x,y]
\int\!dx^\prime dy^\prime P[x^\prime,y^\prime]\cF^2[\sqrt{Q};x^\prime,y^\prime]
\right]
\label{eq:dPdtfurther}
\ee
with $\Delta=1$ for on-line learning and $\Delta=0$ for batch
learning. 
Again we observe that the difference between the two modes of learning
is reflected only in the presence/absence of the $\eta^2$ terms in the
dynamic laws. All complications are contained in the function
$\A[x,y;x^\prime,y^\prime]$, which plays the role of a Green's
function, 
and is given by
\bd
\A[x,y;x^\prime,y^\prime]=\lim_{n\to 0}\lim_{N\to\infty}\bigbra
\int\! \prod_{\alpha=1}^n\left\{d\bJ^\alpha 
\delta\left[Q\minus Q[\bJ^\alpha]\right]
\delta\left[R\minus R[\bJ^\alpha]\right]
\prod_{\mu}\delta\left[P[x_\mu,y_\mu]\minus
P[x_\mu,y_\mu;\bJ^\alpha]\right]
\right\}\right.
\ed
\be
\left.
\times
\bra\bra\delta[x\minus\bJ^1\cdot\bxi]\delta[y\minus\bB\cdot\bxi]
(\bxi\cdot\bxi^\prime)[1\minus\delta_{\bxi\bxi^\prime}]
\delta[x^\prime\minus\bJ^1\cdot\bxi^\prime]\delta[y^\prime\minus\bB\cdot\bxi^\prime]
\ket_\Omega\ket_{\Omega}
\room
\bigket_{\rm sets}
\label{eq:closedgreensfunction}
\ee
After a number of manipulations we can perform the average over the
training sets and write  
(\ref{eq:closedgreensfunction}) 
ultimately in the form 
\bd
\A[x,y;x^\prime,y^\prime]
=\int\!\frac{d\hat{x}d\hat{x}^\prime d\hat{y}
d\hat{y}^\prime}{(2\pi)^4}e^{i[x\hat{x}+x^\prime\hat{x}^\prime+y\hat{y}+y\hat{y}^\prime]}
\times 
~~~~~~~~~~~~~~~~~~~~~~~~~~~~~~~~~~~~~~~~~~~~~~~~~~~~~
\ed
\bd
\lim_{n\to 0}
\lim_{N\to\infty}
\int\! d\bq d\hbq d\hbQ d\hbR\! \prod_{\alpha x^\pprime
y^\pprime}d\hat{P}_\alpha(x^\pprime,y^\pprime)~
e^{N\Psi[\bq,\hbq,\hbQ,\hbR,\{\hat{P}\}]}
\cL[\hat{x},\hat{y};\hat{x}^\prime,\hat{y}^\prime;\bq,\hbq,\hbQ,\hbR,\{\hat{P}\}]
\ed
with 
\bd
\Psi[\ldots]=
i\sum_\alpha\hQ_\alpha(1\minus q_{\alpha\alpha})+iR\sum_\alpha \hR_\alpha 
+i\sum_{\alpha\beta}\hat{q}_{\alpha\beta}q_{\alpha\beta}
+i\sum_\alpha\int dx^\pprime
dy^\pprime~\hat{P}_\alpha(x^\pprime,y^\pprime)P[x^\pprime,y^\pprime]
\ed
\be
+~\alpha \log \cD[\bq,\{\hat{P}\}]
+\lim_{N\to\infty}\frac{1}{N}\sum_i
\log \int\!d\bsigma
~
e^{-i\tau_i\sqrt{Q}\sum_\alpha \hR_\alpha\sigma_\alpha
-i\sum_{\alpha\beta}\hat{q}_{\alpha\beta}\sigma_\alpha\sigma_\beta}
\label{eq:finalpsi}
\ee
The functions $\cL[\ldots]$ and  $\cD[\ldots]$ are given by 
complicated 
integrals. 
The term in the expression for $\cA[\ldots]$ involving $\lim_{n\to 0}$
and $\lim_{N\to\infty}$ will be given by the intensive part
$\cal{L}[\ldots]$ 
evaluated in the dominating saddle-point of $\Psi$, and finally  
we get 
\be
\A[x,y;x^\prime,y^\prime] 
=\int\!\frac{d\hat{x}d\hat{x}^\prime d\hat{y}
d\hat{y}^\prime}{(2\pi)^4}e^{i[x\hat{x}+x^\prime\hat{x}^\prime+y\hat{y}+y\hat{y}^\prime]}
\lim_{n\to 0} 
\cL[\hat{x},\hat{y};\hat{x}^\prime,\hat{y}^\prime;\bq,\hbq,\hbQ,\hbR,\{\hat{P}\}]
\label{eq:finalA}
\ee
in which the order parameters $\{\bq,\hbq,\hbQ,\hbR,\{\hat{P}\}\}$ 
are calculated by extremisation of the function $\Psi[\ldots]$ 
(\ref{eq:finalpsi}). 
The meaning of the order parameters $q_{\alpha\beta}$ in the relevant
saddle point at any time $t$ is given in
terms of the (time-dependent) averaged probability distribution $\bra P_t(q)\ket_{\rm
sets}$ for the mutual overlap between the weight vectors $\bJ^a$ and
$\bJ^b$ of two independently evolving
learning processes with the same realisation of the training set
$\set$. One can show (for $N\to\infty$):
\be
\bra P_t(q)\ket_{\rm sets}
=\bigbra \bra\bra \delta\left[q\minus
\frac{\bJ^a\inn\bJ^b}{|\bJ^a||\bJ^b|}\right]
\ket\ket\bigket_{\rm sets}
~~~~~~~~~~
\bra P_t(q)\ket_{\rm sets} =\lim_{n\to 0}\frac{1}{n(n\minus
1)}\sum_{\alpha\neq \beta}\delta[q\minus q_{\alpha\beta}]
\label{eq:physicalmeaning}
\ee 
\vsp

At this stage one usually makes the so-called replica symmetric (RS) 
ansatz in the extremisation 
problem, which in view of (\ref{eq:physicalmeaning}) is equivalent to
assuming the absence of complex ergodicity breaking (simple ergodicity
breaking, i.e. with only a finite number of ergodic components, is
still possible via the existence of multiple solutions for the replica
symmetric saddle-point equations).  This replica symmetric ansatz 
is usually correct in the transient stages of the
dynamics. If with a modest amount of foresight we put
\bd
q_{\alpha\beta}=q_0\delta_{\alpha\beta}+q[1\minus
\delta_{\alpha\beta}],~~~~~
\hat{q}_{\alpha\beta}=\frac{1}{2}i[r\minus r_0\delta_{\alpha\beta}],
~~~~~
\hat{R}_\alpha=i\rho,~~~~~
 \hat{Q}_\alpha=i\phi,~~~~~
\hat{P}_\alpha(u,v)=i\chi[u,v]
\ed
we end up, after a modest amount of algebra and after elimination of
most of the scalar order parameters via the saddle-point equations, 
 with an extremisation problem for a quantity $\Psi_{\rm
RS}$ involving only the function $\{\chi\}$ and the scalar $q$:
\bd
\Psi_{\rm RS}[q,\{\chi\}]=
\frac{1\minus R^2/Q}{2(1\minus q)} -\frac{\alpha}{2(1\minus q)}
+\frac{1}{2}\log(1\minus q) -\int\!dx^\prime dy^\prime~P(x,y)\chi(x,y)
\ed
\be
+\alpha \int\!Dy Dz~\log\int\!dx~e^{-\frac{1}{2}x^2/Q(1\minus
q)+x[Ay\plus Bz]+\frac{1}{\alpha}\chi(x,y)}
\label{eq:RSsaddle}
\ee
with the short-hands $A=R/(1\minus q)Q$ and $B=\sqrt{qQ\minus
R^2}/(1\minus q)Q$ and the short-hand for the Gaussian measure
$Dz=(2\pi)^{-\frac{1}{2}}e^{-\frac{1}{2}z^2}dz$ (similarly for $Dy$).  
This result is surprisingly simple, compared to similar results for
other complex systems of this class (such as spin-glasses and
attractor neural networks near saturation). Firstly, it involves just a small number of 
order parameters to be varied (just $q$ and the function $\chi$). 
Secondly, if one works out the saddle-point equations one recovers 
from the formalism convenient 
relations such as $\int\!dx~P(x,y)=1$ for all $x$ 
(this makes sense: the distribution of $y=\bB\inn\bxi$ is Gaussian
since the components $B_i$ are statistically independent of the
vectors in the training sets). 

The final solution provided by dynamical replica theory 
thus consists of the equations 
(\ref{eq:dQdtfurther},\ref{eq:dRdtfurther},\ref{eq:dPdtfurther}),
which are to be solved numerically, in which at each infinitesimal time-step 
one has to solve the saddle-point problem for
(\ref{eq:RSsaddle}). 
The training- and generalisation errors are then at any time simply given by:
\bd
\bra\bra E_{t}\ket\ket_{\rm sets}=\int\!dxdy~\theta[- xy]P[x,y]
~~~~~~~~~~
\bra\bra E_{g}\ket\ket_{\rm sets}=\frac{1}{\pi}\arccos[R/\sqrt{Q}]
\ed
The need for solving a complicated saddle-point problem at each
infinitesimal  
time-step explains why working out the predictions of
the theory for very large times requires a prohibitively large amount
of CPU time. 
However, the simple form 
of the present saddle-point equations hints at the possibility to introduce 
a more basic distribution $P(x|y,z)$ for which the saddle-point
problem is sufficiently  trivial to allow for analytical solution, 
and which thus obeys a diffusion-type 
equation given in {\em explicit} form. 
The intuition developped in using this formalism for other systems
with a comparable complex dynamics suggests that  the equations
resulting 
from this formalism will be either exact or a reliable approximation,
especially in the transient stages of the learning process.

\pagebreak
\section{Bibliographical Notes}

The application of statistical mechanical tools to learning processes
in artificial neural networks was mainly initiated by the hugely
influential study \cite{gardner}. It is impossible to list even a fraction of the
papers that followed.  Those interested in early applications of
statistical mechanics 
to neural network learning can find their way
into the literature via the dedicated (memorial) issue \cite{memorial} of Journal
of Physics A (mostly on statics) and the early 
review paper \cite{kinzelopper}. The approaches and styles of many subsequent  
statistical mechanical 
studies of learning dynamics were generated by the two
influential papers \cite{seungetal} and \cite{kroghhertz}. 
More recent reviews of the general area of the statistical mechanics
of learning and
generalisation (including both statics and dynamics) 
are \cite{watkinetal,opperkinzel}. 

The on-line learning algorithms studied in section two were
first introduced/studied  
in \cite{rosenblatt} (perceptron rule), 
\cite{vallet} (Hebbian rule) and \cite{anlaufbiehl} (AdaTron rule), 
although 
at the time these algorithms were not yet studied with the methods
described here.  The convenient 
expression for the generalization error of binary perceptrons in
terms of the inner product of the student and teacher
weight vectors $E_g=\frac{1}{\pi}\arccos(\bJ\inn\bB/|\bJ|)$ appeared
first in \cite{gyorgyi,opperetal,kinzelrujan}. In the latter,
\cite{kinzelrujan}, one first finds in an embryonic form the set-up of
deriving closed macroscopic equations for the observables $\bJ^2$ and
$\bJ\inn\bB$. 
Many of the results we described 
on on-line learning with complete training sets 
in perceptrons with fixed rules and fixed learning rates can be found in
\cite{kinouchicaticha,biehlschwarze,biehlriegler,sompolinskyetal}. 
The general lower bound on the generalization error that can be
achieved for a given number of question/answer pairs, translating into the
lower bound $E_g\sim 0.44\ldots t^{-1}$ for on-line learning rules,  
was derived in
\cite{opperhaussler}. Calculations involving on-line rules with 
 time-dependent learning rates
can be found in \cite{barkaietal,sompolinskyetal}. The systematic 
optimisation of learning rules to achieve the fastest decay of the 
generalization error in perceptrons was introduced already in
\cite{kinouchicaticha}. 

In \cite{heskeskappen} one first finds the method to derive exact stochastic
differential equations describing learning dynamics (an application of
\cite{bedeauxetal}), followed by several studies aimed at extracting 
information from the microscopic dynamics directly\footnote{This line
of research, termed `stochastic approximation theory', is sometimes
presented as opposite to the approach based on 
deriving macroscopic equations, with only little scientific
justification. The two approaches are  
mutually consistent and  complementary; they simply concentrate on 
different levels of description and are (sometimes) worked out  
in different limits.  In the
present paper we used both, and switched from one to another
whenever necessary.}. 
The differences between batch learning and on-line learning appear so
far to have been addressed mainly in equilibrium calculations 
\cite{kinouchicaticha95,vandenbroeckreimann,opper}. 
There is not yet much literature on learning 
dynamics with incomplete training sets, apart from simple cases and
linear models such as \cite{kroghhertz}.  
The generating function approach to the learning
dynamics for incomplete training sets was elaborated for perceptrons
with binary weights in
\cite{horner}. The version of the dynamical replica theory
calculations described in section five was developed in \cite{LCS},
and is
only now being applied to learning dynamics \cite{coolensaad}. 
\vsp

Finally, even within the already confined area of statistical mechanical
studies of the dynamics of learning we have specialised to the
simplest models (binary perceptrons) and the simplest types of tasks (those
generated by a noise-free and realisable teacher). As a result there
are many interesting areas which we had to leave out, such as e.g. 
the dynamics of learning in the presence of noise, for unsupervised 
learning rules or for 
non-stationary teachers
\cite{biehlschwarze,kinouchicaticha93}. 
The most important areas we were forced to leave out, however, are  
the large bodies of work done 
on different classes of learning rules, e.g. those involving continuous rather 
than binary neurons or those in the form of microscopic Fokker-Planck 
equations, as well as (and especially) on various families of  
multilayer networks (mostly so-called committee machines, in which
the weights connecting the hidden layer to the output neuron(s) are
fixed). 
Relevant recent papers in these areas are 
e.g. \cite{biehlschwarze95,saadsolla1,saadsolla2,copellicaticha,kimsompolinsky,simonetticaticha,biehletal96}. 

The techniques described in this review can be applied with only minor
adjustments and extensions to layered networks of graded response
neurons,
provided the number of neurons in the hidden layer(s) remains finite
in the limit $N\to\infty$. As soon as we move to layered networks in
which the number of hidden neurons scales proportional to $N$, on the
other hand,
we again face the problem of macroscopic dynamic equations which fail to
close. Solving this problem, and the one of handling 
incomplete training sets, are the key objectives of most present-day
research efforts in the research area of the statistical mechanics of
learning. 

\subsection*{Acknowledgement}
It is our pleasure to thank Dr D Saad for many interesting discussions
on the dynamics of learning.


\pagebreak
\appendix
\section{Appendix: Integrals}

In this appendix we give brief derivations of those integrals
encountered throughout this paper that turn out to be easy, and give
the appropriate reference for finding the nasty ones. 
All involve the following Gaussian
distribution:
\[
\bra f(x,y)\ket=\int\!dxdy~f(x,y)P(x,y) ~~~~~~~~~~~~
P(x,y)=\frac{1}{2\pi\sqrt{1\minus\omega^2}}e^{-\frac{1}{2}[x^2+y^2-2xy\omega]/(1-\omega^2)}
\]
\begin{description}
\item[I:] $I_1=\langle|y|\rangle$\\
\bd
I_1=\int\! \frac{dy}{\sqrt{2\pi}}e^{-\frac{1}{2}y^2}|y|
=\sqrt{\frac{2}{\pi}}
\ed
\item[II:] $I_2=\langle x\,\sgn(y)\rangle$\\
\bd
I_2=-\int\!\frac{dxdy}{2\pi \sqrt{1\minus
\omega^2}}\sgn(y)e^{-\frac{1}{2}[y^2-2\omega xy]/(1-\omega^2)}(1\minus\omega^2)\frac{\partial}{\partial x}e^{-\frac{1}{2}x^2/(1-\omega^2)}
\ed
\bd
=\int\!\frac{dxdy}{2\pi\sqrt{1\minus
\omega^2}}e^{-\frac{1}{2}[x^2+y^2-2\omega
xy]/(1-\omega^2)}\sgn(y)\omega y
\ed
\bd
=\omega\langle|y|\rangle=\omega\sqrt{\frac{2}{\pi}}
\ed
\item[III:] $I_3=\bra \theta[-xy]\ket$\\  
\bd
I_3=\int_0^\infty\!\!\int_0^\infty\!\!\frac{dxdy}{\pi\sqrt{1-\omega^2}}e^{-\frac{1}{2}[x^2+y^2+2\omega xy]/(1-\omega^2)}
=\frac{\sqrt{1\minus\omega^2}}{\pi}\int_0^\infty\!\!\int_0^\infty\!dxdy\,e^{-\frac{1}{2}[x^2+y^2+2\omega xy]} 
\ed
Introduce polar coordinates $(x,y)=r(\cos\phi,\sin\phi)$:
\bd
I_3=\frac{\sqrt{1\minus\omega^2}}{\pi}\int_0^{\pi
/2}d\phi\int_0^\infty dr\,re^{-\frac{1}{2}r^2 [1+\omega\sin(2\phi)]}
\ed
\bd
=\frac{\sqrt{1\minus\omega^2}}{2\pi}\int_0^\pi \frac{d\phi}{1\plus \omega\sin(\phi)}
=\frac{1}{\pi}\left[\frac{\pi}{2}\minus\arctan\left(\frac{\omega}{\sqrt{1\minus\omega^2}}\right)\right]
\ed
(the last integral can be found in \cite{GR}). 
Finally, using $\cos[\frac{\pi}{2}-\psi]=\sin\,\psi$, we find
\bd
I_3=\frac{1}{\pi}\arccos(\omega)
\ed
\item[IV:] $I_4=\langle x\,\sgn(y)\,\theta[-xy]\rangle$\\
\bd
I_4
=-\frac{1\minus\omega^2}{\pi}\int_0^\infty\!dx\,x\,e^{-\frac{1}{2}x^2}\int_0^\infty\!dy\,e^{-\frac{1}{2}[y+\omega
x]^2+\frac{1}{2}\omega^2x^2}
\ed
\bd
=\frac{1}{\pi}\left[e^{-\frac{1}{2}x^2}\int_{\omega
x/\sqrt{1-\omega^2}}^\infty
dy\,e^{-\frac{1}{2}y^2}\right]_0^\infty-\frac{1}{\pi}\int_0^\infty\!dx\,e^{-\frac{1}{2}y^2}\frac{\partial}{\partial x}\int_{\omega x/\sqrt{1-\omega^2}}^\infty dy\,e^{-\frac{1}{2}y^2}
\ed
\bd
=-\frac{1}{\sqrt{2\pi}}+\frac{\omega}{\pi\sqrt{1-\omega^2}}\int_0^\infty \!dx\,e^{-\frac{1}{2}x^2/(1-\omega^2)}=\frac{\omega-1}{\sqrt{2\pi}}
\ed
\item[V:] $I_5=\langle|y|\theta[-xy]\rangle$\\
\bd
I_5=\int_0^\infty\!\!\int_0^\infty\!dxdy~y[P(x,-y)+P(-x,y)]
=\frac{1-\omega}{\sqrt{2\pi}}
\ed
\item[VI:] $I_6=\langle x^2\theta[-xy]\rangle$\\
\bd
I_6=\frac{1}{\pi\sqrt{1\minus \omega^2}}\int_0^\infty\int_0^\infty
dxdy\,x^2\,e^{-\frac{1}{2}[x^2+y^2+2xy\omega]/(1-\omega^2)}
\ed
\bd
=\frac{1}{2\pi\sqrt{1\minus\omega^2}}\int_0^\infty\int_0^\infty dxdy\,(x^2+y^2)\,e^{-\frac{1}{2}[x^2+y^2+2xy\omega]/(1-\omega^2)}
\ed
We switch to polar coordinates $(x,y)=r(\cos\theta,\sin\theta)$, and
subsequently substitute
$t=r^2[1\plus\omega\sin(2\theta)]/[1\minus\omega^2]$:
\bd
I_6=\frac{1}{2\pi\sqrt{1\minus\omega^2}}\int_0^{\pi/2}d\theta\int_0^\infty dr\,r^3\,e^{-\frac{1}{2}[r^2+2\omega r^2\cos\theta\,\sin\theta]/(1-\omega^2)}
\ed
\bd
=\frac{(1\minus\omega^2)^{3/2}}{4\pi}\int_0^{\pi/2}\frac{d\theta}{(1+\omega\sin(2\theta))^2}\int_0^\infty dt\,t\,e^{-\frac{1}{2}t}
\ed
\bd
=\frac{(1-\omega^2)^{3/2}}{2\pi}\int_0^\pi \frac{d\phi}{(1+\omega\,\sin\phi)^2}
\ed
To calculate the latter integral we define
\[
\tilde I_n=\int_0^\pi\frac{d\phi}{(1+\omega\,\sin\phi)^n}
\]
These integrals obey
\bd
\omega\frac{d}{d\omega}\tilde I_n-n\tilde I_{n+1}=-n\tilde I_n
\ed
so 
\[
\tilde I_2=\tilde I_1+\omega\frac{d}{d\omega}\tilde I_1
~~~~~~~~~~
\tilde I_1=\frac{2}{\sqrt{1-\omega^2}}\arccos(\omega)
\]
(where we used the integral already encountered in {\bf III}). We now find
\bd
I_6=\frac{(1-\omega^2)^{3/2}}{2\pi}\tilde{I}_2=
\frac{(1\minus\omega^2)}{\pi}\arccos(\omega)-\frac{\omega \sqrt{1\minus\omega^2}}{\pi}+\frac{\omega^2}{\pi}\arccos(\omega)
\ed
\item[VII:] $I_7=\langle|x|\,|y|\,\theta[-xy]\rangle$\\
\bd
I_7=
\int_0^\infty\int_0^\infty \frac{dxdy}{\pi\sqrt{1-\omega^2}}xy\,e^{-\frac{1}{2}[x^2+y^2+2xy\omega]/(1-\omega^2)}
\ed
We use the relation 
\bd
xe^{-\frac{1}{2}[x^2+y^2+2xy\omega]/(1-\omega^2)}=-(1\minus\omega^2)\frac{\partial}{\partial x}e^{-\frac{1}{2}[x^2+y^2+2xy\omega]/(1-\omega^2)}-\omega ye^{-\frac{1}{2}[x^2+y^2+2xy\omega]/(1-\omega^2)}
\ed
to give us, using {\bf VI}:
\bd
I_7=
\frac{\sqrt{1\minus\omega^2}}{\pi}\int_0^\infty\!dy\,y\,e^{-\frac{1}{2}\frac{y^2}{(1-\omega^2)}}-\omega\langle y^2\theta[-xy]\rangle
\ed
\bd
=\frac{(1\minus\omega^2)^{3/2}}{\pi}-\frac{\omega(1\minus\omega^2)}{\pi}\arccos(\omega)+\frac{\omega^2\sqrt{1\minus\omega^2}}{\pi}-\frac{\omega^3}{\pi}\arccos(\omega)
\ed
\item[VIII:] $I_8(x)=\int\!dy~\theta[y]P(x,y)$\\
\bd
I_8(x)=\int_0^\infty\!\frac{dy}{2\pi\sqrt{1\minus \omega^2}}
e^{-\frac{1}{2}[x^2+y^2-2xy\omega]/(1-\omega^2)}
\ed
\bd
=\frac{e^{-\frac{1}{2}x^2}}{2\pi\sqrt{1\minus\omega^2}}\int_0^\infty
\!dy~e^{-\frac{1}{2}[y-\omega x]^2/(1-\omega^2)}
=
\frac{e^{-\frac{1}{2}x^2}}{2\sqrt{2\pi}}\left[1\plus
\erf\left[\frac{\omega
x}{\sqrt{2}\sqrt{1\minus\omega^2}}\right]\right]
\ed
\item[IX:] $I_9(x)=\int\!dy~\theta[y](y\minus \omega x)P(x,y)$\\
\bd
I_9(x)=-\frac{\sqrt{1\minus\omega^2}}{2\pi}
\int_0^\infty\!dy~\frac{\partial}{\partial
y} e^{-\frac{1}{2}[x^2+y^2-2xy\omega]/(1-\omega^2)}
=\frac{\sqrt{1\minus\omega^2}}{2\pi} e^{-\frac{1}{2}x^2/(1-\omega^2)}
\ed
\end{description}

\end{document}